\documentclass[aps,pra,twocolumn,superscriptaddress, nofootinbib]{revtex4-1}
\usepackage{graphicx} % needed for figs
\usepackage{dcolumn}  % needed for some tables
\usepackage{bm}    % for math
\usepackage{amssymb}  % for math
\usepackage{amsfonts, amsmath, amssymb, mathtools}
\usepackage{amsmath}
\DeclareMathOperator*{\argmax}{arg\,max}
\usepackage{lipsum}
\usepackage[normalem]{ulem}
\usepackage{commath}
\usepackage{xcolor}

\emergencystretch=\maxdimen
\usepackage{scalerel}[2016-12-29]
\def\scaleint#1{\vcenter{\hbox{\scaleto[3ex]{\displaystyle\int}{#1}}}}
\def\bs{\mkern-12mu} % set amount of backspacing for lower limit of integration

\usepackage[binary-units=true]{siunitx}
\usepackage{hyperref}
\usepackage{cleveref}
\RequirePackage{textcase}
\DeclareUnicodeCharacter{2009}{\,} 

\begin{document}

\newcommand{\condProb}{\operatorname{P}\expectarg}
\DeclarePairedDelimiterX{\expectarg}[1]{(}{)}{%
 \ifnum\currentgrouptype=16 \else\begingroup\fi
 \activatebar#1
 \ifnum\currentgrouptype=16 \else\endgroup\fi
}

\newcommand{\innermid}{\nonscript\;\delimsize\vert\nonscript\;}
\newcommand{\activatebar}{%
 \begingroup\lccode`\~=`\|
 \lowercase{\endgroup\let~}\innermid
 \mathcode`|=\string"8000
}

\newcommand{\todo}[1]{{\color[rgb]{1.0,0.0,0.0}#1}} % gaps to be filled in red
\newcommand{\yg}[1]{{\color[rgb]{0.0,0.0,1.0}#1}} % Yvonne Comments in blue
\newcommand{\cw}[1]{{\color[rgb]{0.2,0.7,0.2}#1}} % CW comments in green
\newcommand{\mar}[1]{{\color[rgb]{0.8,0.4,0.2}#1}} % MAR comments in orange
\newcommand{\st}[1]{{\color[rgb]{0.5,0.1,0.6}#1}} % ST comments in purple
\newcommand{\cut}[1]{\sout{#1}} % Use this line to show deleted text as struck out
\newcommand{\new}[1]{{\color[rgb]{.0,0.0,1.0}#1}} % gaps to be filled in red
\newcommand{\discuss}[1]{{\color[rgb]{0.8,0.0,0.8}#1}} % Use this line to show deleted text as struck out

\newcommand{\err}{\epsilon} % Error generic
\newcommand{\bra}[1]{\left<#1\right|}
\newcommand{\ket}[1]{\left|#1\right>}
\newcommand{\bket}[2]{\left<#1~|~#2\right>}
\newcommand{\tr}[1]{\text{Tr}\left(#1\right)}
\newcommand{\kket}[1]{\left|\left|#1\right>\right>}
\newcommand{\bbra}[1]{\left<\left<#1\right|\right|}
\newcommand{\ba}{\boldsymbol{a}}
\newcommand{\bb}{\boldsymbol{b}}
\newcommand{\bc}{\boldsymbol{c}}
\newcommand{\bd}{\boldsymbol{d}}
\newcommand{\bh}{\boldsymbol{h}}
\newcommand{\bq}{\boldsymbol{q}}
\newcommand{\bp}{\boldsymbol{p}}
\newcommand{\bQ}{\boldsymbol{Q}}
\newcommand{\bP}{\boldsymbol{P}}
\newcommand{\bE}{\boldsymbol{E}}
\newcommand{\mE}{\mathcal{E}}
\newcommand{\Tr}{\text{Tr}}
\renewcommand{\Re}{\text{Re}}
\renewcommand{\Im}{\text{Im}}
\newcommand{\ta}{\tilde{\alpha}}
\newcommand{\bO}{\boldsymbol{\mathcal{O}}}
\newcommand{\br}{\boldsymbol{r}}
\newcommand{\bR}{\boldsymbol{R}}
\newcommand{\bK}{\boldsymbol{K}}
\newcommand{\bJ}{\boldsymbol{J}}
\newcommand{\bH}{\boldsymbol{H}}
\newcommand{\bU}{\boldsymbol{U}}
\newcommand{\bM}{\boldsymbol{M}}
\newcommand{\bX}{\boldsymbol{X}}
\newcommand{\bZ}{\boldsymbol{Z}}
\newcommand{\bY}{\boldsymbol{Y}}
\newcommand{\bI}{\boldsymbol{I}}
\newcommand{\bL}{\boldsymbol{L}}
\newcommand{\bT}{\boldsymbol{T}}
\newcommand{\bD}{\boldsymbol{D}}
\newcommand{\bn}{\boldsymbol{n}}
\newcommand{\bS}{\boldsymbol{S}}
\newcommand{\bsigma}{\boldsymbol{\sigma}}
\newcommand{\bSigma}{\boldsymbol{\Sigma}}
\newcommand{\bDelta}{\boldsymbol{\Delta}}
\newcommand{\bPi}{\boldsymbol{\Pi}}
\newcommand{\red}[1]{\textcolor{red}{#1}}
\newcommand{\green}[1]{\textcolor{green}{#1}}
\newcommand{\blue}[1]{\textcolor{blue}{#1}}
\newcommand{\bphi}{\boldsymbol{\varphi}}
\newcommand{\NN}{\mathcal N}

\raggedbottom
\renewcommand{\thesubsection}{\thesection.\arabic{subsection}}
\renewcommand{\thesubsubsection}{\thesubsection.\arabic{subsubsection}}

\title{A practical guide for building superconducting quantum devices}
\author{Yvonne Y. Gao}
\email[Corresponding author: ]{yvonne.gao@nus.edu.sg}
\affiliation{Centre for Quantum Technologies, National University of Singapore}
\author{M. Adriaan Rol}
\affiliation{Orange Quantum Systems, Vermeerstraat 19, 2612 XJ Delft, The Netherlands}
\author{Steven Touzard}
\affiliation{Centre for Quantum Technologies, National University of Singapore}
\author{Chen Wang}
\affiliation{Department of Physics, University of Massachusetts-Amherst, Amherst, MA, 01003, USA}
\date{\today}

\begin{abstract}
Quantum computing offers a powerful new paradigm of information processing that has the potential to transform a wide range of industries. In the pursuit of the tantalizing promises of a universal quantum computer, a multitude of new knowledge and expertise has been developed, enabling the construction of novel quantum algorithms as well as increasingly robust quantum hardware. In particular, we have witnessed rapid progress in the circuit quantum electrodynamics (cQED) technology, which has emerged as one of the most promising physical systems that is capable of addressing the key challenges in realizing full-stack quantum computing on a large scale. In this article, we present some of the most crucial building blocks developed by the cQED community in recent years and a pr\'{e}cis of the latest achievements towards robust universal quantum computation. More importantly, we aim to provide a synoptic outline of the core techniques that underlie most cQED experiments and offer a practical guide for a novice experimentalist to design, construct, and characterize their first quantum device. \\ \\ 
\end{abstract}

%\pacs{}
\maketitle
\tableofcontents

\section{Introduction}
Over the past decade, quantum computing and quantum information science have gathered tremendous momentum from both academic and industrial research endeavors. With this comes the relentless progress in both the theoretical and experimental front, transforming quantum computing from a mere mathematical curiosity to a rapidly advancing domain of innovation. Remarkable improvements in the creation~\cite{vlastakis2013_deterministically,wang2015_schrodinger, asavanant2019_generation, larsen2019_deterministic,ma2019_dissipatively,gong2019_genuine, sagastizabal2020_variational, ma2020_manipulating}, control~\cite{asaad2020_coherent, figgatt2019_parallel, rol2019_fast, paik2016_experimental, mckay2016_universal, reagor2018_demonstration}, and measurement~\cite{touzard2019_gated, heinsoo2018_rapid, mckay2019_three, erhard2019_characterizing, wright2019_benchmarking} of quantum systems and their entanglement properties have been demonstrated in various different physics systems. Amongst them, the circuit quantum electrodynamics (cQED) technology~\cite{blais2004_cavity, wallraff2004} has emerged as one of the most promising platforms for realizing a robust and scalable universal quantum computer. It has enabled many of the landmark achievements in quantum computing, such as first realizations of quantum error correction~\cite{reed2012_realization, Riste15, kelly2015_state, Corcoles15,ofek2016_extending, hu2019_quantum, campagne-ibarcq2020_quantum, andersen2020_repeated, bultink2020_protecting, marques2021logicalqubit, google_quantum_ai_exponential_2021} and demonstrating the potential advantage of a quantum processor over its classical counterparts~\cite{arute2019_quantum}. 

In a nutshell, cQED describes the interaction of light, typically at microwave frequencies, and matter, composed of superconducting circuit elements. The characteristics of all components in cQED devices are highly configurable. They can be engineered on demand to provide both large, controllable non-linearities for fast quantum operations as well as isolation from the environment for robust quantum coherence in a single hardware. The construction of cQED quantum processors requires both careful considerations in the design and fabrication processes of the device, as well as continual optimization of the measurement and control setup. The key building blocks for realizing a robust full-stack cQED system are summarized in Fig.~\ref{fig:overview}(a). In order to successfully construct a large-scale universal quantum computer, every element in this stack must be constructed with finesse and constantly enhanced through research and innovation. 

As the cQED technology matures, the implementation of each of these elements is becoming increasingly sophisticated and multi-disciplinary. There are many superb review articles that capture the latest advances in key aspects of this technology~\cite{devoret2013_superconducting, wendin2017_quantum, kjaergaard2020_superconducting, blais2020_quantum, blais2020_circuit} geared towards the experts in the field. They offer holistic overviews of the main concepts and key results in this rapidly developing field of research. In contrast, this article aims to provide an introductory tutorial for experimentalists at the early stages of their venture in developing superconducting quantum systems. Compared with the pedagogical review of the cQED knowledge base in Ref.~\cite{krantz2019_engineer}, here we focus on the crucial experimental techniques and practices involved in successfully constructing quantum devices using superconducting circuits. This tutorial aims to provide a useful vantage point for a novice experimentalist to gain practical insights into the various elements required to implement a cQED experiment. 

\subsection{Overview of article}
This article is structured to mimic the full workflow of building a new cQED experimental setup in the laboratory. Each step plays a crucial role in ensuring the eventual realization of a robust quantum device and requires a multitude of careful considerations, which are often skipped over in research articles. In this tutorial, we will aim to provide a step-by-step guide on these practical details, with a specific focus on devices that employ transmon qubits~\cite{koch2007_charge}. 

It is important to highlight that this article is not intended as a comprehensive review of the entire field of cQED or as an exhaustive summary of the wide range of techniques that have been developed by the community. Rather, it aims to discuss a selected set of basic tools and useful intuitions that, in the opinion of the authors, are important for understanding the practical aspects of implementing cQED experiments. In particular, we aim to offer a tutorial style walk-through on the design and characterization of cQED devices while providing more general overviews and useful references on the other aspects of the workflow.

\begin{figure*}
\includegraphics{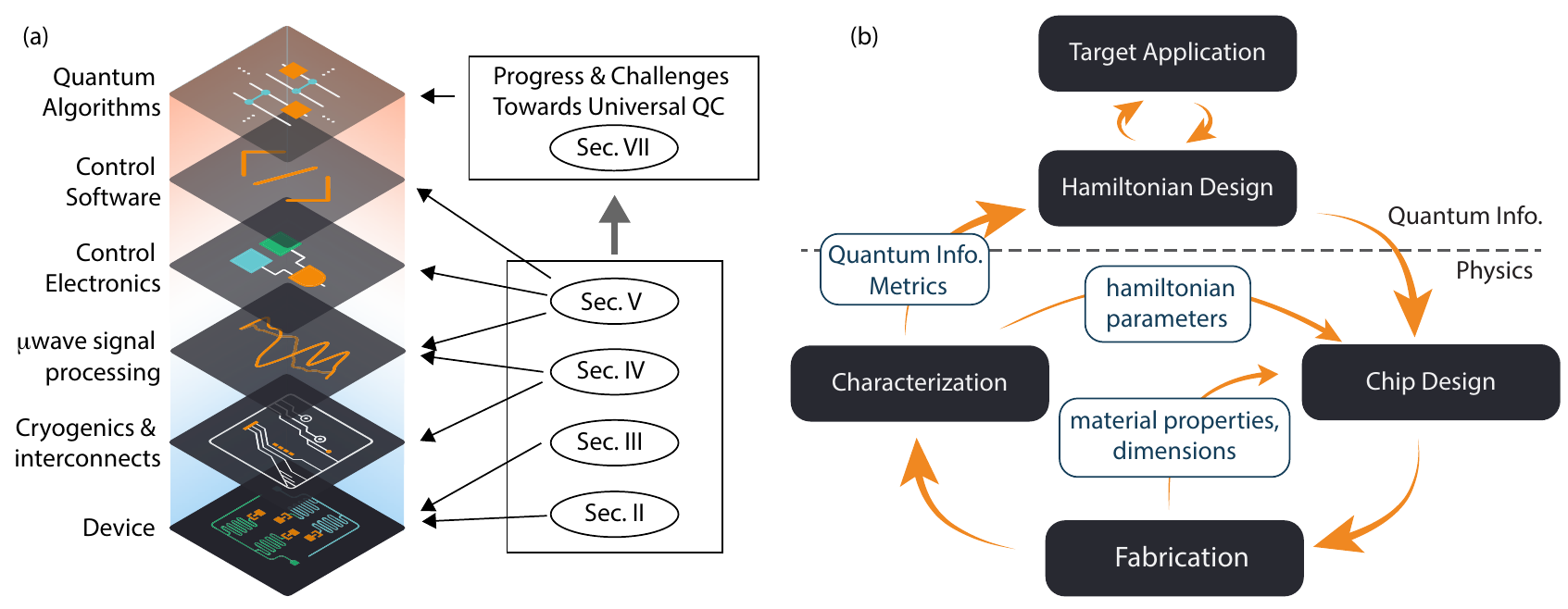}
\caption{
\label{fig:overview}
\textbf{Overview of the crucial hardware building blocks and experimental processes in realizing a quantum system in cQED}. (a) The full-stack of cQED quantum computer. Each layer represents an important element that must be carefully constructed and optimized, in order to construct full-stack quantum processors in cQED. (b) The engineering cycle of a cQED device. The desired target application influences the chip’s Hamiltonian design (\cref{sec:hamiltonians}) in which an equivalent circuit and its target parameters are determined. This design serves as the input for a second design step in which the geometry and layout of the device is determined, taking into account the constraints of the fabrication process (\ref{sec:designfab}). After fabricating the device, the system is carefully characterized (\ref{sec:interacting_with_quantum_circ}) and the resulting information is used to inform the next design iteration to bring the system closer towards the desired Hamiltonian. 
}
\end{figure*}

We start by providing a concise summary of the physics of cQED in Sec.~\ref{sec:hamiltonians}. In addition, we will also present the process of constructing specific Hamiltonians with the fundamental ingredients afforded by the cQED platform. The purpose of this segment is to provide a general basis for the subsequent discussions and provide useful resources for more advanced readers to further investigate the more technically-involved concepts.

In Sec.~\ref{sec:designfab}, we introduce the necessary ingredients and tools for translating the target Hamiltonians into a physical circuit. This requires an iteration cycle consisting of configuring the device layout and circuit components, simulating the electromagnetic mode structures, and extracting of the relevant Hamiltonian parameters. We will further highlight some of the main limiting factors of the coherence of quantum circuits and summarize the known design strategies to mitigate them. This is followed by a short discussion of device fabrication as well as considerations towards achieving better yield and reliability.

In Sec.~\ref{sec:environment}, we move on to provide an overview of the configuration of cryogenic and room temperature environments required to ensure both the coherence and controllability of cQED devices. Here, we will highlight some of the crucial factors involved in the filtering and shielding of the quantum devices in the dilution refrigerator (DR). Furthermore, we will also provide an overview of the typical room temperature microwave signal processing elements and control electronics used in providing sophisticated pulse sequences for quantum operations.

In Sec.~\ref{sec:interacting_with_quantum_circ}, we will present the typical workflow of characterizing and calibrating a cQED device. Here, we cover a series of basic tune-up procedures, from cavity and qubit spectroscopy, coherence measurements, to qubit rotation diagnosis and readout optimization. The information we learn through these measurements offer a comprehensive view of the performance of individual quantum elements, and provide valuable feedback for future iterations of device designs. We then discuss a few more advanced topics including implementing two-qubit gates, characterization of quantum operations, and control of quantum memory cavities, which are crucial in building towards  high-fidelity experiments in multi-qubit systems. 

The building blocks introduced thus far form an iterative cycle (Fig.~\ref{fig:overview}(b)) that enables us to develop cQED devices increasingly robust performance for the desired applications. Collectively, they culminate in the ultimate goal of realizing a robust universal quantum computer capable of tackling real-world challenges. In this tutorial, we do not attempt give a prescription of how to realize and operate a large-scale quantum computer since that remains an open pursuit of the entire community. Instead, in Sec.~\ref{sec:qc}, we will switch gears to provide a brief review of two concurrent research thrusts that aim to develop more robust and powerful quantum devices. First, we discuss the recent developments and main challenges in scaling up planar cQED devices from NISQ era processors to a robust, universal quantum machine. Subsequently, we will illustrate how superconducting cavities can be utilized to realize robust quantum modules with first order quantum error correction. Finally, we will provide some prospective on the challenges and future directions towards using them as building blocks for a fault-tolerant and general-purpose quantum computer.

%%%%%%%%%%%%%%%%%%%%%%%%%%%%%%%%%%%%%%%%%%%%%%%%%%%%%%%%%%%%%%%%%%%%%%%%%%%%%%%%%%%%%%%%%%%%

\section{Physics of cQED}\label{sec:hamiltonians}
The cQED framework describes the dynamics of photon-matter interaction, where devices consisting of engineered quantum circuit elements interact with microwave photons. The first step towards setting up a successful cQED experiment %and successfully achieving the intended outcomes 
is to establish a solid grasp of the fundamental concepts of the main building blocks of the superconducting quantum circuit. There are many excellent materials detailing the general principles of cQED systems~\cite{Girvin12, thesisSchuster07, devoret2004_superconducting, devoret2004_implementing, langford2013_circuit}. In this section, we focus specifically on two key circuit elements, %namely, 
superconducting resonators and transmon qubits, which are the main workhorses for current cQED experiments. We aim to summarize the physical principles governing their quantum dynamics and bring some intuition on how to effectively engineer quantum devices with them.

\subsection{Superconducting resonators}
The simple harmonic oscillator is often used as the basis for formulating the quantum mechanical models of more complex systems. In the cQED framework, the physical manifestation of a simple harmonic oscillator is an LC circuit, which consists of one inductive and one capacitive element. This simple resonant circuit with frequency $\omega$ can be coerced to exhibit fully quantum mechanical behavior if constructed with superconducting materials (with critical temperature $T_c$) and kept at sufficiently low temperature ($T\ll T_c$). This allows such superconducting LC oscillators to achieve lossless conduction, as well as effective suppression of thermal noise ($k T \ll \hbar \omega$, where $k$ is the Boltzman constant). This simple quantum element plays many crucial roles in cQED devices, from readout and mediating quantum interactions to the coherent storage of quantum information. 

\subsubsection{Isolated superconducting resonators}

We often model an isolated LC oscillator as a lumped-element system consisting of an inductor $L$ and a capacitor $C$, as shown in Fig.~\ref{fig:oscillator}(a). However, these notions can be extended beyond lumped element circuits. More generally, the inductive energy stems from charges moving in a conductor and creating a magnetic field which tends to oppose any changes in current. The capacitive energy, on the other hand, arises from the electrostatic energy between two regions of a superconductor that contain different number of charges. A superconducting resonator is simply a collection of such quantum LC oscillators, with each mode of the resonator corresponding to a single oscillator with a well-defined resonance frequency. 

We can capture the dynamics of superconducting resonators with two dimensionless quantities. The first one, $\bphi$, is a result of the quantization of the magnetic flux threaded through the inductor. The total magnetic flux observable is then $\boldsymbol{\phi} = \phi_0\,\bphi$, where $\phi_0 = \hbar/2e$ is the reduced magnetic flux quantum. The second one, $\bn$, is the difference in the number of charges on the two plates of the capacitor in units $2e$, which correspond to the charge of a Cooper pair. It is related to the total charge observable, which is given by $\bn_Q = 2e\bn$. Whlie valuable insight is gained from this lumped-element model, microwave resonators are generally implemented using distributed elements in practice. For such systems, $\bn$ relates to the quantization of the electric energy, caused by an unbalanced charge distribution across different parts of the structure (see Fig.~3(d)). The oscillations of this charge distribution induces a magnetic field flux quantized with $\bphi$, as predicted by Faraday's law.

%\sout{Although microwave LC oscillators are implemented using distributed elements in practice, we can still gain valuable insight about their behavior through the simple lumped-element model. For such systems, $\bn$ relates to the number of Cooper pairs at each point of the distributed element. The oscillations of this charge distribution induces a magnetic field flux, $\bphi$, as predicted by Faraday's law. }}

The two dimensionless quantities, $\bn$ and $\bphi$, obey the canonical commutation relation~\cite{Vool2017_introduction}
\begin{align}\label{eq:phi_n}
  [\bphi, \bn] = i.
\end{align}
With this, the Hamiltonian of a superconducting LC oscillator is given by
\begin{align}
  \label{eq:H_LC}
  \bH = \frac{E_L}{2}\bphi^2 + 4E_C\bn^2,
\end{align}
where $E_L = \frac{\phi_0^2}{L} = \frac{\hbar^2}{4e^2 L}$ and $E_C = \frac{e^2}{2C}$~\cite{manucharyan2009_fluxonium, Vool2017_introduction}. 

Alternatively, we can also express this Hamiltonian using two standard circuit parameters: the angular frequency $\omega$ and the impedance $Z$, defined as
\begin{align}
  \omega = \frac{\sqrt{8 E_L E_C}}{\hbar} = \frac{1}{\sqrt{L C}}, \hspace{1cm} Z_0 = \sqrt{\frac{L}{C}},
\end{align}
such that
\begin{align}
  \bH = \hbar\omega\left[\left(\frac{R_Q}{2Z_0}\right)\bphi^2 + \left(\frac{Z_0}{2R_Q}\right)\bn^2\right].
\end{align}
Here, we have introduced the reduced resistance quantum for superconductors $R_Q~=~\frac{\hbar}{(2e)^2} = $1.027~k$\Omega$, and we can write $R_Q/Z_0 = \sqrt{E_L/8E_C}$

Next, we diagonalize this Hamiltonian by finding ladder operators $\ba$ and $\ba^\dag$ that are linear combinations of $\bphi$ and $\bn$ with $[\ba, \ba^\dag]=1$ . We find
\begin{align}
    \label{eq:annihilation}
  \ba = \sqrt{\frac{R_Q}{2Z_0}}\bphi + i\sqrt{\frac{Z_0}{2R_Q}}\bn
\end{align}
From this, we obtain the flux $\bphi$ and charge $\bn$ observables explicitly, in their respective units and get a physical sense of the zero point fluctuations of the magnetic flux and of the charge by writing $\bphi = \varphi^{\text{ZPF}}(\ba + \ba^\dag)$ and $\bn = n^{\text{ZPF}}(\ba-\ba^\dag)/i$, where
\begin{align}
  \label{eq:phi_ZPF}
  \varphi^{\rm ZPF} &= \sqrt{\frac{Z_0}{2R_Q}}\, \\
  \label{eq:n_ZPF}
  n^{\rm ZPF} &= \sqrt{\frac{R_Q}{2Z_0}}.
\end{align}
We deduce that when the impedance of the resonator is high ($Z_0 \gg R_Q$), the zero point fluctuations of the magnetic flux are large compared to $\phi_0$ and that of charge are small compared to $2e$. The situation is reversed when the impedance is low relative to the flux quantum.

\subsubsection{Resonators in contact with the environment}

\begin{figure}[!htb]
\includegraphics[scale=1]{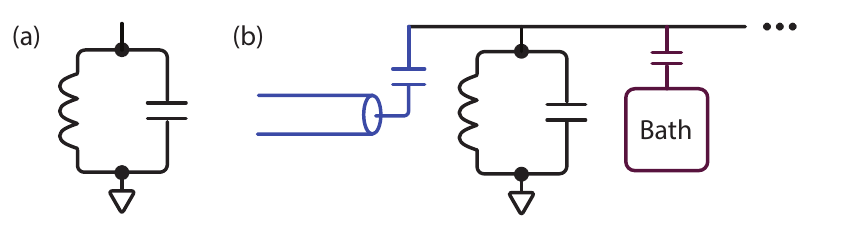}
\caption{\label{fig:oscillator} \textbf{Circuit model of a superconducting resonator.} (a) Lumped element representation of an isolated LC circuit. (b) Circuit of a LC oscillator coupled to a transmission line, which carries the desired microwave probe signals, and a bath, which signifies the presence of other undesired environment couplings that can potentially be noisy.
}
\end{figure}

For any practical applications, the resonator will be in contact with an external environment. This environment consists of both carefully engineered coupling introduced by the observer and undesired interactions with a potentially dissipative bath, as illustrated in Fig.~\ref{fig:oscillator}(b). In this case, we can no longer describe the observables associated with the coupled resonator-environment in the simple Heisenberg picture. Rather, it must be treated as an open system, where the field $\ba(t)$ associated with the resonator receives an incoming field $\ba_{in}(t)$ and emits an outgoing field $\ba_{out}(t)$. Let us consider the example where a resonator is coupled to a microwave transmission line, which provides the drive and measurement tones to the system. Here, $\ba_{in}(t)$ and $\ba_{out}(t)$ represent respectively the incoming and outgoing field of the transmission line where it interacts with our circuit. The fields at different times are not related, such that $\left[\ba_{out}(t), \ba^\dag_{out}(t')\right] = \left[\ba_{in}(t), \ba^\dag_{in}(t')\right] = \delta(t - t')$. This implies $\ba_{in}$ and $\ba_{out}$ have dimension $t^{-1/2}$).

A detailed balance of the field results in the following input-output relation:
\begin{align}\label{eq:in_out}
  \ba_{out} = \ba_{in} + \sqrt{\kappa_c}\ba,
\end{align}
where $\kappa_c$ is defined as the frequency-independent coupling rate at which the oscillator exchanges energy with the transmission line, and can be experimentally characterized for each set-up. Here, we have chosen the sign convention following the approach in Ref.~\cite{steck2020_quantum}. With the incoming and outgoing fields taken into account, we arrive at the following differential equation for $\ba(t)$ in the Heisenberg picture:
\begin{align}
  \label{eq:quantum_langevin}
  \partial_t \ba = - \frac{i}{\hbar}\left[\ba, \bH\right] - \frac{\kappa}{2}\ba - \sqrt{\kappa_c}\ba_{in}.
\end{align}
This expression is called the quantum Langevin equation~\cite{haroche2006_exploring}. It includes two new terms: the first one corresponds to a damping of the field at rate $\kappa/2$, with $\kappa=\kappa_c+\kappa_i$, where $\kappa_i$ is the coupling rate between the system and the uncontrolled environment usually called the internal loss rate; the second term, $\sqrt{\kappa_c}\ba_{in}$, referred to as ``drive" or ``pump", is vital for $\ba$ to obey the same usual commutation relation $\left[\ba, \ba^\dag\right] = 1$ at all times despite the damping term. As an alternative to the quantum Langevin equation, the Lindblad master equation can also be used to describe such dissipative systems~\cite{nielsen2010_quantum, haroche2006_exploring}. However, the quantum Langevin equation is more suited to describe the traveling fields that we consider here.

While $\ba_{in}$ is necessary in order for us to control the state of the resonator, it also introduces undesired fluctuations in its field. To mitigate this, we typically operate in the ``stiff-pump" regime, where $\kappa_c$ is negligible compared to the frequency of the resonators, but the expectation value of $\sqrt{\kappa_c}\ba_{in}$ can be large compared to $\kappa_c$. This way, we have $\ba_{in} = \bar{a}_{in} + \ba_{in}^{0}$, where $\ba_{in}^{0}$ represents the negligible fluctuations of the field and $\bar{a}_{in}$ its average value. In the stiff-pump approximation, a drive is modeled with the Hamiltonian
\begin{align}
  \label{eq:drive}
  \frac{\bH_d}{\hbar} = \epsilon(t) \ba^\dag + \epsilon(t)^*\ba,
\end{align}
with $\epsilon(t) = \sqrt{\kappa_c}\bar{a}_{in}$.

\subsection{Josephson Junction}
Superconducting resonators alone do not provide a useful medium for encoding quantum information. This is because the energy levels of a resonator are separated by an equal spacing of $\hbar\omega$, forbidding us from addressing the transitions individually. Thus, we must introduce a non-linear element in order to achieve universal quantum control of the circuit. 

\begin{figure*}[!bt]
\includegraphics[scale=1]{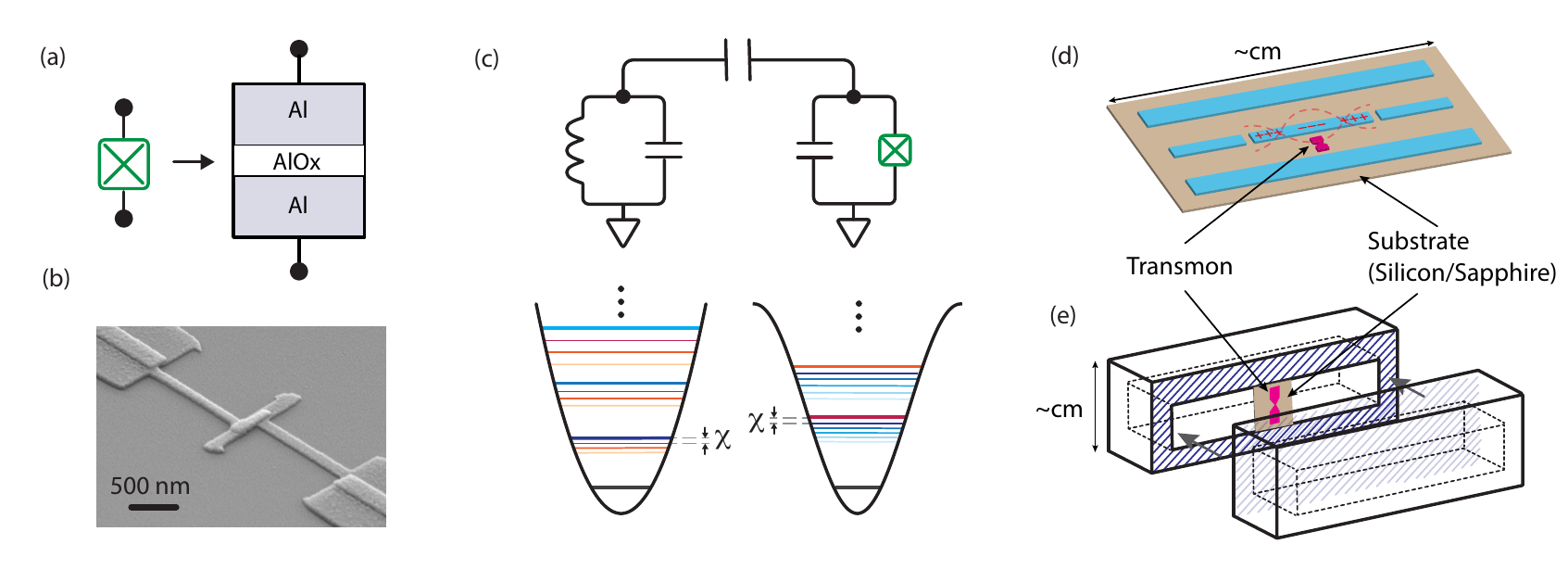}
\caption{\label{fig:dispersive_coupling} \textbf{Dispersive coupling between a transmon and a superconducting resonator.} (a) Lumped element representation of a Josephson junction and a cartoon of its structure, which consists of two layers of aluminium (grey) that are separated by an aluminium oxide tunnel barrier (white). (b) An SEM image of a bridge-free junction. Image credit: Kyle Serniak (Yale University) (c) Lumped element representation of a LC circuit capacitively coupled to a single-junction transmon and the associated the potential of each mode and the dressing of the energy levels due to the dispersive interaction. (d), (e) Two examples of physical realizations of a transmon device dispersively coupled to a superconducting cavity in either the planar (d) and 3D configuration (e). 
}
\end{figure*}

In cQED, the most ubiquitous source of non-linearity is a Josephson junction (JJ), favored for its simplicity and non-dissipative nature. This element is made of two superconducting electrodes separated by an insulating tunnel barrier, represented in Fig.~\ref{fig:dispersive_coupling}(a). In practice, JJs are typically fabricated by overlapping two layers of superconducting films with an oxide barrier in between. The area of the overlap and the properties of the oxide barrier determine the properties of the JJ. %The deposition of such a pattern is usually achieved through double-angle evaporation using either the Dolan bridge~\cite{dolan1977_offset} or the bridge-free technique~\cite{lecocq2011_junction}. 
An SEM image of a typical JJ is depicted in Fig.~\ref{fig:dispersive_coupling}(b).

The detailed theoretical description of the JJ and its underlying physics can be found in Refs.~\cite{wendin2017_quantum, clarke1970_josephson}. Here, we aim to provide a brief intuitive picture that follows the line of Josephson's original work \cite{josephson1962_possible}. Let us consider each side of the JJ as a superconducting island containing a certain number of Cooper pairs. The joint state of the junction is described by the difference in number of Cooper pairs $\bn$, which is the same observable used in our treatment of the simple LC circuit. As Cooper pairs tunnel across the junction, $\bn$ varies accordingly. Therefore, the Hamiltonian of a Josephson junction can intuitively be written in the charge basis as
\begin{align}
  \bH_J/\hbar = -\frac{E_J}{2}\sum^{+\infty}_{n=-\infty}\left(\ket{n}\bra{n+1} + \ket{n+1}\bra{n}\right),
\end{align}
where $E_J$ is the Josephson energy, which is traditionally expressed in units of frequency by dividing it by $\hbar$. We identify that the two parts of $\bH_J$, i.e. $\sum^{+\infty}_{n=-\infty}\ket{n+1}\bra{n}$ and its Hermitian conjugate, are unitary operators that translate the charge by one unit up and down respectively. This translation can also be written in the conjugate basis as $e^{\pm i\bphi}$, similar to how a position translation is written in terms of the momentum operator. Therefore, the Josephson Hamiltonian in the flux basis can be expressed as
\begin{align}
  \bH_J/\hbar = -E_J \cos{\bphi}.
\end{align}
Developing the cosine potential to second order gives rise to a quadratic term that resembles an inductive energy as described in Eq.~\ref{eq:H_LC}. This is commonly referred to as the Josephson inductance $L_J = \hbar^2/(4e^2 E_J)$. As we will show in the case of the transmon, this non-linear inductance affords us the ability to selectively address individual transitions in the system.

\subsection{Transmon}
Building upon the non-linearity afforded by the JJ, we can now construct a variety of circuit elements capable of effectively encoding quantum bits of information. These non-linear components are generally referred to as superconducting qubits. In particular, transmon qubits, which are weakly anharmonic oscillators, are the most widely used in current cQED devices. The transition frequencies between ground ($\ket{0}$) and first excited state ($\ket{1}$) of transmons are typically designed to be around 5 GHz. The choice of this frequency range is two-fold: (1) it is far below the plasma oscillation frequencies of individual electrons in the superconductor, hence, we only need to consider their collective excitations; (2) at the operational temperature of cQED experiments (20\,mK), the probability of supplying sufficient thermal energy to excite this transition is effectively suppressed, such that the qubit can be in or close to its ground state. 

In transmon qubits, the transition between the first two energy levels is detuned from that of its first to second excited state by the anharmonicity, which is typically designed to be 200 - 300\,MHz. With this configuration, transitions between the two lowest levels of the transmon can be addressed with microwave pulses as short as a few nanoseconds~\cite{werninghaus2021_leakage}.

Physically, transmon qubits are constructed with two large superconducting capacitor structures connected by one or more JJs. The capacitor pads provide the capacitive energy $E_C$ and the Josephson junction provides the inductive energy $E_J$. Combining these two, we arrive at the Hamiltonian
\begin{align}\label{eq:transmon_H}
  \frac{\bH}{\hbar} &= 4E_C\bn^2 - E_J\cos(\bphi) \\
  &= 4E_C\bn^2 + \frac{E_J}{2}\bphi^2 + \bH_{4^+}\left(\bphi\right),
\end{align}
where $\bH_{4^+}\left(\bphi\right)/\hbar = -E_J\left(\cos(\bphi) + \bphi^2/2\right)$. Note here that we have added the quadratic term $\frac{E_J}{2}\bphi^2$ and then subtracted it again to eliminate the second order terms after the cosine expansion. The constant term from the expansion is dropped as it simplify contributes to a constant energy offset that does not cause any non-trivial dynamics. With the form of Eq.~\ref{eq:transmon_H}, it becomes apparent that the transmon is effectively a variant of the standard LC circuit with the JJ as a non-linear inductor.

Transmons operate in the regime of $E_J \gg E_C$. In this limit, according to Eq.~\ref{eq:phi_ZPF} and Eq.~\ref{eq:n_ZPF}, the eigenstates of the transmon have suppressed zero-point fluctuations in phase while those in the the charge degree of freedom remain large. More specificially, the sensitivity of the transmon frequency to offset charge drops exponentially in $E_J/E_C$, and is heavily suppressed in the typical parameter regime of $E_J / E_C \approx 50$, while still maintaining a large enough anharmonicity for fast operations~\cite{koch2007_charge, grimsmo2020_quantum}. 

As $\bH_{4^+}$ depends solely on the phase zero-point fluctuations, we  treat it as a perturbation to the linear part of the Hamiltonian. Using this perturbative treatment of $\bH_{4^+}$ and limiting the excitations to only the low-lying states of the transmon, we diagonalize the linear Hamiltonian with the annihilation and creation operators $\bq$ and $\bq^\dag$, as defined in eq.~\ref{eq:annihilation}, with $E_L$ replaced by $E_J$. Therefore, we get

%As $\bH_{4^+}$ depends solely on the phase zero-point fluctuations, we therefore treated it as a perturbation to the linear part of the Hamiltonian. Using this perturbative treatment of $\bH_{4^+}$ and limiting the excitations to only the low-lying states of the transmon, we can diagonalize the Hamiltonian with the annihilation and creation operators $\bq$ and $\bq^\dag$, such that}
\begin{align}
  \label{eq:transmon}
  \frac{\bH_T}{\hbar} &= \sqrt{8 E_J E_C} \bq^\dag \bq +\bH_{4^+}\left(\varphi_T^{\text{ZPF}}(\bq + \bq^\dag)\right) \\
  &\approx \sqrt{8 E_J E_C}\bq^\dag\bq - {E_J \over 24}\left(\phi^{\text{ZPF}}_T\right)^4\left(\bq+\bq^\dag\right)^4 \\
  &\approx \omega_T \bq^\dag \bq - \frac{\alpha}{2}\bq^{\dag 2}\bq^2,
\end{align}
with 
\begin{align}
  \omega_T = \sqrt{8 E_J E_C} - E_C, \quad \alpha = \frac{E_J}{2}(\varphi_T^{\text{ZPF}})^4 = E_C.
\end{align}
In our last approximation we have kept only the resonant terms that have as many ``daggers" as ``non-daggers". This approximation is justified by the fact that, in a frame rotating at $\omega_T$, all the terms that are non-resonant oscillate at an angular frequency at least $\omega_T$, which is much larger than $\alpha/2$. The dynamic produced by such oscillating terms at a given time is cancelled by the dynamic produced by the same terms half an oscillation later, such that their action quickly averages to the identity. This approximation is called the rotating-wave approximation (RWA).

Here, we have chosen to use the ladder operators to describe the transmon qubit. This approach offers a convenient strategy to capture the behaviors of the higher-level transitions in the transmon as well as non-linear frequency mixing in the JJ, which is utilized frequently in parametric conversion processes. An alternative method is to consider the transmon as a discrete two-level system, and use the Pauli operators to describe its properties. We can arrive at an equivalent picture for the dynamics of a transmon and its interactions with other circuit elements using either of these techniques.

So far we have only focused on single-junction transmons whose parameters $E_C$ and $E_J$ are fixed by the design and fabrication of the device. Frequency tunability can also be introduced while maintaining the relative simplicity and robustness offered by the transmon by replacing the single junction with two parallel junctions. With this, it is possible to adjust the device's $E_J$ in-situ~\cite{koch2007_charge}. This configuration of using multiple junctions in a single non-linear element is called a superconducting quantum interference device (SQUID)~\cite{jaklevic1964_quantum}.

To understand this type of SQUID qubits, we consider the simplest case of a system with two junctions of identical Josephson energy $E_J/2$.The system Hamiltonian is given by 
\begin{align}
  \frac{\bH}{\hbar} = 4E_C \bn^2 - \frac{E_J}{2}\left[\cos(\bphi_1) + \cos(\bphi_2)\right], 
\end{align}
where $\bphi_1$ and $\bphi_2$ are the flux operators of each junction. Naively, one would think that having two tunnel-junctions in parallel is equivalent to having a single large junction. However, when an external magnetic field threads a magnetic flux $\phi_{\mathrm{ex}} = \varphi_{\mathrm{ex}}\phi_0$, where $\varphi_{\mathrm{ex}}$ is the number of flux quanta, the behavior of this system qualitatively differs from that of the single-junction devices. In fact, it is favorable for Cooper pairs to tunnel in opposite directions across the two junctions. Thus, the external field results in a flux difference of $\bphi_2 - \bphi_1 = \varphi_{\mathrm{ex}}$. Trigonometric identities allow us to regroup terms such that the Hamiltonian can be written as~\cite{koch2007_charge}
\begin{align}
  \frac{\bH}{\hbar} = 4E_C \bn^2 - E_J\cos\left(\frac{\varphi_{\mathrm{ex}}}{2}\right)\cos(\bphi),
\end{align}
where $\bphi = (\bphi_1 + \bphi_2)/2$ is the average flux between the two junctions. Effectively, a SQUID transmon provides a simple device with in-situ tunability of $E_J$. This on-demand tunability offers new avenues for implementing quantum operations between transmons~\cite{dicarlo2009_demonstration, reed2012_realization, Barends2019, Foxen2020}, probing coherence properties~\cite{Reed10b, Klimov2018, Luthi18} and engineering new Hamiltonians~\cite{Yamamoto2008_fluxpumped, lescanne2020_exponential}. However, any additional noise on the control knob $\varphi_{ex}$ will translate into phase noise for the transmon, which can limit their coherence.

\subsection{Other qubits}
While transmons are currently the most widely used superconducting qubits due to their simplicity and versatility, it is important to note that our rich and diverse cQED toolbox extends much beyond these simple circuit elements. There are a whole host of other superconducting qubits that work in different regimes and employ different junction arrangements~\cite{kjaergaard2020_superconducting, krantz2019_engineer}. Among them, the flux-qubit~\cite{Mooij1999} and the fluxonium ~\cite{manucharyan2009_fluxonium} have been systematically investigated and their performance consistently improved~\cite{yan2016_theflux, Nguyen2019} in recent years. 

In parallel, there are also several promising strategies to construct superconducting circuits that are intrinsically protected against various environmental noise. Such protection schemes are heavily inspired by the continuous-variable quantum-error-correction code designed by D. Gottesman, A. Kitaev and J. Preskill (GKP)~\cite{gottesman2001_encoding}. Recently, the proof-of-principle demonstrations of two of such protected qubits have been reported, namely, the ``cos$2\varphi$'' element~\cite{Gladchenko2009_cos2phi, smith2020_superconducting} and the ``0-$\pi$'' qubit~\cite{brooks2013_protected, gyenis2020_experimental}. These new circuit elements and their intrinsic resilience to noise are valuable building blocks for realizing quantum devices that can potentially achieve the necessary scale and complexity without significant performance degradation.

\subsection{The dispersive coupling}
Having introduced the two key components of cQED systems, superconducting resonators and transmons, we now look into how they can be integrated to form a useful quantum device. 

Physically, when a transmon is placed in proximity to a resonator, the charges of one element can influence those of the other through a coupling capacitor (Fig.~\ref{fig:dispersive_coupling}(c)). We summarize this coupling by a parameter of the Hamiltonian, $g_0$, which can be computed exactly for a given model of the circuit. This leads to an interaction Hamiltonian
\begin{align}
  \bH_{\text{int}} &= g_0\bn_R \bn_T \\
  &= -g \left(\ba^\dag - \ba\right)\left(\bq^\dag - \bq\right)
\end{align}
with the subscripts $T$ and $R$ describing the transmon and the resonator respectively. Note that we have regrouped the charge zero-point fluctuations in $g$. 

The exact diagonalization of the Hamiltonian with this interaction is done in many reviews~\cite{blais2020_quantum, blais2020_circuit, Girvin12}. Here, we present an approximate diagonalization in the most commonly-used parameter regime in cQED, where: 

1) $g\ll|\Delta|=|\omega_T - \omega_R|$, with large qubit-cavity detuning giving rise to a dispersive interaction between the two elements instead of resonant energy exchange;

2) $E_C \ll |\Delta|$ (weak anharmonicity);

3) $|\Delta| \ll \omega_T + \omega_R$, implying that the counter-rotating terms are negligible.

In this regime, we apply the Rotating Wave Approximation (RWA) to reduce the interaction Hamiltonian to $\bH_{\text{int}}/\hbar = g\left(\ba\bq^\dag + \ba^\dag\bq\right)$.

Following this, we write a set of Langevin equations for $\ba$ and $\bq$ in the Heisenberg picture
\begin{align}
  \partial_t\ba &= - i \left(\omega_R - i\frac{\kappa}{2}\right) \ba - i g\bq\\
  \label{eq:q_langevin}
  \partial_t\bq &= -i \omega_T \bq - i g \ba - \frac{i}{\hbar}\left[\bq, \bH_{4^+}\left(\varphi^{ZPF}_T(\bq + \bq^\dag)\right)\right],
\end{align}
where we have included the dissipation of the resonator but neglected the dissipation of the transmon, such that we can analyze the impact of a readout cavity, typically with low or moderate Q, on the relaxation of the transmon (see Sec.\ref{sec:interacting_with_quantum_circ}3). We treat the non-linear part of Eq.~\ref{eq:q_langevin} as a perturbation and diagonalize the linear part given by the matrix
\begin{align}\label{eq:dispersive_mat}
  M = \left(\,
  \begin{matrix*}[c]
    ~(\omega_R - i\kappa/2) & g~ \\ 
    ~g & \omega_T~ 
  \end{matrix*}\,\right),
\end{align}
whose eigenvalues are given to the first order by
\begin{align}
  \tilde \omega_R &= \omega_R + \frac{g^2}{\Delta} - i\frac{\kappa}{2} + i\left(\frac{g}{\Delta}\right)^2\kappa \\
  \tilde \omega_T &= \omega_T - \frac{g^2}{\Delta} - i\left(\frac{g}{\Delta}\right)^2\kappa.
\end{align}
The frequencies $\tilde \omega_R$ and $\tilde \omega_T$ of the resonator and the transmon in the dispersively coupled system are shifted with respect to the uncoupled system by the Lamb-shift 
\begin{equation}
\label{eq:lamb_shift}
\Lambda_\chi = g^2/\Delta. 
\end{equation}
The transmon also inherits some dissipation from the cavity given by the Purcell rate
\begin{align}
  \label{eq:purcell_limit}
  \Gamma_P = 2\left(\frac{g}{\Delta}\right)^2\kappa.
\end{align}
The Purcell rate could lead to a trade-off between a large transmon-resonator coupling for fast operations and the lifetime of the qubit. However, in practice, the Purcell effect can be significantly suppressed with carefully designed Purcell filters~\cite{Reed10b}.

We can now define a new set of annihilation and creation operators that correspond to the eigenvectors of the matrix given by Eq.~\ref{eq:dispersive_mat}. To do so, we first define a mixing angle of $\theta = \frac{1}{2}\arctan\left(\frac{2g}{\Delta}\right)$. With this, the new eigenstates, also called dressed modes, are given by
\begin{align}
  \tilde \ba &= \cos(\theta) \ba + \sin(\theta) \bq \approx \ba + \frac{g}{\Delta}\bq \\
  \tilde \bq &= -\sin(\theta)\ba + \cos(\theta)\bq \approx -\frac{g}{\Delta}\ba + \bq.
\end{align}
A detailed derivation can be found in Ref.~\cite{grimsmo2020_quantum}. With these relations, the linear part of the Hamiltonian is now proportional to $\tilde \omega_R \tilde\ba^\dag\tilde\ba + \tilde \omega_T \tilde \bq^\dag \tilde \bq$. 

To treat the non-linear part of the Hamiltonian $\bH_{4^+}$, we invert the definition of $\tilde \ba$ and $\tilde \bq$ to obtain
\begin{align}
  \varphi_T^{\text{ZPF}}\left(\bq + \bq^\dag\right) &=\varphi_T^{\text{ZPF}}\left(\cos(\theta)(\tilde\bq + \tilde\bq^\dag) + \sin(\theta)(\tilde \ba + \tilde \ba^\dag)\right) \nonumber\\
  &\approx \varphi_T^{\text{ZPF}}\left(\tilde\bq + \tilde\bq^\dag\right) + \varphi_T^{\text{ZPF}}\frac{g}{\Delta}\left(\tilde \ba + \tilde \ba^\dag\right)
\end{align}
It is useful to note that we often combine the ZPF terms with the associated mixing angles into a single coefficient, i.e. $\varphi_T = \varphi_T^{\text{ZPF}}\cos(\theta)$; $\varphi_R = \varphi_T^{\text{ZPF}}\sin(\theta)$, to describe the $(\tilde\bq + \tilde\bq^\dag)$ and $(\tilde\ba + \tilde\ba^\dag)$ terms. This notation is commonly adopted in the treatment of dynamics of transmon-cavity systems in cQED.

With $\varphi_T^{ZPF} \ll 1$, the largest elements in this non-linear Hamiltonian are of the 4th order. We can then perform the same RWA as in Eq.~\ref{eq:transmon} to eliminate the non-resonant terms. Now working exclusively in the new frame, we drop the tilde and simplify the notation to get the transmon-cavity Hamiltonian in the dispersive regime in the following form
\begin{align}
  \label{eq:dispersive_hamiltonian}
  \frac{\bH}{\hbar} =& \omega_T \bq^\dag \bq + \omega_R \ba^\dag\ba \\ \nonumber&-\frac{\alpha}{2}\bq^{\dag 2}\bq^2 - \frac{K}{2}\ba^{\dag 2}\ba^2 - \chi\left(\bq^\dag\bq\right)\left(\ba^\dag\ba\right),
\end{align}
with
\begin{align}
  K &= \frac{E_J}{2} \left(\varphi^{\text{ZPF}}_T\right)^4\left(\frac{g}{\Delta}\right)^4 = E_C\left(\frac{g}{\Delta}\right)^4, \\
  \chi &= E_J\left(\varphi^{\text{ZPF}}_T\right)^4\left(\frac{g}{\Delta}\right)^2 = 2 E_C\left(\frac{g}{\Delta}\right)^2. \label{eq:chi}
\end{align}
Typically, $K$ and $\chi$ are referred to as the self-Kerr and the cross-Kerr non-linearity between the resonator and the transmon, respectively. Physically, we can interpret the effect of a dispersive coupling as a state-dependent frequency shift between the two modes. The coupling is considered to be in the strong dispersive regime if this state-dependent shift is larger than the linewidth of the resonator.

In practice, this type of dispersively coupled transmon-resonator system can be realized by placing them in physical proximity such that their respective fields overlap, as illustrated in Fig.~\ref{fig:dispersive_coupling}{d} and (e). More details regarding the exact configuration of such devices will be discussed in Sec.~\ref{sec:designfab}.

Despite its relative simplicity, the dispersive Hamiltonian is the origin of many interesting phenomena and crucial tools in cQED. Generally, the resonator-transmon system offers us a platform with three main capabilities. First, by coupling the resonator strongly to a transmission line, we can implement fast and efficient readout of the transmon state. Alternatively, two transmons can interact via a mediating resonator mode (quantum bus) to realize quantum operations. Finally, the transmon can act as a non-linear ancilla to perform universal control on the resonator state. 

Going beyond the natural interactions, the addition of microwave drives at well-chosen frequencies also offers the possibilities of engineering new dynamics in the cQED system. This capability is afforded by the non-linear frequency-mixing in the Josephson junction, akin to frequency conversion processes in non-linear optical mediums~\cite{grynberg2010_introduction}. To the first order, a single JJ can be used to enact four-wave mixing in cQED devices.

Such driven interactions are highly useful tools for engineering new dynamics or operations in cQED devices. One of the key applications of such driven operations is to engineer quantum-limited amplifiers, which will be discussed in Sec.~\ref{sec:environment}. Additionally, a wide range of different Hamiltonians have been demonstrated using such drives, enabling a large repertoire of new experimental achievements. This includes stabilization and protection of Schrodinger cat state~\cite{mirrahimi2014_dynamically, leghtas2015_confining, lescanne2020_exponential, grimm2020_stabilization}, remote entanglement of superconducting qubits~\cite{Roch14, Narla16, campagne-ibarcq2018_deterministic, axline2018_ondemand}, cooling of nanomechanical devices~\cite{Teufel2011_sideband}, engineered quantum gates between microwave cavities~\cite{rosenblum2018_cnot, gao2018_programmable, gao2019_entangling}, detection of single microwave photons~\cite{Lescanne2019_irreversible}, longitudinal coupling of a cavity and a transmon~\cite{touzard2019_gated}, and so on.

%%%%%%%%%%%%%%%%%%%%%%%%%%%%%%%%%%%%%%%%%%%%%%%%%%%%%%%%%%%%%%%%%%%%%%%%%%%%%%%%%%%%%%%%%%%%

\section{Designing quantum circuits}\label{sec:designfab}
After we establish an ideal target Hamiltonian for an intended application, the first step in the workflow of a cQED experiment is to translate the targeted set of abstract system parameters (number of qubits, transition frequencies, connectivity and coupling strengths, etc.) into a physical device with appropriately designed circuit elements. 

\begin{figure}
\includegraphics{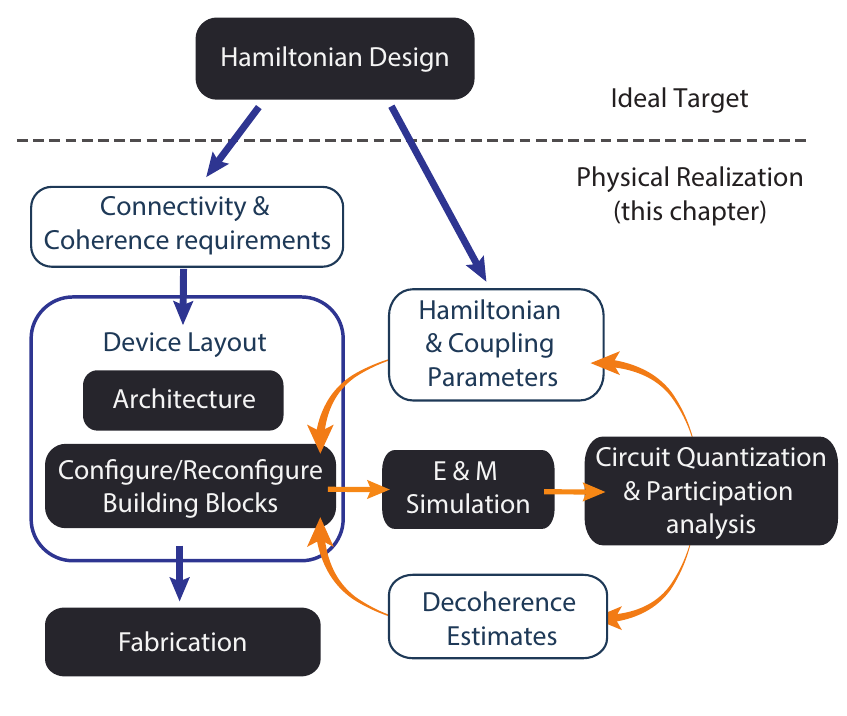}
\caption{
\textbf{Design cycle of a circuit QED device.} To translate a target Hamiltonian into a physical device, we first lay out the main building blocks for the circuit in a suitable architecture. Then, we compute the electromagnetic mode structures of the device reliably, often using numerical simulations. Subsequently, we extract the relevant Hamiltonian parameters from this simulation using circuit quantization techniques. We then compare the result with the target Hamiltonian, and can adjust the design until we converge to a configuration that faithfully reflects our target system. 
}
\label{fig:design_cycle}
\end{figure}

This translation is often an iterative process that consists of several key steps including architectural layout, configuring building blocks, electromagnetic simulation, and circuit quantization, as illustrated in Fig.~\ref{fig:design_cycle}. In this section, we will guide the reader through such a process to highlight some of the common strategies and the main considerations involved in building up the desired cQED device. 

\subsection{Planar and 3D architectures}
Currently, there are two main types of device architectures in circuit QED: colloquially referred to as ``planar", where all quantum modes (e.~g.~qubits and linear resonators) are lithographically defined on chip; and ``3D", where some modes are native to three dimensional structures. In both cases, however, a 3D package (a ``sample box") is engineered to enclose the electromagnetic modes of the circuit and prevent decoherence from microwave radiation loss. 

\begin{figure*}
\includegraphics{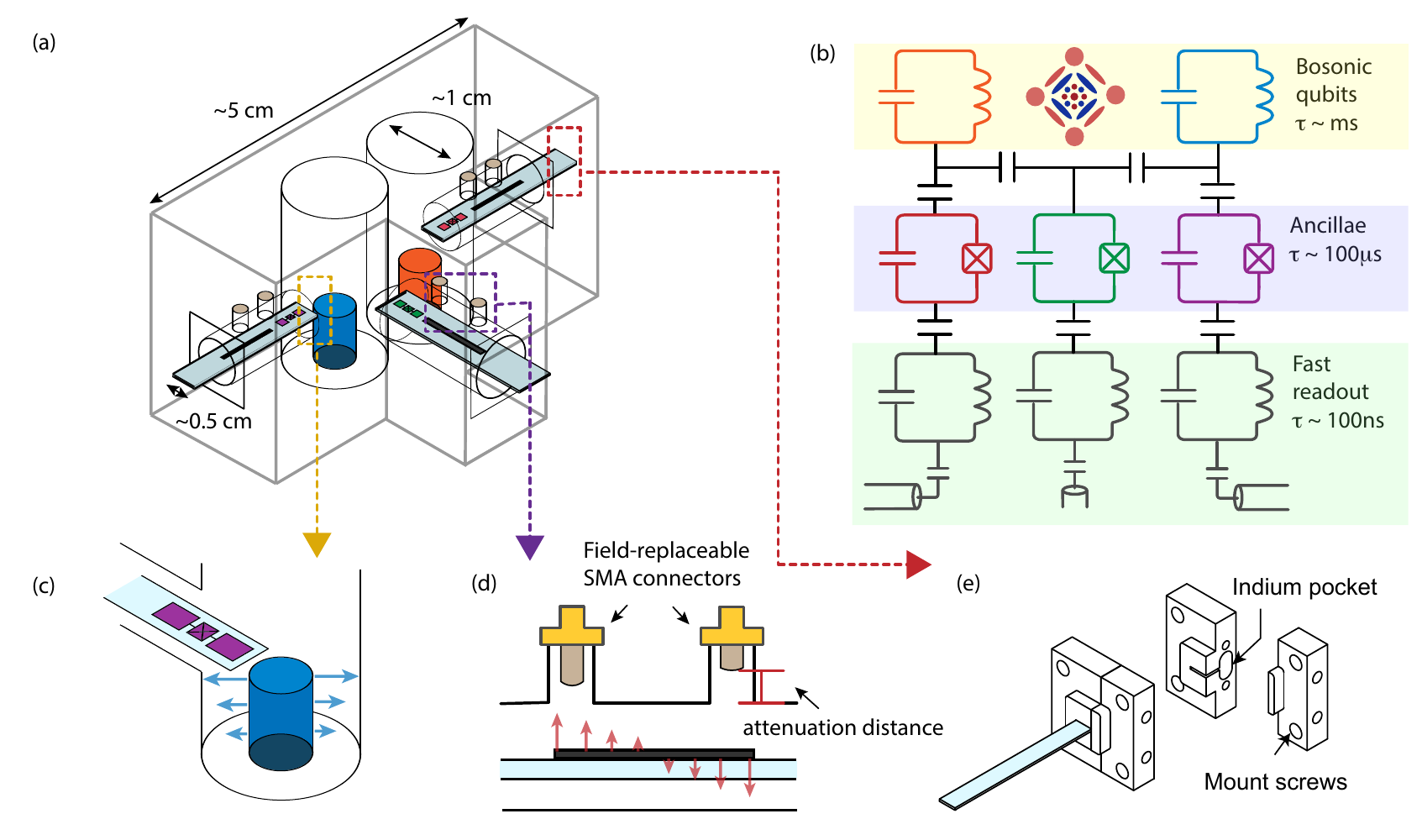}
\caption{
\label{fig:device_3d}
\textbf{Main elements in 3D cQED devices} (a) A sample device consisting of two 3D cavities (blue and orange), 3 transmon ancillae (purple, green, and red) together with their respective planar readount resonators (grey). All components are housed in a high-purity aluminium package. (b) The equivalent circuit representation of the device. (c) A more detailed illustration of the transmon and cavity interface. The coupling strength between the two is determined by the overlap between the transmon capacitor and the field (blue arrows) of the cavity. (d) A cross-sectional view of the coupling mechanism to the transmon and the readout reasonator via field-replaceable pins. (e) A schematic of the clamping structure for chips containing the transmon and its readout resonator. 
}
\end{figure*}

In the 3D architectures (Fig.~\ref{fig:device_3d}), the large volume of vacuum inside the package (i.e.~the 3D cavity), typically a few cm$^3$ in size, is used as the linear modes in the cQED Hamiltonian. They couple weakly to transmission lines via coaxial center pins that are extended into apertures through the cavity wall (Fig.~\ref{fig:device_3d}(d)). These cavities also couple to qubits or other resonator modes, which are typically constructed using lithographically-patterned planar structures (Fig.~\ref{fig:device_3d}(c)) with large capacitor structures to provide effective coupling to the 3D cavity mode. They can be fabricated on one or more chips, which are mechanically clamped in place by nuts and bolts (Fig.~\ref{fig:device_3d}(e)). This architecture has several advantages, especially for a beginner experimentalist working with a small-scale quantum circuit: 
\begin{enumerate}
 \item It is simple to assemble: No peripheral structures and materials, such as printed circuit boards and wire-bonds, are needed to bridge the quantum modes and the external cables. 
 \item It provides a clean electromagnetic environment: The absence of peripheral structures saves the engineering effort to suppress spurious modes in the device package.
 \item It exhibits good coherence times for the same material quality due to the small surface-to-volume ratio of the quantum modes. In particular, the long-lived 3D cavity modes prove to be a valuable quantum resource.
\end{enumerate}
These appealing features make such 3D cQED devices powerful experimental testbeds for developments in quantum error correction~\cite{joshi2021_quantum} and bosonic quantum simulations~\cite{hu2018_simulation, wang2020_efficient}.

However, there are also several important limitations associated with the current generation of devices based on machined cavities. The large spatial extent of the 3D cavity electromagnetic field makes it difficult to achieve good mode isolation within the same cavity. Furthermore, the lack of DC connection to the chip limits tunability and integration with circuit elements that require flux biasing. Active research is underway to develop techniques that could potentially resolve these challenges~\cite{gargiulo2021_fast}.  

\begin{figure}
\includegraphics[width=0.5\textwidth]{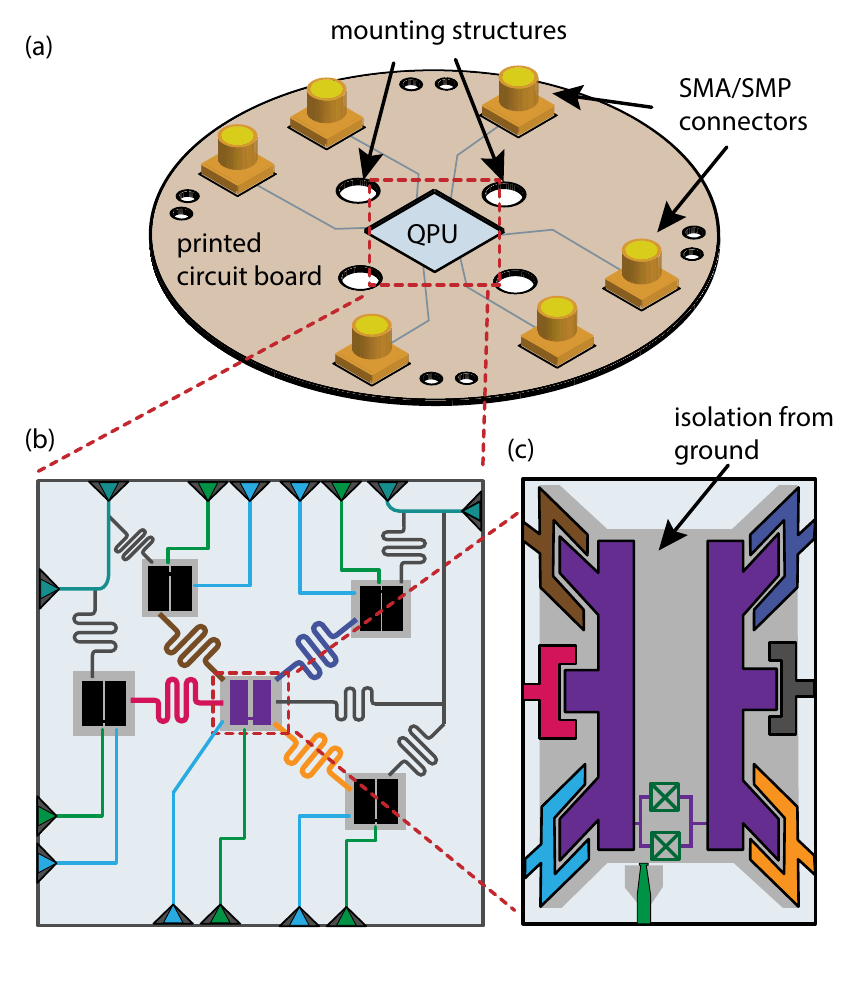}
\caption{
\textbf{A typical example of a planar cQED system.} (a) A superconducting chip embedded in a printed circuit board (PCB) via wire bonds. The PCB provides the structures for grounding, microwave interconnects, and mounting elements to the sample box. (b) A more detailed view of a typical multi-qubit chip together with resonator buses and readout modes (dark grey). Each qubit can be controlled via on-chip flux (green) or microwave (blue) drive lines. (c) A typical flux-tunable transmon qubit capacitively coupled to several resonator modes. 
}
\label{fig:device_2d}
\end{figure}

The planar architecture (Fig.~\ref{fig:device_2d}) is centered around a lithographically-patterned cQED chip. The chip defines all the quantum modes desirable for the application, although the packaging around it may affect the fine details of the Hamiltonian parameters. Other than all the quantum circuit elements, the chip typically also includes feedlines, which are co-planar waveguides (CPW) impedance-matched to the input/output coaxial cables.  They are weakly coupled to the quantum circuit elements to supply and extract microwave signals. The chip is then embedded on a printed circuit board (PCB), with their corresponding CPW ports connected and their respective ground planes tied together, often using a large amount of wire-bonds. Bulk superconducting covers are attached to the ground plane of the PCB to enclose the chip into an area as small as possible (limited to less than $\sim$1 cm$^2$ without on-chip 3D interconnects) to suppress the box modes. The PCB routes the input/output signal lines to a set of planar-to-coax transitions for external cable connections. More details on the design of microwave packaging for planar qubits can be found in Ref.~\cite{Dickel18_thesis, huang2021_microwave}. The electromagnetic modes and the coupling schemes in such systems are relatively local, therefore, well-engineered planar cQED devices afford many favorable features such as
\begin{enumerate}
 \item Fast magnetic flux control of qubits, resonators or coupling elements,
 \item Independent design of large numbers of quantum elements on the same chip with low cross-dependence of parameters, %(thanks to geometric isolation of mode structures between remote elements) 
 \item Flexible access to individual elements locally.
\end{enumerate} 
that are not readily available in the 3D architecture yet. 

Due to these advantages, recent experiments with more than 10 qubits and resonators have generally adopted the planar approach. However, further scaling of planar devices beyond 50 qubits will likely require 3D integration of on-chip elements, employing techniques such as flip-chip indium bonding and through-silicon vias. Therefore, the two disparate device architectures, signified by the difference in their treatments of the box modes and ground planes, may eventually converge into a similar technological platform. 

\subsection{Configuring the building blocks}
The very first step in the cQED design flow is to ask the following questions: How many non-linear elements (e.g.~qubits) and linear elements (e.g.~resonators) are required to realize the desired device functionality? What are the roles, mode frequencies, tunability, and required coherence properties of each of them? What is the required connectivity of the device and how are the different elements coupled to one another? %While the design layout can be highly device-specific, there are some design intuitions that could help us sketch out the initial blueprint of the system.
These basic considerations often dictate whether the device is better (or easier) built with the planar or the 3D approach under the current technology. 

In either planar or 3D cQED devices, linear modes are realized with superconducting resonators whose frequencies are dictated by the dimensions of the electromagnetic structures. While lumped-element LC resonators (with interdigitated capacitors and meandering wires) are valuable for their small footprint, the most widely-used superconducting resonators in the field are the CPW resonators for the planar approach, and variations of coaxial resonators for the 3D approach~\cite{reagor2016_quantum,axline2016_architecture}. Physically, they are equivalent to an infinite series of LC oscillators, but usually only the fundamental mode is employed in cQED experiments. Ideally, the higher-frequency modes are sufficiently detuned from resonance frequency of interest in the system to not interfere with the intended use of the device. However, we must be aware of the existence of these modes and their potential impact on qubit coherence (e.g.~thermal shot noise dephasing~\cite{Sears12}, see Sec.~\ref{sec:environment}), and exercise special care to prevent spurious multi-photon transitions. 

CPW resonators are typically realized with a section of CPW transmission line terminated with open or short boundary conditions on the two ends. The CPW has a geometric capacitance and inductance per unit length $C_g$ and $L_g$ which can be calculated from the cross-sectional dimensions of the CPW (e.~g.~widths of the center conductor and the dielectric gap). Furthermore, the kinetic inductance of the superconducting thin film per unit length $L_k$ also influences the wavelength ($\lambda=2\pi/\omega\sqrt{C_g(L_g+L_k)}$) of the propagating wave. The value of $L_k$ depends on the cross-sectional dimensions and the material properties, but it can be estimated from the room temperature resistivity of the film. The contribution from $L_k$ is typically small but not negligible in standard planar devices. 

Having taken these factors into consideration, the length of the CPW section can then be designed based on the targeted resonator frequency and the type of termination. Recall that a CPW terminated by two open ends forms a $\lambda/2$ resonator, and a CPW terminated by an open and a short end forms a $\lambda/4$ resonator. A typical operating frequency of $\sim$5 GHz translates to CPW lengths on the orders of 1\,cm, but the total footprint can be reduced by using meandering patterns, as depicted in Fig.~\ref{fig:device_2d}(b).

Similarly, the coaxial resonator can be understood as a section of coaxial transmission line terminated with open or short boundary conditions on the two ends. A $\lambda/4$ version of the coaxial resonator can be constructed by conventional machining from a single piece of bulk superconductor (Fig.~\ref{fig:device_3d}(c)), eliminating the ohmic dissipation from the imperfect seams in bulk metal assembly~\cite{brecht2016_multilayer}. This advantage leads to reliably high coherence times, which has made this type of ``reentrant post cavity" a go-to design for high-Q quantum memory in cQED~\cite{reagor2016_quantum}. A $\lambda/2$ version of the coaxial resonator uses a machined tube as the ground shield and an on-chip superconducting strip as the center conductor (Fig.~\ref{fig:device_3d}(d)), affording the flexibility of having multiple resonators, qubits and Purcel filters on the same chip with mostly lithographically defined couplings~\cite{axline2016_architecture}. Again, the required length of the resonators can be estimated from the wavelengths of coaxial travelling modes at the target frequency. 

The design of non-linear modes such as qubits, couplers, and parametric mixers typically begins with lumped-element approximations. The inductive circuit component is provided by the Josephson junctions, whose non-linear inductance can be easily varied via the tunnelling area of the junctions. The capacitance of the mode is tuned by adjusting the dimensions of the superconducting leads and pads ( purple-colored structures in Fig.~\ref{fig:device_2d}(b) and Fig.~\ref{fig:device_3d}(c)). %In the transmon-regime, the high capacitance is typically achieved by physically extending the dimensions of the pads or using inter-digitated structures. 
For transmon qubits, the desired frequency is typically in the range of $4-8$ GHz. This frequency range is sufficiently to suppress thermal excitations at 20\,mK, low enough to mitigate couplings to higher-frequency spurious package modes, and is supported by existing  high-performance and relatively cost-effective microwave electronics. Inductance and capacitance of the transmons are usually designed in the range of $6-16$ nH and $50-125$ fF (with $2-5$ fF from the tunnel barrier of the junction itself), corresponding to $E_J\sim10-25\,$GHz and $E_C\sim160-400$\,MHz, respectively. This range of $E_C$ is chosen to effectively suppress charge noise (which grows exponentially with $E_C$~\cite{koch2007_charge}) while still preserving a sizeable anharmonicity $\alpha\approx E_C/\hbar$ (which is crucial for fast qubit operations).

Next, let us consider the necessary coupling between different elements in the device. The dispersive coupling between a transmon and a resonator, for instance, can be achieved by ensuring a sufficient overlap between their electromagnetic fields. In other words, a strong coupling, or a large $g$, can be realized if there is a high mutual capacitance between the two modes. The coupling capacitance in planar designs (such as the fork-shaped structures in Fig.~\ref{fig:device_3d}(c)) can be calculated either using conformal mapping techniques~\cite{goano2001_general} or analytical expressions for certain specific use cases~\cite{martinis2014_calculation}. Alternatively, an inductive coupling can be achieved by introducing a mutual inductance between the two modes. Moreover, the magnitude of the dispersive interaction $\chi$, depends not only on $g$ but also on the frequency detuning $\Delta$ between the two elements (Eq.~\ref{eq:chi}), which can be modified by adjusting the dimensions of individual elements and hence, their individual frequencies. If the target Hamiltonian requires multiple qubits and gates between them, it is important that we arrange the geometric configurations such that their resonance frequencies are well-separated and their mutual capacitance or inductance is kept small to avoid unwanted couplings. 
Finally, we consider the roles that each element in the device must play, which is crucial for setting up their input/output coupling. For resonators, there are generally three main functions: readout, memory, or bus. A readout resonator should be relatively strongly coupled to a transmission line to allow efficient extraction of information, with an output coupling Q typically on the order of $10^3$ to $10^4$. In planar designs, the coupling Q between a CPW resonator and a CPW feedline can be controlled by a conveniently parameterized coupling capacitance at the end of the resonator. In 3D designs, the coupling Q of the resonator is often controlled by the length of the coupling pin~\cite{sears2012_photon} (extension of the coaxial center conductor) inserted into the cavity wall (Fig.~\ref{fig:device_3d}(d)). On the other hand, a long-lived quantum memory resonator should be designed with an input drive line with coupling Q sufficiently high to not compromise its total quality factor.

Similarly, we also need to evaluate what functionalities each qubit should be endowed with.
A fixed-frequency transmon simply requires a single junction connected to two superconducting pads, whicle a flux-tunable qubit requires more JJs to form flux loops. This can be implemented in configurations such as the symmetric/asymmetric SQUID transmons, SNAIL~\cite{grimm2020_stabilization}, flux qubits~\cite{yan2016_theflux} and fluxonium~\cite{manucharyan2009_fluxonium}. In addition, the use of flux-tunable elements also entails that we must include flux-bias lines. This can use a short-ended CPW in close vicinity to the flux loop in planar circuits. In 3D architecture, a centimeter-scale external coil can be used to provide a global DC flux to flux-tunable devices inside a copper package~\cite{Nguyen2019}, but fast and local flux-biasing remains under development~\cite{stammeier2018_applying, gargiulo2021_fast}.

Controlling these quantum elements requires carefully designed microwave feed-lines. For fixed-frequency transmons, this translates into the addition of capacitively coupled drive line for X/Y control of the qubit. On the other hand, flux-tunable elements require inductively coupled flux-bias lines. Adding these features necessarily introduces a coupling to the (dissipative) environment and could result in elevated dissipation rates in the device. 

The reduction in the coherence (relaxation) time due these additional elements can be calculated 
with a circuit model using the dissipative environmental impedance $\mathrm{Re}[Z(\omega)]$~\cite{Houck08}:
\begin{equation}
    \Gamma = \frac{1}{\mathrm{Re}[Z(\omega)] C_\Sigma }, 
\end{equation}
where $C_\Sigma$ is the effective capacitance of the transmon. 

For a transmon coupled to a resonator, this reproduces the Purcell limit (~\ref{eq:purcell_limit}). For a transmon qubit capacitively coupled to a microwave drive line, this results in an effective decay rate of 
\begin{equation}
    \Gamma_{\mathrm{D}}  \simeq \omega_q^2 Z_0 C_c^2 / C_\Sigma, 
\end{equation}
where $\omega_q$ is the resonance frequency of the qubit, $Z_0$ is the characteristic impedance (typically $50~\Omega$), $C_c$ is the coupling capacitance. Assuming the typical system parameters of $\omega = 2\pi\cdot 5\,\mathrm{GHz}$, $C_\Sigma=65\,\mathrm{fF}$, and $C_c=0.1\,\mathrm{fF}$, we get an effective energy relaxation time of $T_1 = 1/\Gamma_{\mathrm{D}}\approx 132\,\mu \mathrm{s}$. 

Similar analysis for a flux-bias line, modelled as a capacitance $C_c$ in series with the transmon and a shunt inductance $L_c$, gives an effective decay rate~\cite{thesisChow10}:
\begin{equation}
    \Gamma_{\mathrm{FBL}} \simeq \frac{1}{Z_0  C_\Sigma} \left(\frac{\omega_q}{\omega_c}\right)^4 ,
\end{equation}
where $\omega_c= \frac{1}{\sqrt{L_c C_c}}$~\cite{thesisChow10}. Filling in typical values of $\omega_q = 2\pi\cdot 5\,\mathrm{GHz}$, $C_\Sigma=65\,\mathrm{fF}$,  $C_c=10\,\mathrm{fF}$, and $C_c=20\,\mathrm{pH}$ results in an effective $T_1$ $1/\Gamma_{\mathrm{FBL}}\approx 83\,\mu \mathrm{s}$. 

Practically, superconducting quantum devices often more complex than such a simple model could capture. Hence, these equations can only serve as a rule-of-thumb in the design process. Many advanced software tools, such as QuCAT~\cite{Gely_2020} and scqubits~\cite{groszkowski2021scqubits}, can help in calculating the device parameters and dissipation rates more accurately.

\subsection{Extracting system parameters}
After laying out the basic structures of the device, we can now perform numerical simulations to extract the relevant parameters. There are several commercial software capable of simulating electromagnetic structures at GHz frequencies, such as the Ansys HFSS, COMSOL, and Sonnet. These solvers provide a useful tool to visualise the distribution of the various electromagnetic resonance modes in the device and analyze their respective coupling strengths to other modes, as well as to the environment. 

In these simulations, the junction is treated as a simple lumped-element structure consisting of an inductor and a relatively small capacitor (whose values are determined from the specifications of the junction outside the simulation). The kinetic inductance of the tunnel junction accounts for nearly all of the inductance of the transmon~\cite{minev2020_energy}. The rest of the transmon features and planar cavity modes are treated as perfectly conducting metal sheets, as their kinetic inductances are small compared to that of the junction~\cite{Gao08_thesis}. 3D cavities are defined by vacuum boxes whose dimensions define the resonance mode of the package. 

In order for the simulation to reliably and accurately capture the various resonance modes in the system, the mesh seeding and convergence criteria must be set with careful considerations. A finely meshed surface allows more accurate evaluation of the electromagnetic field distributions but imposes a significant penalty on computational time. This can typically be optimised by applying a non-uniform mesh, which places a finer mesh in smaller features and along the edges where the electric fields tend to focus, but creates coarser mesh surfaces on large, regular features with a low concentration of electric fields. The choice of convergence criteria is made based on the goal of the simulation. Typically, we apply a frequency convergence limit of at least 0.5\% or less per pass to capture the resonance frequencies of each mode accurately. 

Based on the results of these simulations, we can apply numerical algorithms to extract the relevant Hamiltonian parameters of the system. There are two main strategies: the Blackbox Quantization (BBQ)~\cite{nigg2014_blackbox} and Energy Participation Ratio (EPR)~\cite{Minev18_thesis, minev2020_energy} approach. 

The BBQ model treats the Josephson junction as a small non-linear perturbation to a network of linear inductors and capacitors. The locations of each of the zeroes and poles of the impedance reveal the various Hamiltonian parameters. This has later been improved to more effectively account for multi-mode systems involving several non-linear modes~\cite{solgun2015_multiport}. While BBQ offers a comprehensive tool for analyzing the key Hamiltonian parameters for cQED systems, it requires additional simulation steps to acquire the numerical impedance associated with the circuit. 

The EPR approach, on the other hand, offers a simpler alternative which computes the Hamiltonian parameters based on the distributions of the electromagnetic field that can be obtained directly from the finite-element simulation. More specifically, the EPR of a junction element $j$ in the mode $m$, denoted by $p_{mj}$, is defined as the fraction of the total inductive energy of $m$ that is stored in $j$. In this framework, $p_{mj}$ is a bounded real number from 0 to 1, where 0 indicates that junction $j$ does not share any energy with mode $m$ while $p_{mj} = 1$ means that $j$ is the only element excited by $m$. Furthermore, we can directly extract the various Hamiltonian parameters, such as the mode frequencies, $\omega_{i}$ and non-linear coupling terms, $\chi_{ik}$, to $p_{ik}$, where $i,k \subseteq\{m, j\}$. Overall, the EPR method reduces the number of simulations needed and readily generalises to arbitrary circuit configurations consisting of one or more non-linear elements. The algorithm gained significant traction over the past few years and has been made open-source by its developers~\cite{minev2020_git}.

\subsection{Design considerations for optimal coherence %Geometry considerations of qubits for optimal coherence
}\label{subsec:devicedesign}

\begin{figure}
\includegraphics[width=0.5\textwidth]{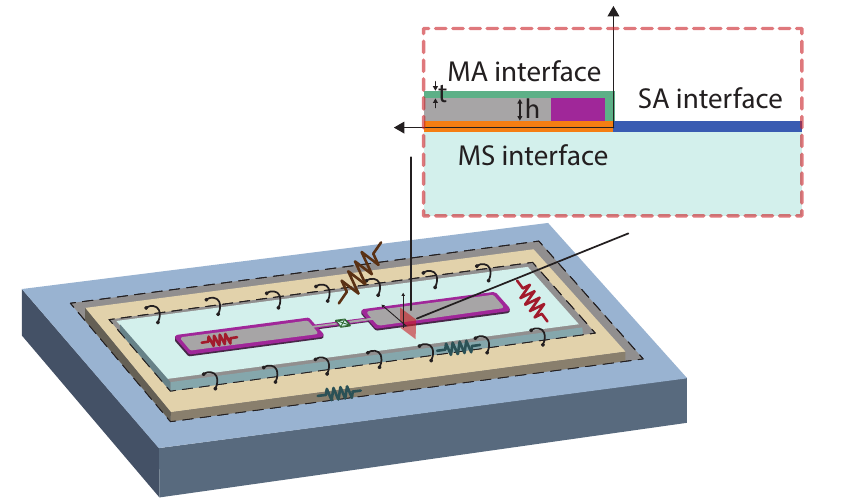}
\caption{
\label{fig:losses}
\textbf{Potential losses in a planar cQED device.} An illustration of various noise channels that are present in a device. Here, a lithographically patterned chip is positioned within a circuit board holder (light yellow), which in turn sits within a metal enclosure (blue). This environment can be lossy (modeled by resistors) due to effects such as unwanted couplings (teal), spurious modes (red), material dissipation (maroon), and radiation (dark brown), etc. Inset: A cross section of an edge of the lithographic structure to highlight the surfaces losses due to different interfaces. The three interfaces of interest with thickness $t$ are shown are MS (orange), MA (green), and SA (blue). The divided regions of the superconductor (thickness $h$) refer to the interior regions (grey) and perimeter regions (purple). Their individual contributions have been simulated and compared across different designs in Ref.~\cite{wang2015_surface}.
}
\end{figure}

While the Hamiltonian parameters and input/output coupling parameters are largely determined by the circuit design to a good accuracy, decoherence effects in quantum circuits are dictated by small residual interactions with a large number of environmental degrees of freedom that are not under direct control of the circuit designer. Some of these interaction channels produce noise ``transverse" to the quantization axis of the qubit, leading to energy relaxation, or $T_1$ processes, and some produces ``longitudinal noise" that leads to dephasing, or $T_{\phi}$ processes. (See Ref.~\cite{krantz2019_engineer} for a more detailed review of these concepts.) An illustration of some of the potential sources of decoherence in a typical planar cQED device is shown in Fig.~\ref{fig:losses}. A quantitative and microscopic understanding of many of these decoherence channels remains an active research frontier of fundamental importance. Nevertheless, based on phenomenological models, there are some overarching design principles for optimizing the coherence properties. 

For each type of physical decoherence mechanism, there are three factors that are multiplied to dictate the magnitude of the decoherence rate:
\begin{enumerate}
\item The matrix element that governs the susceptibility of the qubit Hamiltonian to the qubit-environment coupling operator. The simplest example is the transmon, whose susceptibility to charge noise is diminished by adopting a carefully chosen ratio of $E_J /E_C$. In more advanced circuits, the suppression of this matrix element requires fundamental changes to Hamiltonian itself, such as the case of a ``protected $0$-$\pi$ qubit"~\cite{gyenis2020_experimental}.  
\item The intensity of microscopic noise sources in the solid-state environment: examples include the density of two-level systems (TLS) or quasiparticles. Improvements on this factor hinge on the property of the materials, which depends on the techniques of device fabrication and cryogenic measurement setups. 
\item The geometric coupling efficiency between the noise sources in a given material and the quantum mode of interest: This factor is best captured by the concept of participation ratios, which provides a convenient framework for analyzing and optimizing the device layout for decoherence considerations, and is the focus of this subsection. 
\end{enumerate} 

Here, we focus on mitigation of energy relaxation due to dielectric loss, which is one of the most prominent decoherence mechanisms in cQED devices, and use it as an illustration for the design considerations of transmon qubits. Dielectric loss is widely attributed to microscopic two-level-system (TLS) defects in material interfaces and (to a lesser degree) the bulk materials. They couple to the charge degree of freedom, $\bn$, of the resonator or qubit mode via an electric dipole moment. Macroscopically, the intensity of dielectric loss in a given material can be described by adding an imaginary part to the material's dielectric constant: $\epsilon\rightarrow\epsilon(1+i\tan{\delta})$, where $\tan{\delta}\ll1$ is the loss tangent. The decay rate induced by dielectric (i.e.~capacitive) loss, $\Gamma_{cap}$, can be decomposed into contributions from various materials or components:
\begin{equation}
  \Gamma_{cap} = \eta_{n} \omega\sum_i p_i \tan\delta_i,
  \label{eq:p_ratio}
\end{equation}
where $p_i$ is the capacitive participation ratio of the $i^{th}$ material, defined as the proportion of electric field energy of the quantum mode stored in this material. $\delta_i$ is the loss tangent, $\omega$ is the angular frequency of the mode, and $\eta_n$ is an adjustment factor based on the charge-coupled transition matrix element: $\eta_n=1$ for a harmonic oscillator and is approximately 1 for a transmon.

Typically, circuit elements with planar geometries have the majority ($\approx90\%$) of their electric energy stored in the bulk substrate. Therefore, having a substrate with low loss tangent, such as crystalline sapphire or silicon (tan\,$\delta_i\lesssim10^{-7}$), is a prerequisite to achieve high coherence. Most deposited dielectric materials tend to have a much higher loss tangent, and their presence, if absolutely necessary, must be strictly limited and their participation ratios kept relatively low. Even without deposited dielectric materials, amorphous layers often form at the metal-substrate (MS), substrate-air (SA), and metal-air (MA) interfaces, as shown in the inset of Fig.~\ref{fig:losses}. These few-nanometer-thick interface layers store a small fraction $p_i$ of the field energy but are often the limiting factor of qubit coherence due to their high tan\,$\delta_i$ (on the order of $10^{-3}-10^{-2}$). It is therefore helpful to minimize the surface participation ratios for these layers when choosing qubit geometries whenever possible. 

The rule of thumb for estimating surface dielectric participation ratios is that it scales inversely with the feature size of the capacitive elements. This can be conceptually understood by conformal mapping of complex geometries into parallel-plate capacitors~\cite{murray2018_analytical}. While the absolute magnitude of the surface participation ratios depend on rather subjective assumptions of the thickness $t_{i}$ and dielectric constant $\epsilon_i$ of the surface layers, it allows relative comparison of surface-loss sensitivity across different devices. For example, one may crudely estimate that $p_{MS}\approx \frac{\epsilon_b}{\epsilon_{MS}}\frac{2t_{MS}}{d}$ for a pair of capacitive electrodes separated by a (loosely-defined) distance $d$ on a substrate with dielectric constant $\epsilon_b$. For a transmon with a CPW-style capacitor which has a gap width of 20 $\mu$m, if we assume $d=20$ $\mu$m, $t_{MS}=3$ nm, $\epsilon_{MS}=\epsilon_b=10$, we find $p_{MS}\approx3\times10^{-4}$. 

For more detailed calculation of surface participation ratios, finite-element numerical simulations are often used. To address the huge span of length scales (from millimeter for the size of a typical transmon to the sub-nanometer spatial resolution needed to compute the electric field inside the surface layers correctly), a common strategy is to focus on a 2D cross-sectional or a 3D regional (small-volume) electrostatic simulation, where the surface layers can afford to be explicitly represented in the geometry. One can then use symmetry-based arguments~\cite{wenner2011_surface} to extend these results to the full qubit, or use scaling properties to embed these results into a full-scale high-frequency simulation with lower resolution~\cite{wang2015_surface}. One observation from numerical calculation is that the concentrated field energy near the edges and corners of the device structures contribute to surface participation ratios quite substantially. This has prompted the practices of using rounded corners or circular shapes in the device layout, and etching trenches into the substrate for a depth on the scale of $\mu$m's~\cite{Bruno15Reduced, vissers2012_reduced,calusine2018_analysis}. 

Minimizing surface loss generally calls for larger feature sizes for the shunting capacitors of qubits, but there are a few trade-offs and limitations. One trade-off is radiation loss: the larger footprint for coplanar electrodes results in a larger electric dipole that radiates more effectively into the 3D space inside the device package. If this radiation is coupled to low-Q package modes or a continuum of modes, the qubit mode will incur additional relaxation. A related problem in multi-qubit devices is unintended cross-talk between distant qubits. Therefore, the use of geometrically larger qubits poses more stringent requirements on the microwave hygiene of the device package. Furthermore, even with full control of the package modes such as in a 3D cQED architecture, geometric minimization of surface participation ratio is ultimately limited by the presence of the (nano-scale) Josephson junctions and the narrow superconducting leads near the junctions. The ``3D transmon" design~\cite{paik2011_observation} is approaching such a limit, where the contributions from the leads (e.g.~$p_{MS,leads}\sim2\times10^{-5}$ under the same assumptions of $\epsilon_{MS}=10$ and $t_{MS}=3$ nm) are no longer negligible~\cite{wang2015_surface}. % and Eq.~(\ref{eq:pppr}) no longer applies.

For inductive dissipation mechanisms, a similar analysis can be applied. Part of the total inductive energy is associated with the geometric inductance of the circuit and is present in the form of magnetic field energy in free space and in the dielectric materials. It is usually assumed that non-magnetic materials do not contain a meaningful imaginary permeability component to dissipate magnetic field energy. The other part of the inductive energy is associated with the kinetic inductance of the Cooper pairs, and its ratio to the total inductive energy is known as the kinetic inductance fraction $\alpha$. For a superconducting resonator, the kinetic inductance energy is stored in the surface current of the superconductor and can be computed from the surface integral of magnetic field energy and the penetration depth of the superconductor (20-50 nm for Al~\cite{maloney1972_superconducting} and Nb~\cite{nazaretski2009_direct}). In this case, $\alpha$ can be interpreted as an inductive surface participation ratio, which is inversely proportional to the feature size of the resonator. Interestingly, $\alpha$ can be measured experimentally using the temperature dependence of the resonator frequency based on the Mattis-Bardeen's formula for AC conductivity of a BCS superconductor~\cite{reagor2013_reaching}, which provides a useful tool to quantitatively investigate this effect in addition to finite element simulations. 

If we decompose the kinetic inductance fraction $\alpha$ into contributions $\alpha_i$ from different superconducting materials or different spatial regions $i$, the quality factor of the resonator due to inductive loss can be written as: 
\begin{align}
  \Gamma_{ind}=\eta_{\phi} \sum_i \omega \frac{\alpha_i} {Q_{s,i}},
  \label{eq:ind_p_ratio}
\end{align}
where each region of the material has its ``surface Q", $Q_{s,i}$. The finite surface Q indicates the presence of Ohmic loss in the surface current, which can originate from the presence of quasiparticles~\cite{catelani2011_quasiparticle}, vortices~\cite{nsanzinesa2014_trapping}, or normal regions that might be caused by imperfect superconductor-to-superconductor interfaces~\cite{dunsworth2017_characterization, brecht2015_demonstration}. Furthermore, there is also an intrinsic limit on the surface Q due to dissipation from excitation of phonons in the presence of an AC fields~\cite{Tinkham96}. 

Resonators made from Al or Nb films on the order of 100 nm thick can be made with low kinetic inductance ($\alpha <0.01$) and hence relatively insensitive to inductive loss. However, superconducting qubit modes often have $\alpha$ approaching 1, because the inductive energy of the Josephson junction dominates the magnetic field energy. Therefore, as in the case of a transmon, the inductive loss is usually dominated by quasiparticle tunneling across the junction~\cite{catelani2011_quasiparticle, serniak2019_direct}. As a result, reduction of inductive loss in transmons cannot be achieved by reducing surface participation ratios through variations of electrode geometry. Instead, quasiparticle traps realized with vortices or normal metals can be employed to reduce non-equilibrium quasiparticle density and consequently, the total inductive loss~\cite{wang2014_measurement, hosseinkhani2017_optimal}. Although such measures degrade the average $Q_s$ of the superconducting film, it is often a winning trade because the kinetic inductance fraction of the tunnel junction far exceeds the surface contribution from the film.

\subsection{Device fabrication}
One of the most crucial aspects in constructing cQED devices is the ability to fabricate JJs with precise and reproducible parameters. Generally, JJs are fabricated on either silicon or sapphire substrates using standard electron-beam lithography techniques, followed by double-angle evaporation using either the Dolan-bridge~\cite{dolan1977_offset} or the bridge-free method~\cite{lecocq2011_junction}. Additional techniques that potentially allow more flexible junction geometries and potentially better yield are also being explored~\cite{mamin2021_merged, kreikebaum2020_improving, costache2012_lateral, wu2017_overlap, stehli2020_coherent, bilmes2021_in-situ}.

Although the fabrication techniques are well-established, there are a few key considerations specific to producing reliable and reproducible quantum devices. In particular, the presence of spurious TLSs in the device can lead to lossy interfaces between the substrate and the superconducting film, as discussed in the preceding subsection. Furthermore, parasitic  coupling  from  the circuit elements to these spurious TLSs in the substrate or the superconducting film also results in fluctuations in system parameters and deterioration of their performance~\cite{steffen2017_recent}. One commonly employed technique to reduce TLSs level in typical cQED devices is to ensure that the substrate-metal interface is pristine. This is achieved by performing \textit{in-situ} isotropic oxygen-plasma cleaning in the evaporator before metal is deposited. The intensity of this process must be optimized carefully because excessive cleaning could damage the substrate surface while insufficient cleaning will leave undesired residuals, both would result in a deterioration in the coherence properties of the device. In addition, the appropriate cleaning procedure has also been shown to reduce the aging of JJs and enhance the stability of the devices~\cite{pop2012_fabrication}. For silicon substrates, cleaning in hydrofluoric acid (HF) and terminating the surface with hydrogen prior to film deposition has been shown to reduce losses due to native oxides, residues, and surface defects~\cite{muller2019_towards}. Apart from surface treatment, less reactive superconducting materials have also been adopted to avoid amorphous surface oxides, and fabricating electrode and junction barriers from crystalline materials. A detailed discussion on the various sources of TLS and their respective mitigation techniques is provided in Ref.~\cite{muller2019_towards}.

After JJs are fabricated, we typically want to have a quick and simple way to probe their key characteristics before cooling down the device to 20\,mK. At room temperature, the tunnel junctions behave like a normal resistor. We can perform a quick verification of the Josephson energy, $E_{J}$, by measuring the normal state resistance of the junction, $R_{n}$, which is related to $E_{J}$ by the Ambegaokar-Baratoff relation~\cite{ambegaokar1963_tunneling}, 
\begin{equation}\label{eq:AB}
I_{C}R_{n} = \Phi_0\Delta(T)\tanh{\frac{\Delta(T)}{2k_{B}T}}.
\end{equation}
Here, $\Delta(T)$ is the superconducting energy gap at a certain temperature, $T$. This relation indicates that the product of the critical current, $I_{C}$, and normal state resistance, is an invariant quantity depending only on the material property and the temperature. 

At $T=0$, Eq.~\ref{eq:AB} reduces to 
\begin{equation}\label{eq:AB_EJ}
E_{J} = \frac{\Phi_{0}\pi\Delta}{2eR^{'}_{n}},
\end{equation}
where $R'_{n}$ is the normal-state resistance at $T=0$ and $\Phi_0\equiv h/2e$ is the flux quantum. %This expression implies that we are able to predict the junction characteristics using its normal resistance before cooling the device down. 

It is important to note that $R'_{n}$ typically increase by $10$-$20\%$ compared $R_{n}$ due to the change in effective barrier height~\cite{gloos2020_wide}. In a typical device fabricated with thin-film aluminum, standard critical temperature measurements of the deposited film indicate that $\Delta\approx150$-$200\,\mu$eV. Combining these parameters, we find that a junction with a room-temperature resistance of 6\,k$\Omega$ corresponds approximately to an inductance of $L_J = 8\,$nH when it is cooled to 20\,mK. 

Another useful component in cQED devices is the 3D superconducting cavity, which is often employed to realize high-Q quantum memories or bosonic logical elements. They are usually made with high-purity (4N) aluminium through conventional machining techniques. To enhance their intrinsic quality factor, chemical etching is performed to remove residues and machining defects on the metal surface~\cite{reagor2013_reaching}. In addition, techniques such as annealing are also used to further improve the consistency and reproducibility~\cite{kudra2020_high} of these 3D cavities. By combining the coherence of 3D systems with the small form-factors of planar devices, micromachining~\cite{brecht2016_multilayer} provides another promising strategy for realizing robust and compact superconducting cavities. Although a relatively recent development, remarkable progress has been made in improving the performance of these micromachined cavities~\cite{lei2020_high} and the effective integration of transmon devices~\cite{brecht2017_micromachined}. 

%%%%%%%%%%%%%%%%%%%%%%%%%%%%%%%%%%%%%%%%%%%%%%%%%%%%%%%%%%%%%%%%%%%%%%%%%%%%%%%%%%%%%%%%%%%%
\section{Creating an effective quantum environment}\label{sec:environment}
The performance of a quantum device relies heavily on the operating environment it is embedded in. In general, an effective quantum environment must provide thorough isolation from the various sources of noise both from the free space and the transmission lines while allowing coherent transmission of control signals to implement fast quantum operations. For cQED systems, creating such an environment entails careful considerations in the design of the cryogenic set-up as well as the room temperature microwave signal processing chain. Here, we aim to provide some practical guidelines on the techniques and practices used in achieving a robust quantum environment. A more advanced discussion can be found in Ref.~\cite{kinner2019_engineering}.

\begin{figure*}[!bt]
  \centering
  \includegraphics{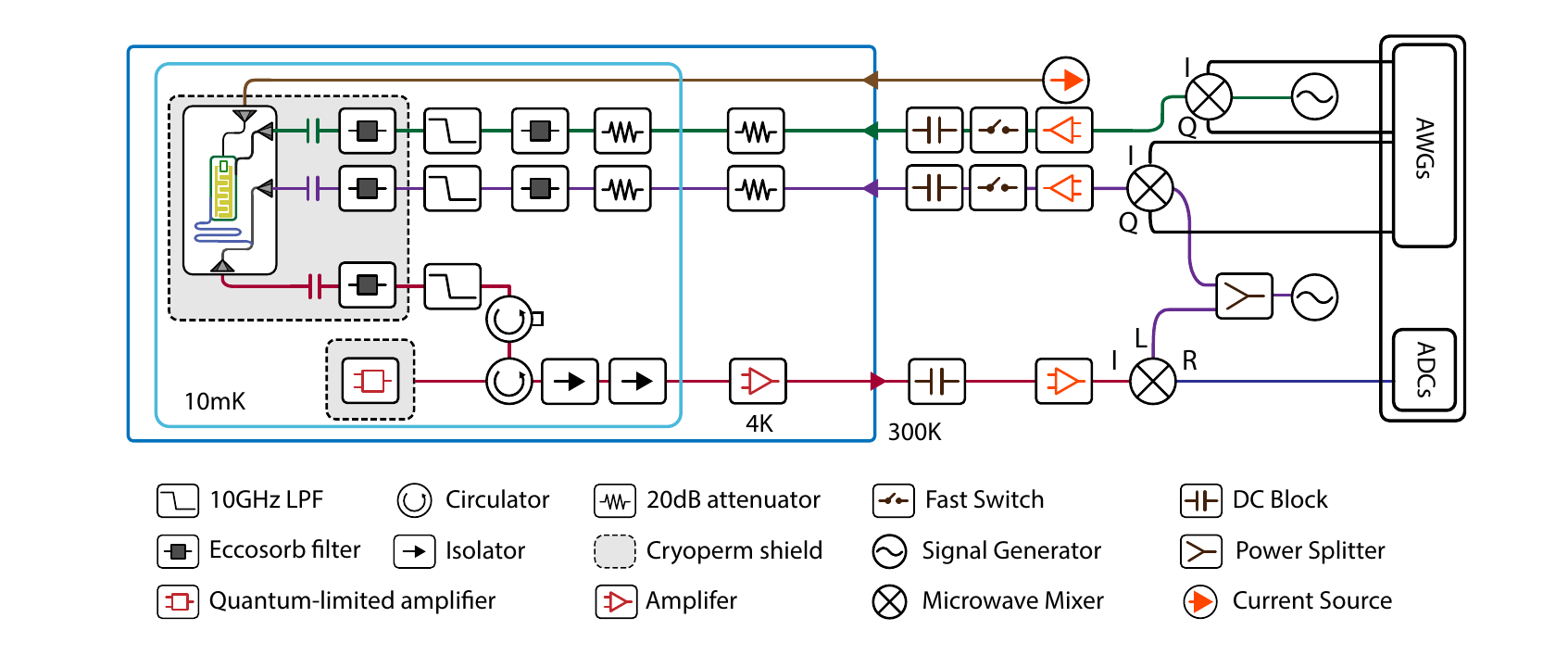}
  \caption{\label{fig:example_device}
  \textbf{A typical microwave signal processing chain for a flux-tunable transmon coupled to a readout resonator.} The input lines (green and purple) to the sample are usually wired with 20\,dB cryogenic attenuators at both the 4K and 20\,mK stage. This level of attenuation, together with the intrinsic insertion loss incurred along the coaxial cables, ensures that the thermal photons reaching the device are kept below $10^{-3}$. Commercial low-pass filters (LPF) and custom-made Eccosorb filters~\cite{Santavicca08, Slichter2012_measurement}are also introduced to suppress high-frequency noise. The output signal passes through a single Eccosorb filter, located inside the Cryoperm, and a low-loss filter. This typically incurs a $<2\,$dB insertion loss on the signal, which then travels through two cryogenic circulators that provide -20\,dB isolation with an additional 0.3 dB loss. The signal is routed by the two circulators to a quantum-limited amplifier which is usually tuned to operate at 20\,dB gain. Here, we have omitted the pump line to the quantum amplifier. The amplified microwave signals then pass through one or two isolators to the 4K stage where they are further amplified by a wideband high-electron-mobility transistor (HEMT) amplifier before exiting the fridge to the room temperature signal processing elements. At room temperature, the drive signals to the tranmon and resonator are generated by mixing a LO tone from a microwave signal source with DAC outputs from a quantum controller. These signals are then amplified and filtered before travelling to the qubit drive line (green) and the resonator drive line (purple), respectively. The returning signal from the resonator (red) is demodulated by mixing with the same input carrier tone and then sent to the ADC channels on the controller. The flux control is performed via a flux drive line (brown) with a current source.
  }
\end{figure*}

\subsection{Cryogenic configurations}
Superconducting circuits are operated at 20\,mK where the quantum system can be initialized in its ground states and avoid spurious thermal excitations. While such temperatures can be attained reliably using commercially-available Dilution Refrigerators (DR), careful shielding and filtering considerations must also be put in place to minimize the exposure to the residual thermal noise and stray electromagnetic radiations. In this segment, we will focus our discussion on three aspects in particular, namely, arrangement of the input microwave lines for low-noise control, configuration of the output lines to allow optimal extraction of the signal, and finally, shielding of the device itself at 20\,mK. 

To configure the microwave lines for our room-temperature control signal to reach the device at 20\,mK, we must consider thermal noise arises from both passive and active heat loads propagating down the DR. The former is caused by the flow of energy from upper to lower temperature stages in the DR, whereas the latter is due to the dissipation of applied control signals en-route to the device. 

Overall, residual thermal photons result in an elevated effective temperature of device. In particular, the readout resonators, which are typically strongly coupled to the transmission lines and the environment, can be prone to a non-negligible thermal photon population, $\bar{n}_\mathrm{th}$. This, in turn, lead to dephasing, and a reduced $T_2$, of the qubit that couples to the resonator, at a rate given by~\cite{Gambetta06, sears2012_photon, schoelkopf2003_qubits}:
\begin{equation}
  \Gamma^{\mathrm{th}}_{\phi} = \frac{\bar{n}_\mathrm{th}\kappa\chi^2}{\kappa^2 + \chi^2},
\end{equation}
where $\kappa$ is the linewidth of the resonator and $\chi$ is its dispersive coupling strength to the qubit. Therefore, it is crucial to to reduce the amount of residual thermal photons in our quantum system in order to achieve better coherent times. 

In order to suppress thermal noise, attenuators must be introduced along the microwave lines in the DR. To illustrate the importance of choosing the appropriate attenuation, let us consider the effect of the heat load at each temperature stage. The thermal energy at a given temperature, $T_i$, translates into an influx of photons propagating down to the device, with an average number, $\langle n_i\rangle$, given by
\begin{equation}\label{eq:nbar}
\langle n_i\rangle = \frac{1}{e^{\hbar\omega/k_B T_i} - 1}.
\end{equation}
It must be suppressed to or below the temperature of the subsequent cooling stage in order to minimize the propagation of thermal noise. This is achieved by introducing appropriate cryogenic attenuators at each temperature stage. An attenuator can be modeled as a beamsplitter which transmits a small fraction of the incident power while at the same time, adds a certain amount of thermal photons due to the black-body radiation at the temperature stage where it is located. To illustrate this, let us consider a system with an attenuator $A_1$ placed at a higher temperature stage and $A_2$ on a lower stage. The resulting average photon number at is given by
\begin{equation}\label{eq:propagation_nbar}
\langle n_{f}\rangle = \frac{\langle n_i\rangle}{A_1A_2} + \left(1-\frac{1}{A_1}\right)\frac{\langle n_1\rangle}{A_2} + \left(1-\frac{1}{A_2}\right)\langle n_2\rangle,
\end{equation}
where $n_i$ is the average number of photons going into the system and $n_{1, 2}$ the photons at each of the temperate stages as given by Eq.~\ref{eq:nbar}. Note that the attenuations, $A_{1,2}$, are in linear units which are related to the typical units of attenuators by $A=10^{A'/10}$ where $A'$ is the level of attenuation given in dBs. 

From Eq.~\ref{eq:propagation_nbar}, we infer that attenuators are more effective in suppressing thermal noise when they are placed on the lower temperature stage. However, most attenuators employ dissipative elements which could cause additional heating if the dissipation exceeds the cooling power of the DR at the specific temperature. In order to balance both of these effects, reflective elements such as custom reflective attenuators~\cite{wang2019_cavity} have been developed to achieve sufficient attenuation without significant dissipation at the 20\,mK. Additionally, directional couplers can also be used to route the excess microwave power back to the higher temperature stage (typically 4\,K) where there is significant higher cooling power such that the dissipation does not have any measurable impact on the temperature. A typical configuration of the transmission lines connecting the device in a DR to room temperature is shown in Fig.~\ref{fig:example_device}. 

In addition to measures to reduce the thermal noise due to active heat loads, precautions should also be taken to filter out spurious frequency components that could cause unwanted transitions or couplings. This is achieved by adding cryogenic filters at the 20\,mK plate of the DR. Low pass filters that can provide more than 40dB suppression of frequencies above 10GHz or 12GHz are now commercially available. Another widely adopted element is the Eccosorb filter, which employs a castable silicone rubber material capable of effectively absorbing microwave signals (20\,GHz and above) to protect the device from high-frequency radiation.

The output lines are responsible for carrying the resulting signals from the device from base temperature back to room temperature for analysis. These signals, typically only on the single photon level, can be easily degraded by noise or dissipation. Hence, they must be handled carefully to ensure clean, stable, and effective collection of quantum information from the device.

First, it is crucial to adequately suppress the thermal noise propagating from the higher temperature stages while allowing the signal to propagate without significant attenuation. To balance these two requirements, directional components such as circulators and isolators are employed on the output lines, instead of attenuators. These non-reciprocal elements provide a low-loss path for the precious quantum signal extracted from the device to travel to room temperature while effectively reducing thermal noise flowing in the opposite direction. 

Furthermore, we also need to enhance the weak outgoing signal through a well-designed chain of amplifiers such that it can be distinguished from noise as it propagates back to room temperature. In other words, the role of each amplifier along the outgoing signal path is to make the signal stand out against the noise added by the subsequent amplifier as well the thermal photons from the higher temperature stages. 

When designing amplification configurations, it is important to note that amplification always comes at the cost of elevated noise temperature. This causes a degradation of the signal-to-noise ratio (SNR) despite the overall increase of the signal power. The noise temperature, $T_N$, provides a measure of the noise added by the amplifier. For an amplifier with a power gain of $G$, we can relate the amplified signal, $P^{S}_{out}$, and noise, $P^{N}_{out}$ to the input powers $P^{S, N}_{in}$, by
\begin{align}\label{eq:amp_powers}
P^{S}_{out} &= GP^{S}_{in} ,\\
P^{N}_{out} &= G(T_{in} + T_{N})k_{B}B, 
\end{align}
where $B$ is the bandwidth of the noise received by the amplifier, $T_{in}$ and $T_N$ the effective temperatures associated with the input signal and the noise of the amplifier, respectively. Here, the effects of quantum fluctuations are negligible as $k_B T \gg \hbar\omega$. With this, we can now write the SNR at the output of the amplifier as $\mathrm{SNR}_{out} = P^{S}_{in}/P^{N}_{out} = P^{S}_{in}/(T_{in} + T_N)k_{B}B$. Comparing this with the input SNR, given by $\mathrm{SNR}_{in} = P^{S}_{in}/T_{in}k_{B}B$, we arrive at the relation
\begin{equation}\label{eq:snr}
\frac{\mathrm{SNR}_{in}}{\mathrm{SNR}_{out}} = 1 + \frac{T_N}{T_{in}} > 1, 
\end{equation}
indicating that the SNR after the amplifier is always going to be worse than that at the input. Therefore, in general, amplification should be applied with caution and avoided if possible. On the incoming control lines, this means maximizing IQ modulated drives within the operational range of the mixer. More importantly, on the output line carrying information from the devices, it is crucial to minimize the noise associated with the first gain stage. Practically, this is achieved using a quantum limited amplifier connected to the output of the readout resonator with minimal attenuation in between.

In standard cQED measurement set-ups, the amplification chain usually begins with a quantum-limited amplifier. As the name suggests, these amplifiers only add the minimum amount of noise allowed by the laws of quantum mechanics~\cite{Caves1982_quantum, Clerk10}. They are designed to be either phase-preserving or phase-sensitive. While the former has the advantage of preserving the phase of the incoming signal, it must necessarily double the vacuum fluctuations (i.e. add ``half a photon'' of noise). On the other hand, the latter amplifies one quadrature of the signal without adding any noise, at the cost of de-amplifying the other quadrature by the same amount. Experimentally, phase-preserving amplifiers are featured heavily in most applications for their simplicity. In addition, some designs also offer the advantage of an extremely large bandwidth (several GHz) and compatibility with larger input signals without being saturated~\cite{Macklin15, Roy2015_broadband, planat2020_photonic}. Hence, such phase-preserving amplifiers provide the capability to measure several qubits simultaneously through multiplexing~\cite{Kundu2019_multiplexed}. A more detailed discussion about the operating principles and capabilities of quantum-limited amplifiers can be found in Ref.~\cite{aumentado2020_superconducting}. Following the quantum-limited amplifier, a commercial wide-band cryogenic High Electron Mobility Transistor (HEMT) amplifier is also employed at the 4K stage to further boost the signal before it is acquired at room temperature. 

We can experimentally verify that a given amplification chain satisfies its purpose with a spectrum analyzer at room temperature. As the chain amplifies the vacuum fluctuations as well as the signal, the amplified fluctuations should become larger than the thermal noise. Therefore, over the bandwidth of the quantum-limited amplifiers, the noise-floor measured by the spectrum analyzer should be raised when the amplifier is turned on. The amount by which the noise-floor is raised is sometimes referred to as the Noise Visibility Ratio (NVM)~\cite{clerk2010_introduction}.

Finally, the device must also be shielded from stray magnetic fields. In several studies~\cite{wang2014_measurement, vool2014_nonpoissonian, nsanzinesa2014_trapping}, it has been shown that when a device is cooled to 20\,mK in the presence of a magnetic field of more than 0.1 Gauss, vortices can be trapped in the thin film superconductor and cause an appreciable reduction in coherence properties. To mitigate this, we house the sample in Cryoperm shields, which are made out of high permeability nickel alloys and treated specifically to ensure robust magnetic screening properties. However, the best magnetic shielding is achieved with bulk conventional superconductors, which only work at cryogenic temperatures. These two types of shields are often used together to reject any stray magnetic field.%, first when the device is cooled-down, and then when it reaches the critical temperature of the superconducting shield.

It is also crucial that we minimize the residual magnetic field inside the shield by using only components (screws, RF connectors, and cables, etc.) made from non-magnetic materials. The only exception to this is the presence of the Eccosorb filters. Although they are mildly magnetic, it is still common practice to place them within the shield, directly before the coupling port to the device. Some recent works have shown that this effectively reduces the residual thermal population of the devices without any measurable degradation of their coherence properties~\cite{serniak2019_direct}. \\

\subsection{Microwave signal processing}\label{subsec:rfconfig}
To execute the desired experiment on cQED devices, we must be able to both generate the appropriate control pulses to perform quantum operations and digitize the signal returning from the device to obtain information about the quantum system. Traditionally, these tasks are performed using a combination of arbitrary waveform generators (AWG) and digitizers/Analog-Digital Converters (ADC). In recent years, FPGA-based systems have become increasingly popular due to their ability to integrate the functionalities of the AWGs and ADCs in a single compact hardware solution. The current state-of-the-art commercial systems are able to provide multiple well-synchronized output channels, with a sampling rate of at least 0.5-1 GSample/second and a typical vertical resolution of 16 bits in output voltages, well in excess of what is required for high-fidelity control pulses~\cite{van_Dijk_2019}.

To perform high-fidelity gates, we typically want to be able to perform operations as fast as possible. 
In the case of single-qubit gates on transmons, the duration of a $\pi$-pulse, i.e.~a $\pi$ rotation along the X or Y axis that flips the qubit state between $\ket{0}$ and $\ket{1}$, is limited by the presence of the nearby transition to the second excited state. As such, short ($\tau_g \sim 20~\mathrm{ns}$~\cite{Chen16b}) Gaussian pulses with simple DRAG corrections~\cite{motzoi2009_simple, Chow10b} are usually chosen to enact these $\pi$-pulses to minimize unwanted transitions outside the computational space. Following Ref.~\cite{krantz2019_engineer}, we can use the Rabi-driving equation to determine the required pulse amplitude $V_0$: 
\begin{equation}
    \Theta(t) = -\Omega V_0 \int_0^t s(t')dt'
\end{equation}
where $\Theta(t)$ is the rotation angle at time $t$, $s(t)$ the pulse amplitude, and $\Omega$ is the Rabi-frequency, given by 
\begin{equation}
    \Omega = \frac{C_c}{C_\Sigma}Q_{\mathrm{zpf}}.
\end{equation}
Here, $C_c$ is the capacitive coupling to the drive line, $C_\Sigma$ the effective capacitance, and $Q_{\mathrm{zpf}}=\sqrt{\hbar/2Z}$ the zero point fluctuation, and $Z=\sqrt{L/C}$ the impedance of the circuit to ground. Filling in typical values, this results in a required power of $-66~\mathrm{dBm}$~\cite{Krinner19} applied to the qubit, which translates into a peak power of $\sim - 6~\mathrm{dBm}$ or an amplitude of $V_P\approx 160~\mathrm{mV} $ at the input of the dilution refrigerator (assuming typical attenuation of $\sim60~\mathrm{dB}$ in the microwave lines) .

%In order to prevent transitions to this non-computational state, simple Gaussian shaped (DRAG~\cite{motzoi2009_simple, Chow10b}) pulses with a duration of $\tau_g \sim 20~\mathrm{ns}$~\cite{Chen16b} \footnote{It is possible to reduce the pulse duration further using optimal control techniques~\cite{werninghaus2021_leakage}.} setting the required Rabi frequency. 
%Following Ref.~\cite{krantz2019_engineer}, we can use the Rabi-driving equation to determine the required pulse amplitude $V_0$: 
%\begin{equation}
  %  \Theta(t) = -\Omega V_0 \int_0^t s(t')dt'
%\end{equation}
%where $\Theta(t)$ is the rotation angle at time $t$, and $s(t)$ the pulse amplitude. 
%$\Omega$ is given by 
%\begin{equation}
%%    \Omega = \frac{C_c}{C_\Sigma}Q_{\mathrm{zpf}}, 
%\end{equation}
%where $C_c$ is the capacitive coupling to the drive line, $C_\Sigma$ the effective capacitance, and $Q_{\mathrm{zpf}}=\sqrt{\hbar/2Z}$ the zero point fluctuations and $Z=\sqrt{L/C}$ the impedance of the circuit to ground.  Filling in typical values, this results in a required peak power of $-66~\mathrm{dBm}$~\cite{Krinner19}. Combined with a typical attenuation of $\sim60~\mathrm{dB}$ this results in a required peak power of  $\sim - 6~\mathrm{dBm}$ or an amplitude of $V_P\approx 160~\mathrm{mV} $ at the input of the dilution refrigerator.

Next, in order to generate these control pulses, we often combine the microwave signals emitted by a local oscillator (LO) and an intermediate frequency (IF) tone generated by the DAC channels of the quantum controller in a double-balanced IQ mixer~\cite{Pozar05}. Ideally, the mixer combines the LO and IF input to provide the desired radio frequency (RF) tone at a single-sideband (SSB) with frequency $\omega_{RF} = \omega_{LO}+\omega_{IF}$, where the intermediate frequency can be either positive or negative. The applied voltages on the I (in-phase) and Q (out-of-phase) modulation ports afford full control over the amplitude and phase of the resulting RF tone. In practice, there are often several sources of imperfection that must be addressed to achieve a clean SSB signal~\cite{jolin2020_calibration}. 

The first consideration is mixer non-linearity, which leads to distortions of the resulting signal. To prevent this undesirable effect, it is crucial to ensure that IF input powers are kept below the 1dB compression point of the specific mixer employed. 

The next imperfection to address is the presence of LO leakage, which is a result of some residual direct coupling between the LO and RF ports of the mixer. If left uncorrected, this could lead to a continuous drive at the resonance frequency of the quantum element and thus, causing unwanted Stark shifts or even elevated effective temperature of the device. This leakage is suppressed by applying well-chosen voltage offsets to the IF inputs. Experimentally, we can systematically adjust the offsets while monitoring the power at the LO frequency with a spectrum analyzer to find the optimal configurations that most effectively eliminate this leakage tone. 

After tuning the LO leakage, let us know consider the choice of SSB and the technique for eliminating the opposite component. Both SSBs are present in real IQ mixers because the I and Q inputs are not perfectly out of phase with each other (skewness). Furthermore, the same input powers on the I and Q ports do not translate into equal powers at the RF output (amplitude imbalance). Therefore, in order to suppress one of the spurious sideband tone, we must address both the skewness and amplitude imbalance of the mixer systematically. The former relates to a relative phase between the I and Q inputs while the latter is tuned by introducing a scaling factor between the input powers. These parameters are again, algorithmically adjusted to achieve the minimal leakage of the undesired sideband. 

In typical cQED experiments, the lower sideband (LSB) at $\omega_{LO}-\omega_{IF}$ ($\omega_{IF}\sim -$50-100\,MHz) is often employed to drive the qubits. This choice ensures that any residual LO or upper sideband (USB) leakage is far away from the second transition which is lower in frequency than that between the ground and first excited state.

Finally, the resulting SSB-modulated drive signal is sometimes amplified before transmitted to the DR. As discussed in Sec.~\ref{sec:environment}.2, this boots the overall signal level at the expense of the SNR. Hence, it is crucial that the amplified microwave drive is subsequently attenuated and filtered in order to suppress the propagation of the added noise to the quantum device. Furthermore, a fast microwave switch with $<10$\,ns response time is often added after the amplifier to prevent the added noise from propagating to the quantum device during idle times. An example of a standard room temperature RF signal processing chain is shown in Fig.~\ref{fig:example_device}.

%%%%%%%%%%%%%%%%%%%%%%%%%%%%%%%%%%%%%%%%%%%%%%%%%%%%%%%%%%%%%%%%%%%%%%%%%%%%%%%%%%%%%%%%%%%%

\section{Device characterization and calibration}
\label{sec:interacting_with_quantum_circ}

Before a cQED device can be used as a quantum processor, a series of experiments must be performed to fully characterize the system and calibrate the necessary operations. Although the exact details of this process vary with the design of the quantum device and the choice of control electronics, most cQED characterization processes share a similar set of core ingredients and a general workflow. 

In this section, we offer a brief summary of the key \textit{modus operandi} for probing the essential characteristics of a cQED device. We will start by discussing the procedure for tuning up a single qubit coupled to a readout resonantor and use it as an example to introduce the main techniques and practices that have been developed by the community. 

\begin{figure}
  \centering
  \includegraphics{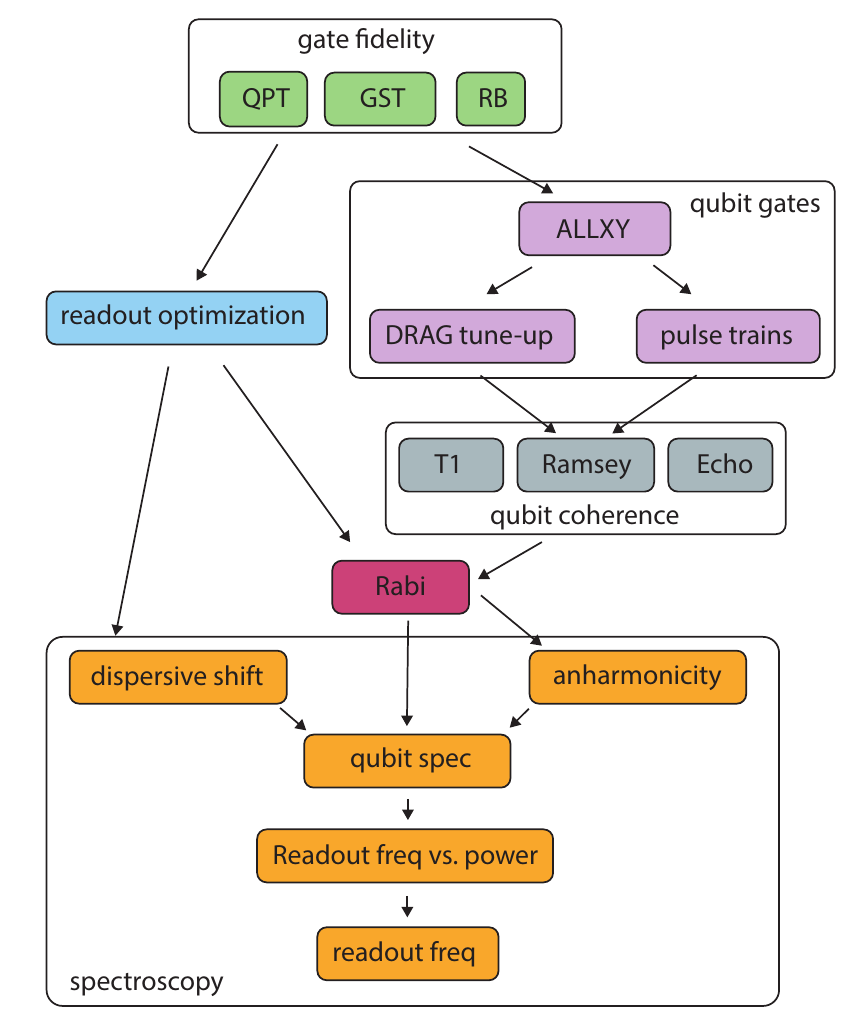}
  \caption{\label{fig:calibration_graph}
  \textbf{Tuneup of a single flux-tunable qubit}. The typical process can be visualized as a dependency graph~\cite{kelly18physical}. Nodes correspond to different calibration experiments, while edges denote dependencies between different experiments. 
  }
\end{figure}

To make this illustration more concrete, we focus on a flux-tunable transmon coupled to a readout resonator as shown in Fig.~\ref{fig:example_device}. An overview of the different calibration experiments is given in Fig.~\ref{fig:calibration_graph} in the form of a dependency graph, similar to the presentation in Ref.~\cite{kelly18physical}. 

\subsection{Spectroscopy experiments}
\label{sec:basic_char}
Spectroscopy generally refers to the measurement of intensity as a function of frequency and is commonly used in cQED experiments to determine resonance frequencies of resonators and qubits. 

Identifying the frequency of the readout resonator is usually the first step in characterizing a device. This is typically done using single-tone spectroscopy, in either the transmission, reflection, or `hanger' configuration in which one or multiple resonators are coupled to a common feedline (see~\cref{fig:example_device}). These techniques have their respective appeals and limitations. A thorough analysis of these different resonator measurement schemes can be found in Ref.~\cite{Gao08_thesis}.

Generally, transmission measurements are simple to set up and allow us to extract the amplitude and phase information of the signal as it passes through the resonator. Reflection measurements offer similar capabilities but with the added advantage of allowing a direct extraction of the internal quality factor. However, it requires a circulator in order to appropriately direct the input and output signals. This can potentially introduce additional losses in the measurement. The hanger configuration employs an unbalanced tee~\cite{Pozar05} in contrast to the circulator used in the reflection circuit. In this case, part of the input signal interacts with the resonator and upon its return, interferes with the other half that is directly transmitted through the tee. One important advantage of the hanger circuit is that it allows for many elements to be multiplexed using the same feedline. It has become a widely adopted technique in the multiplexed measurement of planar resonators.

As an example, let us consider the case of several $\lambda/4$ CPW resonators capacitively coupled to a common microwave feedline. The measured S21 of each resonator in response to a drive at $\omega=2\pi f$ is described by a Lorentzian line shape, which can be written as~\cite{Bruno15}:
\begin{equation}
\label{eq:resonator_S21}
S_{21} = A \left[1+\alpha\frac{\omega-\omega_r}{\omega_r}\right]\left[1 - \frac{\frac{|\kappa_c|}{\kappa}}{1+2i \frac{\omega-\omega_r}{\kappa}}\right] e^{i(\tau_v \omega + \phi_0)},
\end{equation}
where $A$ is the transmission amplitude away from resonance and $\omega_r$ is the qubit-state dependent resonance frequency of the readout mode. 
The parameter $\alpha$ allows for a linear variation in the overall transmission chain in the frequency range around the resonator and $\tau_v$ and $\phi_0$ relate to propagation delays to and from the sample. 

The frequency independent coupling rate $\kappa_c$ denotes the strength with which the field of the resonator couples to the transmission line and is set by the system design, as discussed in Sec.~\ref{sec:hamiltonians}. On the other hand, $\kappa_i$ is indicates the internal coherence properties of the resonator, and it is determined by the material quality and device layout, as discussed in Sec.~\ref{sec:designfab}. Finally, $\kappa$ contains contributions from both $\kappa_i$ and $\kappa_c$, i.e.~$\kappa = \kappa_c + \kappa_i$. Fitting the measured S21 data to Eq.~\ref{eq:resonator_S21} returns the key parameters $\omega_r, \kappa$, and $\kappa_c$, as described in Ref.~\cite{Bruno15, Gao08_thesis}. We can also relate these coupling rates to the associated quality factors of the mode by $Q_{c, i} = \omega_r/\kappa_{c,i}$.

Regardless of the exact measurement circuit, it is important to note that this spectroscopy experiment takes place before the optimization of the various pulses and amplification configurations. Thus, it can be challenging to obtain a clean signal at this stage. A useful technique to circumvent this issue is to perform a spectroscopy experiment in the high-power limit, where the resonator can be driven to a `bright' state regardless of the state of the qubit~\cite{Reed10}. In this regime, the signal-to-noise ratio is boosted significantly as the high microwave power used in the probing tone dwarfs the noise in the system. This provides a convenient tool to quickly locate the bare resonance frequency of the readout mode $\omega_R$ without tuning up the rest of the system. 

\begin{figure}[!t]
  \centering
  \includegraphics[scale=1]{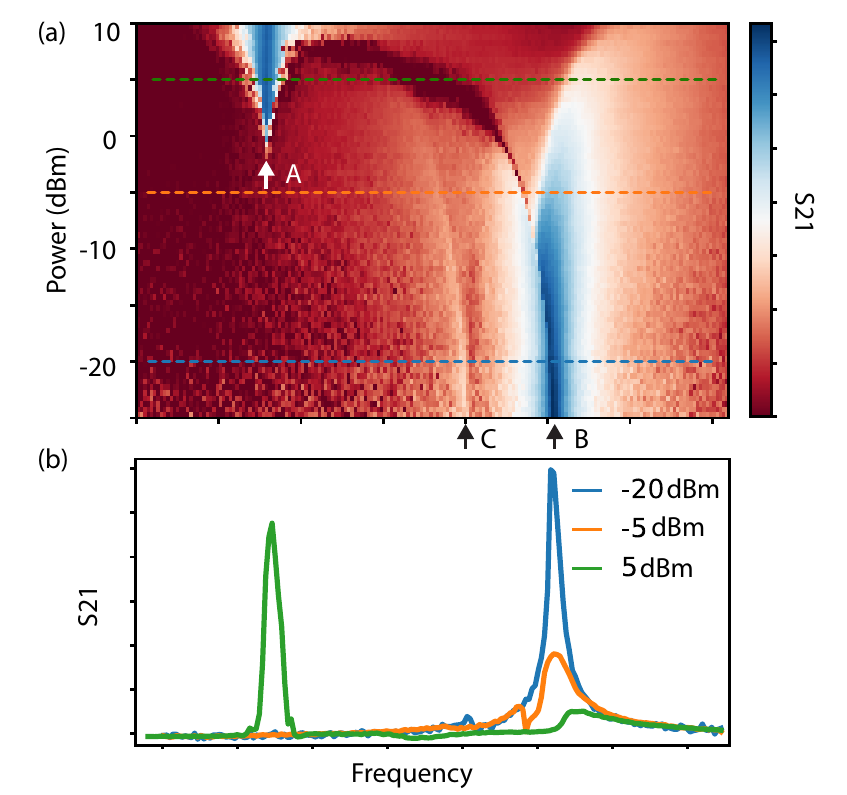}
  \caption{\label{fig:resonator_power}
  \textbf{Resonator power scan}. (a) Outcome cavity spectroscopy experiments repeated at different input powers. The three features of interest are indicated by the arrows, corresponding to the high-power tranmissoin peak (A), lower power transmission peaks with transmon in ground state (B) and residual excited state (C), respectively.  (b) A line cut at three different powers, highlighting the clear shift in resonance frequency of the resonator as it transitions from the high- to low-power regimes. 
  }
  \label{fig:cavspec}
\end{figure}

Subsequently, we can determine the `dressed' frequency of the resonator $\tilde\omega_R$ by repeating the spectroscopy experiment at different powers. This power sweep also provides a convenient test of whether the resonator is coupled to a qubit mode. If a qubit is dispersively coupled to the resonator, the resonance frequency will be different in the low-power (one or few photons) and high-power (hundreds of photons or more) regimes. In the transmission circuit, a typical power-vs-frequency sweep reveals a sharp resonance at the bare cavity frequency, which then undergoes a clear shift as the power is progressively reduced to the few photon limit, as shown in Fig.~\ref{fig:cavspec}(a). 
This power-dependent dispersive shift is the the Lamb-shift $\Lambda_\chi=g^2/\Delta$ [\cref{eq:lamb_shift}].
The profile of the resonance at this power becomes a well-defined Lorentzian with a sufficiently narrow line-width compared to the magnitude of the dispersive coupling (Fig.~\ref{fig:cavspec}(b)). Furthermore, based on this power sweep, we can now choose an appropriate readout power for the subsequent characterization experiments. It should produce a sufficiently high signal without causing non-linear distortions or broadening of the resonator response. In the example shown in Fig.~\ref{fig:cavspec}(a), we could choose $-20\,$dBm and $-25\,$dBm for a high-contrast readout. It is also worth pointing out that in this particular device, the transmon qubit has some residual thermal population, which gives rise to an additional feature (labeled point C in Fig.~\ref{fig:cavspec}(a)) besides the main low-power transmission peak. This corresponds to resonator's frequency when the transmon is in its first excited state, with the magnitude of the shift given by the dispersive coupling strength, $\chi=2E_C (\tfrac{g}{\Delta})^2$ as shown in Eq.~\ref{eq:chi}.

Next, we can use this un-optimized readout to identify the frequency of the qubit in a separate spectroscopy experiment. For flux-tunable qubits, a useful technique prior to qubit spectroscopy is to first move the qubit to the flux-insensitive sweet spot. This can be done by performing a resonator spectroscopy in the low-power regime while changing the current applied to the flux-bias line $I_{\mathrm{FBL}}$. If the qubit couples to this flux line, $\omega_r$ will depend on the flux bias through the dispersive shift $\chi$. The period of $\omega_r$ as a function of $I_{\mathrm{FBL}}$ corresponds directly to the current per flux quantum $I_{\mathrm{FBL}}/\Phi_0$ picked up by the SQUID loop. In the regime of $\omega_q<\omega_r$, the qubit is at the sweet-spot when $\omega_r$ is at a maximum. This also provides an estimate of the dispersive shift and the rough frequency regime of the qubit. While these are rather crude estimates, it is nonetheless helpful to narrow down the possible range of the qubit frequency as it can sometimes be challenging to identify the qubit resonance, which is typically much narrower (higher-Q) compared to the readout mode. 

The technique for measuring the resonance frequency of the qubit is often referred to as a ``two-tone" continuous-wave spectroscopy. In this protocol, a constant microwave tone is applied at $\omega_r$ while a second drive ($\omega_s$), with variable frequency, is employed to probe the state of the qubit. When this second tone comes close to the resonance of the qubit, i.e. $\omega_s\approx \omega_q$, the frequency of the resonator changes due to the dispersive coupling. This shift leads to a change in the resonator's response to the constant drive tone at $\omega_r$ and hence, provides an indication of the qubit state. 

While this is conceptually straightforward, it can sometimes be challenging to implement in practice. One of the difficulties relates to the power that is applied to the qubit via the spectroscopy tone. If the power is too low, the qubit will not get excited even if the drive frequency is close to its resonance and thus, no shift will be observed. On the other hand, if too much power is applied, the line width of the resonance will increase due to power broadening~\cite{Schuster05, schuster2007_resolving}, eventually resulting in the peak becoming so wide that it can no longer be distinguished from the background. It can be hard to estimate what the ``right'' amount of power is for spectroscopy because this depends on how strongly the qubit couples to the drive line, something that is unknown before characterizing the device. An operational approach to this problem is to simply repeat the spectroscopy experiment for different powers until the resonance is found. 

Following this, we can confirm that it is indeed the qubit mode of interest by verifying the flux-dependence of its resonance frequency. Furthermore, in order to obtain a slightly more precise reading of the qubit frequency, we often employ another form of two-tone spectroscopy where both drives are pulsed, rather than continuous. Typically, a (square) pulse is first applied at $\omega_s$, in the vicinity of the previously found qubit frequency, followed by a second pulse at $\omega_r$. The pulsed-spectroscopy experiment removes the AC Stark shift the qubit experiences due to the always-on readout tone~\cite{schuster2007_resolving}. Furthermore, the absence of these continuous drives also prevent spurious heating of the device due to dissipation of microwave power at 20\,mK, where the cooling power of the dilution refrigerator is limited to several $\mu$W. 

After identifying the qubit, the anharmonicity can be determined by a simple variant of a regular two-tone spectroscopy experiment with the introduction of a third drive. This three-tone spectroscopy allows us to have an additional probe that could induce the $\ket{1}\leftrightarrow{\ket{2}}$ transition of the transmon at frequency $f_{12}$ while also performing operations on the readout resonator and the $\ket{0}\leftrightarrow{\ket{1}}$ transition of the qubit. 

In such an experiment, the first spectroscopy tone ($s1$) is varied in frequency $f_{s1}$ around the $\ket{0}\leftrightarrow{\ket{1}}$ resonance of the transmon at frequency $f_{01}$. The frequency, $f_{s2}$, of the second spectroscopy tone ($s2$) is swept around the expected $\ket{1}\leftrightarrow{\ket{2}}$ transition, which is typically $\sim 200-300\,$MHz below the $\ket{0}\leftrightarrow{\ket{1}}$ resonance. The the $\ket{0}\leftrightarrow{\ket{1}}$ transition manifests as a line at $f_{s1}=f_{01}$ while the $\ket{2}$ state appears as a diagonal line corresponding to the condition $f_{s1}+f_{s2} = f_{01}+f_{12}$. The location of this second transition indicates the anharmonicity of the transmon, which can be an important consideration for the design and calibration of subsequent single-qubit operations.

\subsection{Single-qubit experiments}
\label{subsec:onequbit_expt}
Having established a working (but not yet optimized) configuration for the readout and a good approximation of the qubit frequency through spectroscopy experiments, we can move on to more precise characterizations via a series of single-qubit experiments. This task typically involves simple measurements of the qubit state as a function of a certain parameter, such as duration, phase, amplitude, etc, of the preceding control pulses. Here, we will focus on a set of core experiments of this form that are essential for our characterization process. 

\subsubsection{Rabi and DRAG}
Usually, we start this characterization process with the Rabi experiments~\cite{Rabi36}, in which the qubit state oscillates between the $\ket{0}$ and $\ket{1}$ state as a function of the external ``impulse", i.e.~the time-integral of the applied pulse amplitude. Traditionally, this is done by applying a pulse at the qubit frequency with a fixed amplitude and varying its duration~\cite{Wallraff05}. Alternatively, it is also a common practice to choose a fixed pulse duration and vary its amplitude. From the resulting oscillations, we can determine the amplitude or duration required to perform a rotation of the Bloch vector of the qubit along the X or Y axis by any arbitrary angle.

The typical starting point for tuning up single qubit control is to calibrate the $\pi$ pulse. As the rotation angle is only a function of the impulse, there is, in principle, freedom of choice between higher drive amplitude and longer duration, as well as the overall waveform, to achieve the same $\pi$ pulse. 

In practice, an optimal $\pi$ pulse is often determined based on two main considerations. First, we generally want to have fast controls on the qubit state, and therefore, would prefer to minimize the pulse duration. On the other hand, we must also be mindful that transmon qubits are not perfect two-level systems, as described by the simple Rabi model. Instead, they are anharmonic oscillators with more than two discrete energy levels. These higher levels, although detuned from the first transition frequency, must be taken into account in order to achieve a good $\pi$ pulse without causing leakage out of the two lowest energy levels. Hence, the choice of pulse duration for qubit drives is bounded by the anharmoncity. More concretely, we must choose a pulse length such that its spectral components do not overlap with the frequency of the next transition. For a typical anharmonicity of $\alpha/2\pi \approx 200\,$MHz, a gate can be enacted by a Guassian drive with its full width half maximum, $\sigma$, set to 4-6\,ns and its total duration of $\sim 4\sigma\approx 20~\mathrm{ns}$. 

In addition to choosing the appropriate pulse duration, a DRAG (Derivative Removal by Adiabatic Gate) correction is often introduced to further suppress spurious transitions to the higher levels~\cite{motzoi2009_simple, Chow10b}. Intuitively, the effect of this derivative component can (as a crude approximation) be understood as a combination of reducing the spectral overlap between the pulse and the spurious transition, while also correcting for an AC stark shift introduced by the pulse~\cite{Gambetta11}. 

Mathematically, DRAG pulses consist of a standard Gaussian waveform as the in-phase component and scaled derivative of the Gaussian waveform as the quadrature component:
\begin{equation}
  I(t) = G_{amp} e^{- \frac{(t-\mu)^2}{2\sigma^2}},\\
  Q(t) = -D_{amp} \frac{(t-\mu)}{\sigma} I(t), 
\end{equation}
where $\mu$ is the center of the Gaussian pulse, $\sigma$ the standard deviation, and $G_{amp}$ and $D_{amp}$ are system dependent scaling factors for the in-phase and quadrature components that can be determined through calibrations. In practice, $D_{amp}$ is usually be set to zero (making the pulse effectively Gaussian) when performing the initial Rabi experiments to determine $G_{amp}$, and can be added as a correction term in subsequent calibrations.

\subsubsection{Coherence measurements}
After tuning up a $\pi$ pulse using a Rabi oscillation, the next step is to determine the coherence properties of the qubit, namely, the relaxation time $T_1$, the Ramsey time $T_2^\star$, and the Echo dephasing time $T_2^E$. 

To measure the relaxation time, the qubit, initialized in $|{0}\rangle$, is excited to $|{1}\rangle$ with a $\pi$ pulse. This is then followed by a variable wait time $t$, and finally a measurement of qubit state. This results in an exponentially decaying signal, given by : 
\begin{equation}
S(t) = A\cdot e^{-t/T_1}+B,
\end{equation}
where $S(t)$ is the signal as a function of wait time $t$, $A$ and $B$ are scaling and offset factors.

The Ramsey dephasing time can be obtained by initializing the qubit in $\ket{0}$, applying a $\pi/2$ pulse, waiting for a time $t$, applying a final $\pi/2$ pulse and then measuring the qubit. 
The Ramsey experiment will produce an (often exponentially) decaying oscillation described by:
\begin{equation}\label{eq:t2star}
  S(t) = A e^{-(t/T_2^\star) ^ n} \cdot (\cos(2 \pi f t + \phi) + C) +B,
\end{equation}
where $A$,$B$ and $C$ are scaling and offset factors, $n$ describes the profile of the exponential decay, and $f$ the frequency of the oscillation. This oscillation frequency, $f$, corresponds to the frequency detuning between the control pulses and the actual qubit frequency. Because it is difficult to distinguish an oscillation due to a low frequency (small detuning) from an exponential decay, the pulses are often slightly detuned intentionally to make this distinction more obvious. This can be either done by physically detuning the microwave source, or by changing the phase of the second $\pi/2$ pulse (artificial detuning). 

In general, $T_1$ measurement probes the overall energy relaxation times due to all the loss channels in the system, and provides an indication of the internal quality factors of the device. The Ramsey time contains information on both energy relaxation and pure dephasing ($T_{\phi}$) in the qubit, i.e. $1/T_2^\star = 1/(2T_1) + 1/T_{\phi}$. It provides a useful metric for quantifying the effective decoherence time scales of the qubit and a rough bound on the performance of the subsequent single-qubit gates. 

The profile of the decay $n$ in Eq.~\ref{eq:t2star} provides some indication of whether $T_2^\star$ is limited by $T_1$ or another source of incoherent noise $n\approx1$. If there is a coherent noise process, such as a slow drift in the qubit frequency, $n$ will be larger than 1. In this case, it is possible to recover part of the information by adding an echo pulse. The echo dephasing time $T_2^E$ can be measured by adding a single $\pi$ pulse in the middle of a Ramsey experiment. The final experiment consists of preparing the qubit, applying a $\pi/2$ pulse, waiting for a time $t/2$, applying a $\pi$ pulse, waiting for a time $t/2$, and applying a final $\pi/2$ pulse before measuring the qubit state. The resulting dynamics now follows an exponential decay, given by
\begin{equation}
S(t) = A\cdot e^{-(t/T_2^E)^n}+B.
\end{equation} 
Apart from this simple sequence, there are many other different types of echo configurations developed in the field of nuclear-magnetic resonance (NMR) in the early days of quantum computing. By employing the appropriate sequences, it is possible to significantly increase the quality of certain single-qubit operations. A more detailed discussion on these useful techniques can be found in Ref.~\cite{Vandersypen05}.

Additionally, all of these coherence measurements can be applied to characterize the properties of the higher-order transitions in the transmon as well. In particular, the second excited state ($|f\rangle\rangle)$ in the trasmon qubit typically has coherence properties comparable to that of the two lowest levels and can even be employed as a distinct mode for certain specialized tasks~\cite{wang2015_schrodinger, rosenblum2018_fault-tolerant, reagor2018_demonstration}.

In general, the careful characterization of a qubit's coherence provides crucial information about both the design and fabrication of the device as well as the quality of the measurement and control set-up. We refer to the recent review by Krantz et al.~\cite{krantz2019_engineer} for a detailed explanation of these decoherence channels as well the associated experimental protocols. In addition to the basic relaxation and dephasing experiments, valuable information on the mechanisms that are limiting the coherence properties of the device can be gleaned by measuring coherence as a function of flux bias~\cite{Martinis03, yan2016_theflux, Luthi18}.

\subsubsection{Single qubit manipulations}
\label{sec:single_qubit_manipulations}
Tuning up a full set of high-quality operations on a single qubit is a crucial step in the calibration of a cQED device. These operations correspond to rotations of the qubit state on the Bloch sphere. In the previous section, we discussed that a generic $\pi$ rotation can be calibrated using simple Rabi experiments. Naively, one might think that by adding a phase and amplitude scaling factor to this pulse, we can obtain all other single-qubit rotations. However, imperfections, such as detuning or small distortions in the control signal, can cause appreciable differences in the pulses required to implement rotations along different axes or by different angles. Thus, it is important to calibrate each single qubit gate in a systematic and comprehensive manner. 

\begin{figure}[!tbh]
  \centering
  \includegraphics[scale=1]{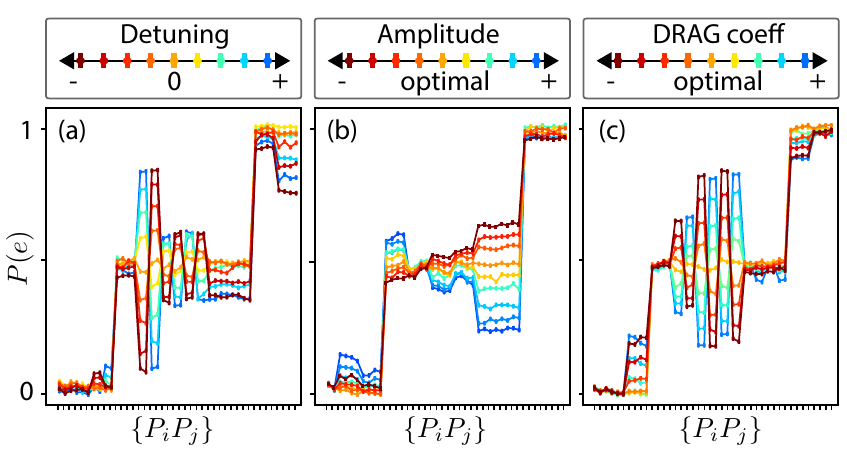}
  \caption{\textbf{Example of an ALLYX measurement}. The outcome of a typical ALLXY experiment with 21 combinations of one or two single-qubit rotations around the $x$- and $y$-axes by an angle of $\pi/2$ or $\pi$ applied on a qubit before the measurement of its state, i.e. $\{P_iP_j\} \subseteq \{I, X(90), Y(90), X(180), Y(180)\}$. The sequence is designed to highlight the different error syndromes of a single-qubit rotation, such as (a) detuning from resonance, (b) variations in drive amplitudes, and (c) variations in the coefficient of the DRAG correction.}
  \label{fig:allxy}
\end{figure}

A commonly used technique to gauge the quality of single qubit rotations in transmon qubits is the ALLXY experiment~\cite{ReedPhD13, Asaad16, Fu17, Bultink16, Fu19}. In this protocol, we perform pairs of single qubit gates chosen from different combinations of $\pi$ and $\pi/2$ rotations along the $x$ and $y$ axis on the qubit. The resulting states of the qubit should ideally form a `staircase pattern' where only $|g\rangle$, $|e\rangle$, or an exactly equal superpositon of the two are present. In practice, different types of imperfections will result in various deviations from this ideal configuration and exhibit qualitatively different syndromes. An example of the measurement outcomes for three types of errors are shown in Fig.~\ref{fig:allxy}. A trained experimentalist can use these patterns to quickly diagnose distinct errors in the control pulse, including pulse amplitude, DRAG correction coefficients, detuning, and basic pulse distortions. A detailed discussion on the physics behind the ALLXY protocol can be found in Ref.~\cite{ReedPhD13}.

Once the errors in each single qubit rotation are identified, we can implement a systematic procedure to correct and improve the fidelities of each operation. This generally consists of three-steps, starting with a fine calibration of the drive frequency, then making adjustments to the amplitude of the derivative component of the DRAG pulse, also known as the Motzoi parameter, and finally the fine-tuning the pulse amplitudes. 

To precisely calibrate the frequency of the qubit, we use the repeated Ramsey protocol. In typical Ramsey experiments, the frequency of the observed oscillation corresponds directly to the detuning of the qubit from the drive frequency. To prevent under sampling of the oscillation due to the Nyquist limit, the first Ramsey experiment should cover a short time span to get a coarse frequency estimate with a large bandwidth, while subsequent iterations can increase the time span to achieve more accurate frequency estimates at a lower bandwidth (for the same number of data points). Furthermore, we can also introduce an artificial detuning in the Ramsey experiment such that any small detuning is effectively extracted by comparing the fitted oscillation frequency with the artificial one. This method allows a more robust calibration of the qubit frequency than spectroscopy experiments, with its accuracy limited by the physical coherence property, $T_2$, of the qubit. 

Next, we calibrate the DRAG coefficient using the method detailed in Ref.~\cite{ReedPhD13}. This involves minimizing the difference in excited-state population produced by the $Y_\pi X_{\pi/2}$ and $X_\pi Y_{\pi/2}$ combinations. As shown in Fig.~\ref{fig:allxy}(a) and (c), errors in detuning and the DRAG coefficient have a similar signatures in the AllXY experiment. Thus, by first calibrating the frequency using Ramsey experiments, which are insensitive to the DRAG coefficient, these two error sources can be decoupled. 

Finally, we tune the drive amplitudes by applying an initial $\pi/2$ pulse followed by $2N$ repeated $\pi$ pulses. If the amplitude is calibrated correctly, the excited-state population should be independent of the number of $\pi$ pulses. Effects of any deviation from the correct amplitude will be exaggerated by this sequence, which is sometimes referred to as a `pulse-train' experiment. Overall, this method provides a simple tool for us to extract and correct any fine amplitude deviations in the $\pi$ pulse.

\subsection{Readout optimization}
\label{subsec:ro_optimization}
Having tuned up the single qubit manipulations, we now are in the position to further optimize the readout configuration of the device. Conceptually, reading out the state of a qubit consists in entangling each one of its two states with certain states of the resonator that can be detected and distinguished at the macroscopic scale. 

In cQED, the general procedure for reading out a qubit is to first entangle it with a coherent state of a readout resonator, from which the signal can quickly ``leak'' out into a coaxial cable. This signal is subsequently amplified and detected at room temperature using a heterodyne measurement technique. Such a process is typically referred to as a dispersive readout as the entanglement occurs as a result of the natural dispersive Hamiltonian.  

Before going into the optimization techniques, we shall first address the question: what constitutes a good readout? In general, we care about two main properties: the measurement must be faithful and must not alter the state of the qubit. The former is quantified by the fidelity $\mathcal{F}$, which gives the probability that we can correctly assign the qubit state based on the measurement outcome. The latter is called the quantum non-demolition-ness (QND-ness), denoted by $\mathcal{Q}$, which gives the probability that a qubit state persists throughout the measurement process. This figure of merit is especially important when qubits are interrogated repeatedly, such as in most quantum error correction schemes.

In order to obtain both a high fidelity and QND-ness, we must carefully optimize the readout configurations. In this subsection, we will first summarize the physical process of the dispersive readout and then present the protocols to extract the key figures of merit. Finally, we will discuss some commonly used strategies for achieving an optimal readout performance. 

\subsubsection{The mechanism of a dispersive readout}

\begin{figure}[t]
\includegraphics[scale=1]{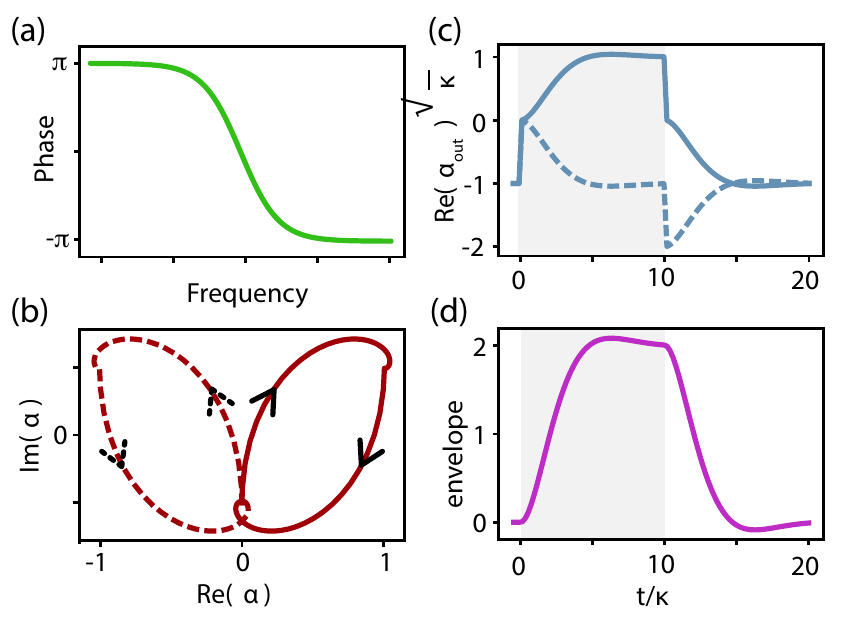}
\caption{\textbf{Readout resonator trajectories for the dispersive readout} (a) The phase response of the resonator in standard dispersive readout in the reflection configuration. (b) Average internal state of the readout cavity represented in phase-space. The arrows indicate how the state evolves in time. (c) Average reflected signal as a function of time. The light-gray area represents the time during which the input field is on. The trajectory corresponding to the qubit being in $\ket{g}$ is a blue dashed line and the one corresponding to the qubit being in $\ket{e}$ is a blue solid line. (d) Average distance between the reflected signal corresponding to the qubit being in $\ket{g}$ and $\ket{e}$. Similarly, the light-gray area represents when the input field is on.
}\label{fig:dispersive_readout}
\end{figure}

Recall that in cQED the dispersive interaction Hamiltonian, $\bH_{int} = -\frac{\chi}{2} \ba^\dag\ba\bsigma_z$, stems from a simple capacitive coupling between an anharmonic mode, i.e. the qubit, and the readout resonator. To obtain the information of the qubit state, we detect the response from the readout resonator one of the measurement circuits discussed in Sec.~\ref{sec:interacting_with_quantum_circ}.1. When the qubit state undergoes a transition, it shifts the resonance frequency of the resonator as a result of the dispersive coupling. This, in turn, leads to a measurable change in the response of the readout mode to the probe signal, regardless of the exact measurement circuit. When this frequency shift is larger than the linewidth of the resonator, the state of the qubit can be detected. Here, we will use the reflection set-up, as shown in Fig.~\ref{fig:dispersive_readout}(a), as an example to illustrate how the signal evolves during the readout process. 

Let us consider the response of a resonator driven by a constant readout pulse, whose field is denoted by $\ba_{in}$. We model the evolution of field of the readout mode with the Langevin equation $\partial_t{\ba} = \left(\pm i\frac{\chi}{2} - \frac{\kappa}{2} \right)\ba - \sqrt{\kappa}\bar{a}_{in}$. Again, we have performed the stiff-pump approximation on the incoming field as discussed in Sec.~\ref{sec:hamiltonians}.1. If the readout resonator is initially in the vacuum state, the solutions to the Langevin equation are coherent states with amplitudes given by:
\begin{align}
   \label{eq:dispersive_response}
   \alpha_{g, e}(t) = \frac{2\sqrt{\kappa}\bar{a}_{in}}{\pm i \chi - \kappa}\left[
   1 - \exp\left(\left(-\frac{\kappa}{2} \pm \frac{\chi}{2}\right)t\right)\right].
\end{align}
In Fig.~\ref{fig:dispersive_readout}(b), we represent the average time-dependent trajectory of the field in phase space according to Eq.~\ref{eq:dispersive_response}. This reveals two important aspects about the readout. 

First, Eq.~\ref{eq:dispersive_response} indicates that the response of the readout mode changes according to the state of the transmon, which allows the state-differentiation. As shown in Fig.~\ref{fig:dispersive_readout}(b), the imaginary parts of the two trajectories of the readout field are the same, but their real parts are opposite. The maximum separation depends directly on the amplitude of the incoming field. Thus, if the readout drive is sufficiently large, we can effectively infer the state of the qubit by simply measuring the real part of the outgoing field. Based on this, one might deduce that it would be optimal to maximize the power of a readout drive. However, in practice, a powerful drive triggers a significant level of qubit relaxation~\cite{Boissonneault2009_dispersive, Slichter2012_measurement, Sank16}. This phenomena is commonly referred to as the $T_1$ vs $\bar{n}$ problem and its exact physical mechanism is still an ongoing topic of research. 

Furthermore, Eq.~\ref{eq:dispersive_response} also highlights that the resonator's response depends closely two coupling parameters, $\chi$ and $\kappa$. To illustrate how this qubit-dependent trajectory affects the response signal, we plot the real part of the averaged outgoing field $\alpha_{out}$ as a function of time in Fig.~\ref{fig:dispersive_readout}(c). When the input field is turned on, it is first completely reflected off the resonator and the output field sharply rises to the value of the input field. Then, the field inside the cavity increases with a timescale $\kappa/2$ and interferes constructively (resp.~destructively) with the input signal when the qubit is in $\ket{e}$ (resp.~$\ket{g}$). Finally, when the input drive is switched off, the output field is composed solely of the field leaking out of the readout cavity with a timescale $\kappa/2$. Overall, the behavior of this field can be described by Fig.~\ref{fig:dispersive_readout}(d), where the separation of the two trajectories corresponding to the qubit in $\ket{e}$ and $\ket{g}$ increase at the rate of $\kappa/2$ at the beginning of the readout process and then diminishes after the readout pulse as photons leak out of the resonator into the tranmission line. It has been shown the optimal choice of the these two parameters occurs at $\chi=\kappa$~\cite{Gambetta08}.

During the readout process, outgoing field from the resonator is entangled with the qubit. As a result, the qubit itself is not in a pure state: it is dephased as the field is measured. Essentially, as the outgoing field provides information about the state of the qubit, any superposition state is destroyed. The dephasing mechanism corresponds to the ``leakage'' of the readout field to the transmission line. This process can be modeled using the Lindblad superoperator~\cite{nielsen2010_quantum} acting on the density matrix $\rho$ as $\frac{\kappa}{2}\mathcal{D}[\ba]\rho = \frac{\kappa}{2}\left(2\ba\rho\ba^\dag - \ba^\dag\ba\rho - \rho \ba^\dag\ba\right)$.

In our framework, the state of the system is always a superposition of the joint states $\ket{\alpha_g(t), g}$ and $\ket{\alpha_e(t), e}$. Thus, we can perform a unitary transformation to better describe the evolution of such states in a frame where the field of the readout cavity is given by
\begin{align}
  \ba \xrightarrow{} \ba + \frac{\alpha_g(t) + \alpha_e(t)}{2} + \frac{\alpha_g(t) - \alpha_e(t)}{2}\bsigma_z.
\end{align}
In this frame, the system is first displaced to the center of mass of the two coherent states, followed by a state-dependent displacement $\pm \frac{\alpha_g(t) - \alpha_e(t)}{2}$. With this unitary transformation, the Lindblad superoperator becomes
\begin{align}\label{eq:ro_dephase}
  \frac{\kappa}{2}\mathcal{D}[\ba]\rho \xrightarrow{} \frac{\kappa}{2}\left(\mathcal{D}[\ba]\rho + \frac{1}{4}\abs{\alpha_g(t) - \alpha_e(t)}^2\mathcal{D}[\bsigma_z]\rho + ... \right),
\end{align}
where the remaining terms are either unitary or neglected under the RWA.

We identify that second term in Eq.~\ref{eq:ro_dephase} causes dephasing of the qubit state due to photons leaving the resonator. The rate at which this takes place is given by $\frac{\Gamma_\varphi}{4}\mathcal{D}[\bsigma_z]\rho$, where $\Gamma_\varphi$ is referred to as the measurement-induced dephasing rate and its time-dependence is given by 
\begin{align}\label{eq:dephase_rate}
  \Gamma_\varphi(t) = \frac{\kappa}{2}\abs{\alpha_g(t) - \alpha_e(t)}^2.
\end{align}
This depends on both the decay rate ($\kappa$) of the resonator and the specific shape of the readout pulse. In a perfect world, due to conservation of quantum information, $\Gamma_\varphi(t)$ should be equal to the rate at which the experimentalist acquires information from the readout.

\subsubsection{Extracting figures of merit}

In order to characterize the performance of a readout configuration, let us first consider how the state-dependent continuous outgoing signal is treated to obtain a binary result that reveals the qubit state. As mentioned previously, a heterodyne detection scheme is used to sample the outgoing signal. While the average of this signal resembles Fig.~\ref{fig:dispersive_readout}(c), for a single readout instance (or ``shot''), each sample has a finite amount of noise due to both the quantum fluctuations of coherent states and the added noise of the electronic components between the readout resonator and the acquisition card. This noisy signal is then integrated over time to give an outcome with a single value. The accumulation of such single-shot measurements will form a Gaussian distribution whose average depends on the initial state of the qubit~\cite{Jeffrey14}. If the readout is good enough, the separation between the averages of these two distributions (``signal'') is larger than their width (``noise''). A good readout must have a signal-to-noise ratio (SNR) much larger than 1. Finally, the outcome of each measurement is digitized by setting a threshold between the two distributions.

\begin{figure}[!th]
   \centering
   \includegraphics{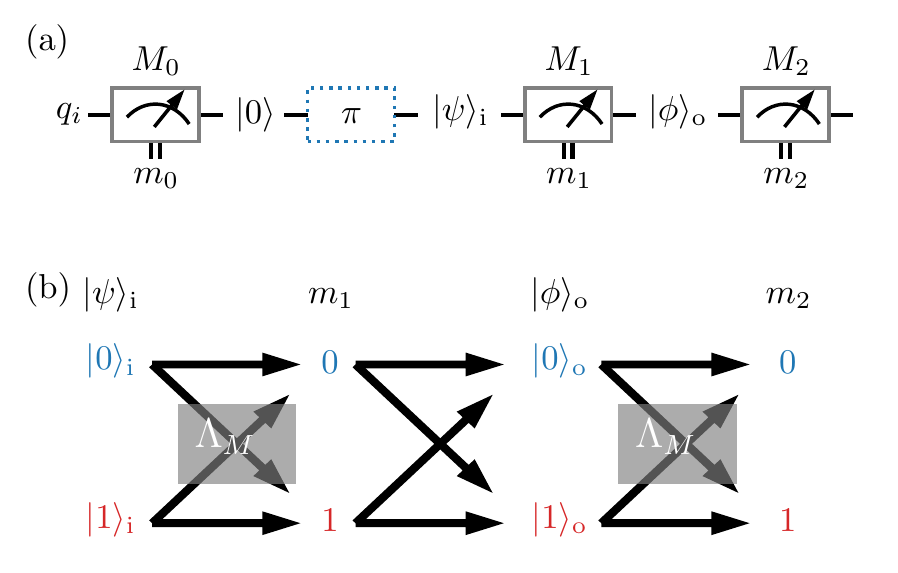}
   \caption{\label{fig:buttrefly_exp}\textbf{The butterfly measurement scheme for readout charaterization.}
   (a) Circuit diagram for the butterfly experiment to characterize measurement $M_1$ on qubit $q_i$. Post-selection based on an initial measurement $M_0$ is used to initialize $\ket{0}$. (b) Correlations between $m_1$ and $\ket{\psi}_{\mathrm{i}}$ are described by $\Lambda_M$
   }
\end{figure}

To quantify the fidelity ($\mathcal{F}$) and QND-ness ($\mathcal{Q}$) of a readout, let us consider the effect of the measurement on two initial states $\ket{0}_i$ and $\ket{1}_i$. The preparation of these initial states must be as accurate as possible so that it does not limit the characterization procedure. This can be achieved by using a first measurement with a stringent threshold combined with post-selection and/or a feedback pulse, as depicted in Fig.~\ref{fig:buttrefly_exp}(a). 

An ideal measurement would yield an outcome $m=0$ (resp. $m=1$) when the initial state is $\ket{0}_i$ (resp. $\ket{1}_i$) and would leave the qubit in a state $\ket{0}_o$ (resp. $\ket{1}_o$). The fidelity and the QND-ness of a realistic measurement are defined as
\begin{align}
  \mathcal{F} &= 1 - \left(\condProb{m=0|\ket{1}_i} + \condProb{m=1|\ket{0}_i}\right)/2 \\
  \mathcal{Q} &= 1 - \left(\condProb{\ket{0}_o|\ket{1}_i} + \condProb{\ket{1}_o|\ket{0}_i}\right)/2,\label{eq:qnd}
\end{align}
where $\text{P}$ is the Bayesian conditional probability. 

These figures of merit can both be deduced from the measurement butterfly experiment represented on Fig.~\ref{fig:buttrefly_exp}(b). After preparing an initial state $\ket{0}_i$ or $\ket{1}_i$, two consecutive measurements $M_1$ and $M_2$ are performed. The mapping between a given initial state $\ket{\psi}_i$ to an outcome $m$ is characterized by the matrix $\Lambda_M$, with
\begin{align}
\Lambda_M =
\begin{bmatrix}
\condProb{m=0 | \ket{0}_i}
  & \condProb{m=0 | \ket{1}_i} \\ \\
\condProb{m=1 | \ket{0}_i}
  & \condProb{m=1 | \ket{1}_i}
\end{bmatrix}
.
\end{align}
We can then write the outcome probabilities of the first measurement as a column vector: $\text{P}(m_1) = \Lambda_M \text{P}(\ket{\psi}_i)$. Each element of $\Lambda_M$ can be measured and the fidelity $\mathcal{F}$ is simply calculated from the off-diagonal elements. 

Based on this, let us now evaluate the conditional probabilities required to determine the QND-ness of the readout. Using the same $\Lambda_M$ to describe both $M_1$ and $M_2$, we can write
\begin{equation}
   \text{P}(\ket{\phi}_{\mathrm{o}}) = \Lambda_M^{-1}\text{ P}(m_2), 
\end{equation}
which leads to 
\begin{align}
\begin{bmatrix}
\condProb{\ket{0}_{\mathrm{o}}, m_1 | \ket{\psi}_i}\\ \\
\condProb{\ket{1}_{\mathrm{o}}, m_1 | \ket{\psi}_i}
\end{bmatrix} =
\Lambda_M^{-1} \begin{bmatrix}
\condProb{m_2=0, m_1 | \ket{\psi}_i}\\ \\
\condProb{m_2=1, m_1 | \ket{\psi}_i}
\end{bmatrix},
\end{align}
and
\begin{align}
&\condProb{\ket{0}_o| \ket{1}_i} \nonumber\\
&= \condProb{\ket{0}_o,m_1=0 | \ket{1}_i}+\condProb{\ket{0}_o,m_1=1 | \ket{1}_i},\\
&\condProb{\ket{1}_o| \ket{0}_i} \nonumber\\
&= \condProb{\ket{1}_o, m_1=0| \ket{0}_i}+\condProb{\ket{1}_o,m_1=1| \ket{0}_i}.
\end{align}
With this, we can now deduce the $\mathcal{Q}$ of the readout (Eq.~\ref{eq:qnd}) from these conditional probabilities. 

Overall, this protocol offers a simple tool for us to obtain two crucial figures of merit for the readout. With this information, we can then further optimize the readout configurations to perform high-fidelity measurements of the transmon state. 

\subsubsection{Optimization considerations}

In general, achieving an optimal readout requires us to carefully balance two contradicting effects. On the one hand, measuring for a longer duration and having the largest probe signal available should lead to a better SNR and thus, to a higher fidelity. On the other hand, this would also result in having more state transitions because of the natural relaxation time of the qubit, as well as induced state transitions due to a strong drive~\cite{Boissonneault2009_dispersive, Slichter2012_measurement, Sank16}. These transitions during the readout process result in a reduction of both the fidelity and the QND-ness. 

To put the optimization considerations more concretely, let us consider the simple example of a square readout pulse. Naively, we just need to progressively adjust the duration and amplitude of the pulse such that the measurement is faster than the effective (both natural and induced) relaxation time scales of the qubit. This should lead to the optimal readout fidelity and QND-ness. However, more advanced pulse shaping techniques is often necessary to tune up a high-quality readout beyond this compromise. 

Let us recall that at the beginning and at the end of the measurement, the separation between the trajectories corresponding to the two qubit states is small, while the noise is constant~\cite{didier2015_fast, Bultink18} due to the internal dynamics of the readout resonator. Hence, it is favorable to reduce the timescale of the initial rise and final fall of the field beyond $1/\kappa$ by modifying the the readout pulse shape. For example, the initial rise can be accelerated if we use a short but large pulse~\cite{Jeffrey14}. Similarly, the final decay of the field can be sped up by using a fast-emptying sequence inspired by dynamical decoupling schemes~\cite{McClure16, Bultink18}.

Furthermore, we can also acquire information more effectively by discarding certain parts of the signal. For instance, the duration over which the average trajectories corresponding to the two qubit states are similar offers very little useful information. Therefore, we can apply an envelope to take into account which parts of the signal are the most valuable, i.e. contains the most information about the qubit state. It has been shown that the optimal envelope is given by the difference between the average signals corresponding to the two qubit states~\cite{Ryan15}, as shown on Fig.~\ref{fig:dispersive_readout}(d).

With this optimized integration weight applied to the signal, the theoretical SNR is given by 
\begin{align}
  \text{SNR}(\tau) = \sqrt{2\scaleint{5ex}_{\bs 0}^\tau
  |\alpha_{out, g}(t) - \alpha_{out,e}(t)|^2~dt},
\end{align}
where $\alpha_{out, i}$ is the average outgoing field that has been acquired when the qubit is prepared in state $\ket{i}$. If the field leaking from the readout resonator is not corrupted on its way to the acquisition apparatus, then $\alpha_{out, i}$ can be obtained with the input-output relation of Eq.~\ref{eq:in_out}. In this case, we get $\text{SNR}_{\text{ideal}}^2(\tau) = 4 \gamma_\varphi(\tau)$, where we define the total dephasing as $\gamma_\varphi = \scaleint{5ex}_0^{\tau}\Gamma_{\varphi}(t)~dt$. This quantity can be directly measured using a Ramsey sequence where two $\pi/2$ pulses are separated by a measurement pulse. This relation indicates that the maximum information acquired, given by the SNR, is equal to the amount of qubit information destroyed, given by the total dephasing $\gamma_\varphi$ (up to a numerical factor). 

For a realistic measurement setup, however, we must take into account of degradation of the signal, either due to loss or added noise as it travels to the acquisition system at room temperature. The resulting SNR will therefore be lower compared to its theoretical maximum. This non-ideality is quantified by the efficiency $\eta$ given by
\begin{align}
  \eta = \frac{\text{SNR}^2(\tau)}{4\gamma_\varphi(\tau)}.
\end{align}
The value of $\eta$ goes between 0, when no information reaches the observer and the qubit is entirely dephased, and 1, when as much information reaches the observer as it is destroyed during measurement. Typically, achieving a high efficiency requires us to minimize the signal loss between the readout resonator and the quantum-limited amplifier. In the case of a phase-preserving quantum-limited amplifier, the maximum efficiency, as defined in our framework, would be 0.5. However, in many published literature, this has often been re-scaled to allow more direct comparison across different experimental set-ups. Furthermore, the gain of this amplifier should also be adjusted such that noise added by the cryogenic HEMT amplifier does not significantly corrupt the signal (see Fig.~\ref{fig:example_device}). 

Several protocols exist to experimentall characterize $\eta$~\cite{Bultink18, Hatridge2013_weak}. They provide a useful way to probe the overall quality of the measurement set-up and ensure that it does not limit the performance of the readout. Currently, new readout techniques and high-performing quantum-limited parametric amplifiers have enabled us to reliably achieve efficiencies between 0.1 and 0.6~\cite{Jeffrey14, Roch14, Hacohen-Gourgy2016, Eddins2018, touzard2019_gated}.

\subsection{Engineering two-qubit gates}
Apart from characterizating and performing operations on physical qubits individually, another key ingredient for a useful quantum system is the ability to implement robust two-qubit operations. Broadly speaking, there are three classes of two-qubit gates for superconducting qubits, namely, fast flux pulsing, microwave driven, and flux modulated. While all of them have demonstrated gate fidelities of $\geq 99\,\%$, their requirements for coupling mechanisms between individual elements in the quantum hardware often differ significantly. Here, we will provide an overview of the main techniques for realizing two-qubit operations in current superconducting quantum devices and highlight some of the key technical pros and cons in each of them. 

In essence, two-qubit gates enacted via fast flux pulsing\footnote{Here, two-qubit gates employing flux-controlled tunable couplers are considered as a special case of flux pulsing based two-qubit gates.} are performed by tuning certain transitions close to resonance~\cite{Strauch03,dicarlo2009_demonstration,Neeley10,barends2014_logic,Chen14,rol2019_fast, sung2020_realization}. Their typical gate times are $<50$\,ns, limited by the effective coupling strength between the transmons. However, the fast gate time afforded by rapidly applying a flux pulse to shift a qubit's frequency into the required resonance conditions often results in detuning from the flux sweetspot. Consequently, this lowers the actual gate fidelity as compared to the naive estimate based only on the gate time over coherence time considerations. Furthermore, this type of operations also require precise pulse shaping in order to to achieve their optimal performance, making them potentially sensitive to the small distortions induced by the various electrical components used for microwave signal processing. 

Alternatively, we can also use fixed frequency qubits to implement universal control with only microwave drives~\cite{Rigetti10,Poletto12,chow2013_microwave,sheldon2016_procedure}. The most well-known example of all-microwave two-qubit operations is the cross-resonance gate~\cite{paik2016_experimental}, which is performed by driving one qubit at the frequency of another. While they can be significantly slower ($\sim\,150-500\,$ns) than flux-based gates, they eliminate the need for fast flux control in the hardware. As a result, the microwave-only operations typically lower the complexity in the fabrication and control aspects of the device. Furthermore, the absence of fast-flux on the device also removes some known noise channels, offering the potential for better intrinsic coherence properties. 

However, by removing the ability to tune the resonance frequencies of each qubit on demand, microwave-based gates encounter the challenge of frequency-crowding and undesired transitions. In particular, frequency crowding~\cite{Schutjens13} becomes increasingly problematic when fixed-frequency transmons are coupled to multiple neighbors~\cite{Brink18} as is often necessary for devices that aim to employ the surface code for error correction. 

A third approach is to use parametric flux modulation to perform two-qubit gates. In this method, we can directly modulate either the frequency of the individual qubits or that of a coupling element in order to induce a specific interaction between two qubits~\cite{mckay2016_universal,Caldwell18,Hong19}. This type of two-qubit gates can be seen as a hybrid between the flux pulsing and all-microwave gates. Some of the frequency crowding issues of the all-microwave gates are alleviated as the coupling terms are only activated when the drive is present. Furthermore, the parametrically driven gates are less susceptible to distortions than flux pulsing ones because the drives operate at a much lower frequency. Consequently, they are significantly slower compared to flux pulsing while still require additional flux-biasing elements, which negates the main advantages of an all-microwave architecture.

\begin{figure}
\centering
\includegraphics[scale=1]{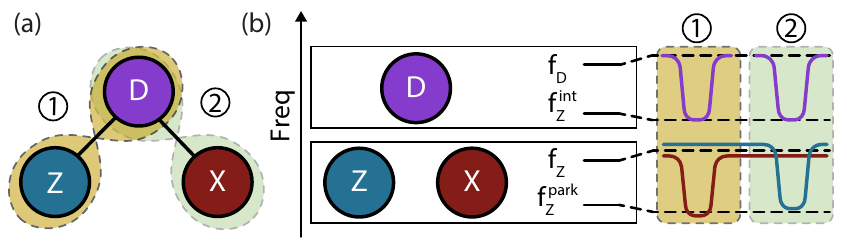}
\caption{
\label{fig:freq_collisions} \textbf{Two-qubit gates in a multi-qubit device}. (a) A high frequency qubit, $D$, is coupled to two low-frequency qubits, $Z$ and $X$. A CZ operation is to be implemented first between $D$ and $Z$ (step 1), followed by $D$ and $X$ (step 2). (b) To perform a CZ gate between $D$ and $Z$, $D$ detunes from $f_D$ to the interaction frequency $f^{\mathrm{int}}$ while $X$ must be moved to a parking frequency $f_X^{\mathrm{park}}$ to avoid an unwanted interaction. Similarly, in step 2, a CZ gate between $D$ and $X$ is performed with $Z$ detuned from its original frequency over the operation. 
}
\end{figure}

All of these techniques have demonstrated promising performance in the context of isolated pair-wise interactions. However, as we incorporate more physical qubits on the quantum device, frequency crowding or collisions can become a major source of imperfection for these two-qubit gates. To illustrate this, let us consider a standard control-Z (CZ) operation implemented using the fast flux pulsing technique. This is typically realized by employing the the avoided crossing between the $\ket{11}$ and the $\ket{02}$ state of a pair of coupled transmon qubits~\cite{Strauch03,dicarlo2009_demonstration}. Now, when we introduce one additional neighbouring qubit to the pair, in a configuration depicted in Fig.~\ref{fig:freq_collisions}(a), it is no longer feasible to naively perform two pair-wise CZ gates. In fact, qubit $D$ cannot be tuned to independently interact with either $Z$ or $X$ without causing spurious interactions with the other. Therefore, although $X$ is simply a spectator qubit during a CZ operation between $D$ and $Z$, a flux drive must be applied to $X$ concurrently to introduce a sufficiently large detuning so as to minimize undesired interaction (Fig.~\ref{fig:freq_collisions}(b)). Keeping track of and avoiding such frequency collisions can quickly become an intractable task as more qubits are coupled together and more frequencies are involved in large quantum devices.

Thus, careful considerations must be put in place when designing the frequency components of all the physical elements on a multi-qubit device. For instance, Ref.~\cite{Versluis17} proposed a scheme to alleviate this challenge by constructing bigger devices with 8-qubit unit cells, each containing only three sets of qubit frequencies. A unique detuning pattern for each qubit in the unit cell, known as the flux dance, allows for a pipelined execution of all parity checks while avoiding all undesired interactions. These 8-qubit unit cells can be incorporated in the design of more complex quantum processors for executing surface code. Alternatively, other strategies such as employing tunable couplers~\cite{kafri2017_tunable, yan2018_tunable} can also be adopted to address the challenge of undesired cross-talk on multi-qubit chips.

\subsection{Characterization of qubit operations}
Another key class of experiments we must perform to fully understand our quantum processor is the characterization of various single and two-qubit gates. In general, the quality of gates depends on the coherence of the qubits, as well as the accuracy and precision of the pulses used to realize these gates. Here, we will introduce several useful frameworks and experimental techniques for gate characterization. 

Mathematically, the effect of a gate can be expressed as a Completely-Positive
Trace-Preserving (CPTP) map ($\Lambda$) that acts on a density matrix ($\rho$)~\cite{Nielsen_2020}. The Pauli transfer matrix (PTM) provides a convenient representation of $\Lambda$ that relates the input and output states of a quantum operation~\cite{chow2012_universal}. The PTM can be written as 
\begin{equation}
\label{eq:PTM}
\left(R_\Lambda\right)_{ij}= \frac{1}{d}\mathrm{Tr}\left\{ P_i \Lambda (P_j)\right\},
\end{equation}
where $\{P_i, P_j\}\in \mathcal P^{\otimes n}$ are elements of the $n$-qubit Pauli set with $\mathcal P = \{ I, X, Y, Z\}$ and $d$ is the dimension of the Hilbert space. Using the superoperator formalism~\cite{Kitaev02}, a density matrix $\rho$ can be expressed as a vector in the Pauli basis $\ket{\rho}\rangle$ with components $\frac{1}{\sqrt{d}}\mathrm{Tr}\{ P_k \rho \}$~\cite{greenbaum2015_introduction}. Applying the map amounts to a simple multiplication
\begin{align}
\ket{\Lambda (\rho)} \rangle = R_\Lambda \ket{\rho}\rangle,\\ 
R_{\Lambda_2 \cdot \Lambda_1} =R_{\Lambda_2} R_{\Lambda_1} .
\end{align}

The main goal of characterizing quantum operations is to capture the imperfections in them. These imperfections can arise from both coherent and incoherent errors. Coherent errors, such as an over-rotation or leakage out of computational basis, are typically related to control non-idealities. They can potentially be addressed by improving the microwave pulses through more thorough calibrations as well as numerical optimal control protocols~\cite{Kirchhoff18}. Incoherent errors, on the other hand, are caused by decoherence effects in the quantum system. For transmon-based systems, the dominant decoherence effects are energy relaxation and dephasing. 
For a single transmon qubit, the PTM for energy relaxation is given by
\begin{align}
R_{T_1} =
\begin{pmatrix}
1 & 0& 0&0 \\
0 & \sqrt{1-p} & 0&0 \\
0 & 0& \sqrt{1-p} &0 \\
p & 0& 0 & 1-p \\
\end{pmatrix}
\end{align}
with $p=1-e^{-t/T_1}$ being the probability of relaxation. Similarly, the single qubit PTM of pure dephasing is given by
\begin{align}
R_{T_\phi} =
\begin{pmatrix}
1 & 0& 0&0 \\
0 & 1-p & 0&0 \\
0 & 0& 1-p &0 \\
0 & 0& 0&1 \\
\end{pmatrix}
\end{align}
where $p=1-e^{-t/T_\phi}$ is the probability of dephasing. 

Using the PTM formalism, we can extract several metrics that are often used to quantify gate performance. For instance, the average process fidelity, which expresses the distance of a gate to the target operation, can be obtained by
\begin{align}
F_{\mathrm{avg}}({\Lambda_{\mathrm{targ}}}, {\Lambda}) = \frac{\mathrm{Tr}\left( R_{\Lambda_{\mathrm{targ}}}^{-1} R_{\Lambda}\right) +d}{d(d+1)}.
\end{align}
In other words, the average error $\err$ associated with an operation can be expressed as $\err = 1-F_{\mathrm{avg}}$. Other metrics, such as the diamond norm~\cite{Sanders15} and unitarity~\cite{Wallman15,Feng16,Dirkse19}, can also be used to quantify the performance of the quantum operations, especially in the presence of coherent errors. 

In practice, PTMs are usually extracted experimentally by performing process tomography~\cite{nielsen2010_quantum}. However, this technique does not provide any means to distinguish the non-idealities due to state preparation and measurement (SPAM) from errors in the operation that we aim to characterize. 

To address this problem, the gate-set tomography (GST) framework~\cite{Blume13,Blume-Kohout17} was developed. In principle, GST provides an accurate and robust characterization of all operations in a gate set, including state preparation and measurement. However, it can be rather demanding to implement and requires a more significant number of experiments.

Alternatively, randomized benchmarking (RB)~\cite{Magesan11,Magesan12} can be used to condense all the imperfections into a single error metric, $\err$. In Clifford-based randomized benchmarking, a series of random gates is sampled from the Clifford group. A final gate is then added that inverts the action of all preceding gates. By changing the number of Cliffords and averaging over many randomizations, we obtain an exponentially decaying curve, whose decay constant is related to $\err$ while the offset and amplitude are related to SPAM errors. If the error rates of different Cliffords are known to differ significantly, such as when characterizing two-qubit gates, one can use interleaved randomized benchmarking~\cite{Magesan12b}. However, this must be performed with ample precautions as certain coherent errors can accumulate and result in a bogus fidelity~\cite{kimmel2014_robust}. Other variants of RB have also been developed to probe different types of gates and errors~\cite{Wallman15,Feng16,Dirkse19, Boixo18}. 

In general, when characterizing quantum operations, we are often faced with a difficult trade-off between information gain verses time/resource consumption. For instance, full process tomography or GST offers the complete information about our system but can be extremely time-consuming to implement. This increases the susceptibility to parameter drifts and environmental instabilities, which in turn, degrades the quality of our characterization. On the other hand, RB can be highly time-effective, but it only provides limited information on the gate performance without providing details on the underlying mechanisms.

\begin{figure}[ht]
\centering
\includegraphics[scale=1]{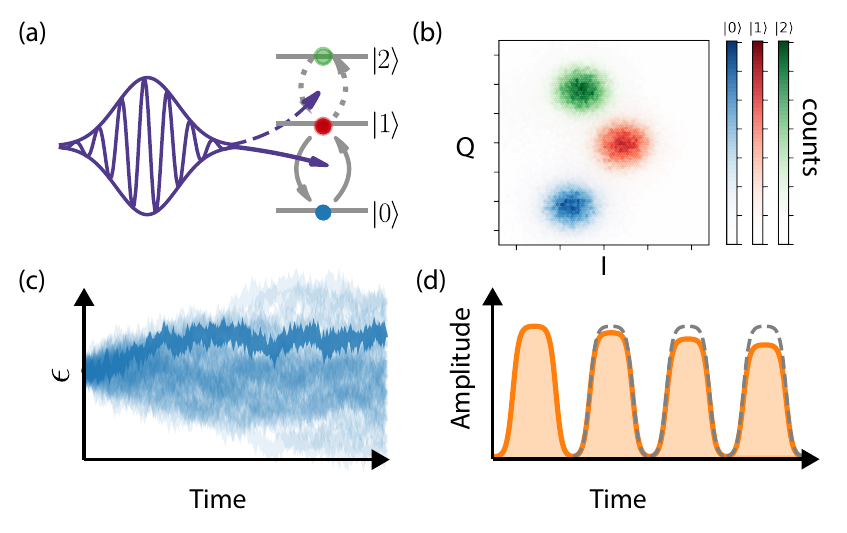}
\caption{\label{fig:other_errors} \textbf{Examples of errors in cQED devices that violate the typical assumptions in characterization protocols}. (a) Microwave control pulses can drive transitions to the second excited state. (b) Performing a measurement can induce transitions to a non-computational state. (c) Low-frequency noise on control parameters or coherence can cause fluctuations in error rates. (d) Operations can be history dependent due to e.g., pulse-distortions acting on timescales longer than the pulse duration.
}
\end{figure}

Aside from the practical limitations of current characterization techniques, a more fundamental problem for the above-mentioned protocols is that they are based on the assumptions that all processes are Markovian~\cite{sarovar19} and that the qubits are two-level systems. Unfortunately, these assumptions are violated in most realistic cQED devices. In transmon qubits, in particular, leakage out of the computational subspace and non-Markovian errors such as gate-bleedthrough, have significant impacts on the quality and repeatability of quantum operations. A few examples of errors that are not accounted for in most of the current gate characterization schemes are illustrated in Fig.~\ref{fig:other_errors}.

To address these limitations, modifications have been made to existing protocols to extend their capabilities. For example, enhancements of the RB scheme such as interleaved randomized benchmarking (IRB)~\cite{Magesan12b}, character benchmarking~\cite{Helsen19_characterbenchmarking, Xue18}, and speckle purity benchmarking~\cite{arute2019_quantum} have been developed to measure gate-specific error rates. Furthermore, simultaneous randomized benchmarking (SRB)~\cite{Gambetta12} has been proposed to measure cross-talk, and more specialized tactics that address the leakage problem~\cite{Wallman16,Wood18} have also been investigated. Similar modifications exist for GST that allow characterization of drift~\cite{proctor19,Nielsen16} or idle crosstalk~\cite{Nielsen16}. A good open-source library containing a wide range of characterization protocols is pyGSTi~\cite{Nielsen16, Nielsen_2020}, which is maintained by Sandia National Laboratories. A commercial alternative is provided by Quantum Benchmark, a Canadian startup company. However, understanding the subtle differences between the various assumptions made in these protocols and implementing them in the appropriate system requires in-depth knowledge of the hardware.

We should also keep in mind that although these protocols have a firm theoretical basis, there are often hidden assumptions that can limit their practical applicability in realistic cQED devices. As an example, consider the leakage protocol of Wood and Gambetta~\cite{Wood18}, which works without the need to directly measure leakage. The population out of the computational subspace can be estimated using the fact that probabilities must sum to 1. However, in most cQED experiments, this property is already invoked implicitly in the readout process where a state is assigned $\ket{0}$ if the measured signal is below some threshold with probability $p$, and otherwise assigned to be $\ket{1}$, with probability $1-p$. This makes it incompatible with the convenient leakage treatment technique given in Ref.~\cite{Wood18}. While it can be relatively straightforward to introduce a modification to enhance this protocol~\cite{Asaad16,rol2019_fast}, it is often system-specific and not readily generalizable to other devices. 

In order to give a meaningful interpretation to characterization results, it is often desirable to express them in terms of a parameter that has physical significance and units such as $T_1~(\mathrm{s})$ or an effective coupling strength $J~(\mathrm{Hz})$. Although these parameters can in principle be extracted from GST results, the experiments tend to be very costly in terms of acquisition time and analysis to get sufficiently accurate results. 

A useful technique to tackle this characterization challenge is to adopt a more operational approach. We can often make system-specific assumptions and simplifications based on physical reasoning to devise more efficient experiments tailored to characterize certain effects. To do so, we start by constructing a model that describes how the system, including the effect of interest, behaves. These models usually contain certain physically-motivated assumptions which are supported by the literature or verified in other independent control experiments. Next, we design a set of simple experiments which are particularly sensitive to the effect of interest. Finally, the experiment is compared to the model and based on the level of consistency between them, we either reject, accept, or iteratively improve the original model based on the measurement outcomes. 

Over the years, many such models have been constructed and tested. In particular, a lot of techniques have been devised to independently measure specific sources of cross-talk between qubits and implement the necessary corrections. As an example, let us consider the cross-talk caused by the residual coupling between two flux-tunable transmon qubits connected by a bus resonator. 

The coupling between two nearest-neighbor qubits $q_i$ and $q_j$ is often employed to implement a controlled-phase gate. During idle times, the qubits are ideally far detuned such that there are no spurious interactions between them. However, in practice, a residual always-on Hamiltonian of the form $H=\zeta_{ij} \ket{11} \bra{11}$ is typically present, causing unwanted evolution of the qubit states even during the idle times. 

\begin{figure}
\centering
\includegraphics{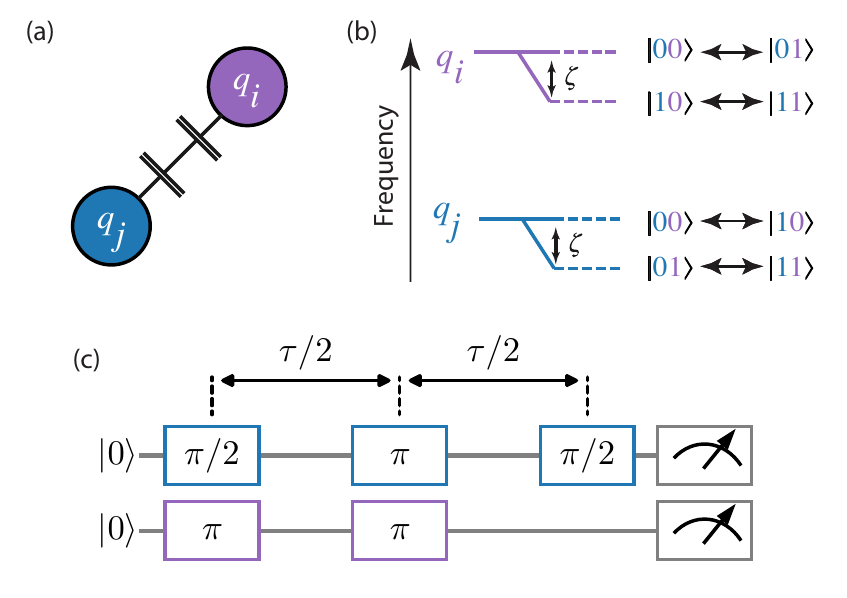}
\caption{\label{fig:residual_ZZ} \textbf{Residual ZZ interactions in multi-qubit devices.} (a) Two neighboring qubits $q_i$ and $q_j$ are coupled through a coupling resonator. (b) At their idle frequencies, both qubit frequencies are slightly dependent on the state of the other qubit with strength $\zeta$. (c) The protocol developed to probe the strength of the always-on ZZ-coupling between two qubits.}
\end{figure}

Since the interaction is constant, the single-qubit terms can be taken as a renormalization of the qubit frequency (Fig.~\ref{fig:residual_ZZ}(a,b)). The resulting state-dependent frequency shift $\zeta_{ij}=\hbar\omega_{11} - \hbar\omega_{01}-\hbar\omega_{10}$ is approximated as
\begin{equation}
\label{eq:residual_ZZ_approx}
\zeta_{ij} = - J_2^2 \left(\frac{1}{\hbar\omega_{20}-\hbar\omega_{11}}+\frac{1}{\hbar\omega_{02}-\hbar\omega_{11}}\right),
\end{equation}
where the subscripts in $\omega_{kl}$ are used to label the states based on the number of excitations in qubits $i$ and $j$ and $J_2 \approx \sqrt{2} J_1$ is the effective coupling between them. 

The residual idle Hamiltonian is then modelled as an additional ZZ-coupling in the system, given by
\begin{equation}
H=\zeta_{ij} \ket{11} \bra{11}=-\frac{\zeta_{ij}}{2}\left(1-Z_i-Z_j+Z_iZ_j\right).
\end{equation}
Following this definition
\begin{equation}
\zeta_{ij} = E_{11} - E_{01} - E_{10}
\end{equation}
where $E_{kl}$ corresponds to the energy of the state with $k$ $(l)$ excitations on qubit $i$ ($j$). This results in coherent correlated $ZZ$ errors with angle
\begin{equation}
\theta=\frac{\zeta_{ij}\tau_I}{4},
\end{equation}
where $\tau_I$ is the total interaction time. 

In other words, such an always-on ZZ coupling induces a frequency shift on $q_i$ that is correlated with the state of its neighbor, $q_j$. The strength of this coupling corresponds to the difference in frequency $\Delta \omega_i^{(j)}$ of Ramsey oscillations of $q_i$ with $q_j$ in states $\ket{1}$ and $\ket{0}$.
\begin{align}
\Delta \omega_i^{(j)} &= \left(E_{11} - E_{01}\right) - \left(E_{10} - E_{00}\right),\\
\Delta \omega_i^{(j)} &= \left(\zeta_{ij} + E_{10}+ E_{01}- E_{01}\right) - \left(E_{10}\right),\\
\Delta \omega_i^{(j)} &= \zeta_{ij}.
\end{align}
We can accurately measure this frequency difference through the residual-ZZ-echo experiment as described in Fig.~\ref{fig:residual_ZZ}(c). In this protocol, an echo experiment over a time $\tau$ is performed on $q_i$ while an excitation is added and subsequently removed from $q_j$. In the first arm of the echo experiment, $q_i$ will acquire a phase $\varphi_A =(\omega_i + \zeta_{ij})\cdot \tau/2$ which is partially canceled by the phase acquired in the second arm $\varphi_B=(-\omega_i)\cdot \tau/2$. This results in oscillations with frequency $\zeta_{ij}/2$ in the measured signal of $q_i$:
\begin{align}
\label{eq:res_zz_echo_phases}
\varphi =& \varphi_A + \varphi_B,\\
\varphi =& (\omega_i + \zeta_{ij})\cdot \tau/2 + (-\omega_i)\cdot \tau/2,\\
\varphi =& (\omega_i + \zeta_{ij} - \omega_i) \cdot \tau/2,\\
\varphi =& (\zeta_{ij}/2) \cdot \tau.
\end{align}
%Note the factor 2.

Compared to using two consecutive Ramsey measurements, this particular technique ensures that the experiment is robust against slow fluctuations in qubit frequency. Furthermore, from a practical perspective, all data is acquired in a single experiment and the desired information can be extracted directly from the outcomes, with only a small number of fit parameters.

It is beyond the scope of this work to give an exhaustive overview of all of these techniques, many of which date back to the days of NMR~\cite{Vandersypen05}. A more extensive overview of error sources affecting transmon systems as well as the proposed methods to characterize them can be found in Ref.~\cite{Rol20_thesis}.

\subsection{Cavity Characterization}

\begin{figure*}
\includegraphics{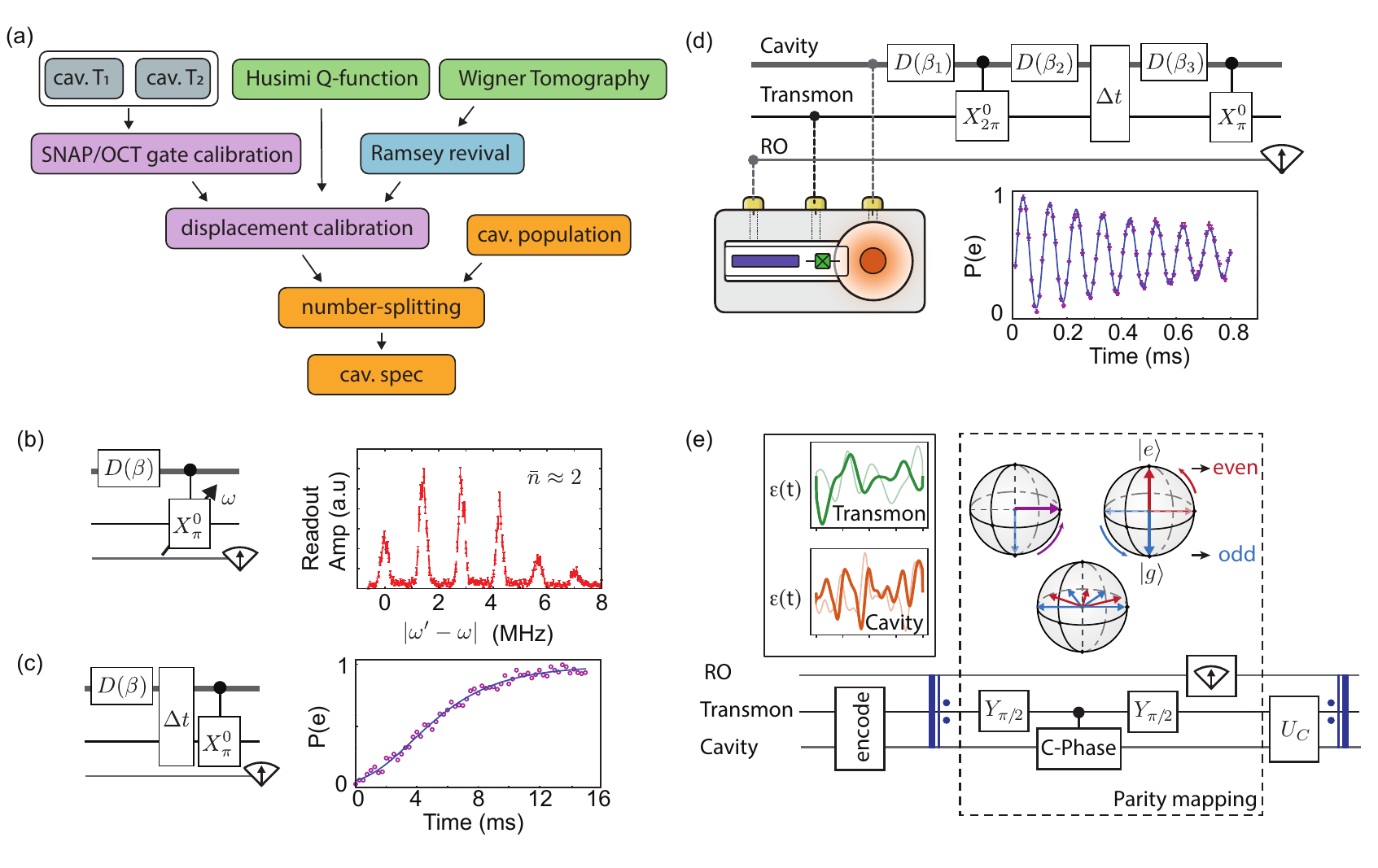}
\caption{
\label{fig:cav_tuneup}
\textbf{Standard procedures for characterizing a high-Q cavity mode.} (a)A typical cavity characterization process presented as a dependency graph. (b) The gate sequence and outcome of a number-splitting experiment where the transmon’s frequency is shifted when a coherent state is initialized in the cavity. Separation between the peaks corresponds to $\chi$. (c) Energy relaxation time scale of the cavity obtained by initializing it in a large coherent state, followed by a flip of the transmon state conditioned on the cavity in vacuum after a variable delay. (d) The typical cQED hardware for operating a bosonic qubit is composed of a high-Q superconducting cavity (e.g.~a 3D coaxial $\lambda/4$ resonator, orange) for storing the logical qubit, a transmon ancilla (green) for universal control, and a low-Q resonator (e.g.~a $\lambda/2$ stripline resonator, blue) for readout (RO). Universal control of the cavity state can be performed via a transmon ancilla (green) and read out via a planar resonator (purple). (e) A QEC protocol realized by mapping the parity of the logical qubit onto the state of the transmon, followed by the appropriate correction unitary. The logical state is initialized using numerical optimal control pulses~\cite{heeres2017_implementing}.
}
\end{figure*}

In addition to qubits and readout resonators, high-Q superconducting cavities are also frequently featured in cQED devices. Endowed with both low intrinsic decoherence rates and large Hilbert space for compact information storage, these cavities can play the role of a coherent quantum memory in cQED systems. In recent years, they have also been frequently employed for encoding logical qubits. More detailed discussions about the developments on this front are presented in Sec.~\ref{sec:qc}.2. 

Here, let us focus on the techniques for characterizing both the Hamiltonian and coherence properties of such high-Q superconducting cavities, as well as the tomography of the quantum states stored in them. The typical workflow for the characterization of the cavity is described in Fig.~\ref{fig:cav_tuneup}a. 

A typical 3D cQED unit consisting of a storage cavity, a readout cavity, and a transmon qubit dispersively coupled to the storage~\cite{kirchmair2013_observation} is depicted in Fig.~\ref{fig:cav_tuneup}(d). The transmon affords universal control of the storage cavity, allowing effective characterization the cavity by mapping the property of interest onto a particular transmon state. The transmon state is subsequently extracted via the readout resonator as described in the standard dispersive readout configuration~\ref{sec:interacting_with_quantum_circ}.3.

In order to map the property of the cavity on the transmon state, conditional operations between the two elements are required. In particular, there are two crucial conditional operations in cQED deivces, both arise from the natural dispersive interaction between the transmon and cavity. 

The first operation is a rotation on the transmon conditioned on a specific photon number in the cavity, $R_{\theta}^{n}$, also commonly referred to as a photon number selective rotation. Here, $R$ is a rotation along either the $x$ or $y$ axis of the transmon Bloch sphere, $\theta$ is the angle of the rotation, and $n$ refers to the photon number in the cavity. Under the dispersive interaction, the transmon frequency is naturally dependent on $n$. Hence, this conditional rotation is simply realized by using long, spectrally-selective drives with duration $\tau>1/\chi$ such that the transmon state is only affected when the cavity contains the correct photon number state. 

The second conditional operation is a cavity-transmon conditional phase gate:
\begin{align}
  \mathbf{C}_\Phi = \mathcal{I} \otimes \ket{g}\bra{g} + e^{i\Phi\mathbf{a}^\dagger\mathbf{a}}\otimes\ket{e}\bra{e}
\end{align}
This C-Phase gate arises naturally during an idle time $t$, with $\Phi=\chi t$. 

Equipped with these two useful conditional operations between the cavity and the transmon ancilla, we now proceed to characterize the storage cavity. Similar to that for a transmon qubit, we start the characterization process of a high-Q cavity with a spectroscopy experiment that allows the extraction of its resonance frequency. In this case, a strong drive with variable frequencies is first applied on the cavity, followed by the $X_{\pi}^{0}$ gate on the transmon ancilla. When the drive frequency on the cavity matches that of its resonance, the cavity is displaced far from the vacuum state. As a result, the transmon state will be unaffected by the selective $\pi$ pulse. Hence, by measuring the resulting state of the transmon, we can extract the resonance frequency of the cavity. 

Next, we proceed to characterise the exact dispersive coupling strength, $\chi$, via the number-splitting experiment~\cite{schuster2007_resolving}. The typical measurement procedure and outcome is shown in Fig.~\ref{fig:cav_tuneup}b, where the transmon spectrum is probed after a large displacement on the cavity. The resulting frequency separations, which correspond to different photon number states in the cavity, indicate the value of $\chi$.

Another useful technique to extract the value of $\chi$ is a Ramsey-type measurement, commonly referred to as Ramsey revival. This experiment uses the coherent transmon precession, which depends on the cavity state, to probe the interaction strength between the two elements. It starts with an initial state $\ket{\psi(0)} = \ket{\beta}\ket{g}$, followed by two transmon rotations, $Y_{\pi/2}$, separated by a variable delay time. 

The resulting evolution can be captured by measuring the state of the transmon ancilla, which is given by~\cite{vlastakis2013_deterministically_supp}
\begin{equation}
%  P_e = \frac{1}{2}\{1 + e^{|\beta|^2(\cos{\chi t}-1)}\cos{(|\beta|^2\sin{\chi t}})\}.
  P_e = \frac{1}{2}\{1 + e^{-2|\beta|^2\sin^2(\chi t/2)}\cos{(|\beta|^2\sin{\chi t}})\}.
\end{equation}
For a large coherent state $|\beta|\gg 1$, $P_e\approx \frac{1}{2}$ during most of the evolution time due to the entanglement of the cavity-transmon system, which results in an apparent mixed state when only the qubit is measured. However, at times $t = 2n\pi/\chi$ ($n\in\mathbb{Z}^+$), the transmon and cavity are no longer entangled, resulting in a ``revival" of the transmon excited state. Hence, the times at which such revival occurs provides a precise measure of $\chi$. 

Next, let us consider the experimental process to extract the coherence times of the cavity. Analogous to the transmon qubits, the performance of a high-Q cavity can be captured by its $T_1$ and $T_2$ Ramsey times. These measurements often require the preparation of Fock states in the cavity. This can be implemented via numerically optimized drives, such as the selective number-dependent arbitrary phase (SNAP)~\cite{krastanov2015_universal, heeres2015_cavity} or optimal control theory gate (OCT)~\cite{heeres2017_implementing} gates.

Physically, the $T_1$ of a cavity indicates the photon-loss rate. This is measured using a simple protocol described in Fig.~\ref{fig:cav_tuneup}(c), where we probe the probability of the cavity returning to its vacuum state from a large coherent state after a variable delay. Alternatively, cavity $T_1$ can also be extracted by measuring the time-scale over which it returns to vacuum from Fock state $|1\rangle$. 

The $T_2$ Ramsey time of an harmonic oscillator mode is defined as duration over which the phase between its lowest two energy levels can be coherently preserved. To probe this, we prepare a state along the equator of the pseudo-Bloch sphere of the resonator’s subspace spanned by $\{|0\rangle, |1\rangle\}$ such as
\begin{equation}
  |\psi_0\rangle \approx \frac{1}{\sqrt{2}}(|0\rangle + |1\rangle).
\end{equation}
The time evolution of this state is similar to that of a two-level system initialized on the equator of its Bloch sphere. After a variable wait time, we perform another displacement operation on the cavity such that the interference effect results in the cavity returning to $|0\rangle$ if the initial phase of the prepared superposition state is unchanged. A comprehensive discussion of these protocols can be found in Chapter 6 of Ref.~\cite{reagor2015_thesis}. 

Finally, for any arbitrary state $\ket\psi$ in the storage cavity, we can also implement QND measurement of photon number parity with the protocol shown in Fig.~\ref{fig:cav_tuneup}(e). In this process, the transmon is first rotated to the equator state $(\ket{g}-\ket{e})/\sqrt{2}$ with a $Y_{\pi/2}$ pulse. After an idle time of $t=\pi/\chi$, under the effect of the $\mathbf{C}_\Phi$ gate, the even-parity and odd-parity photon states accumulate phases of 0 and $\pi$ (modulo 2 $\pi$) respectively. After another transmon $Y_{\pi/2}$ pulse, these unitary operations complete entanglement of the transmon with the parity operator of the cavity:
\begin{align}
  \ket{\psi}\otimes \ket{g} \longrightarrow \mathbf{\Pi_e}\ket{\psi}\otimes\ket{e} + \mathbf{\Pi_o}\ket{\psi}\otimes\ket{g}
\end{align}
where $\mathbf{\Pi_e}$ and $\mathbf{\Pi_o}$ are projectors to the even and odd parity subspace. Each single-shot QND measurement of the transmon state reveals the parity of the cavity state, which informs the binary status of ``error" or ``no-error" on the logical qubit. Repetitive monitoring of the error syndrome using this technique was first demonstrated in Ref.~\cite{sun2014_tracking}, and was subsequently used in a growing number of bosonic QEC experiments~\cite{ofek2016_extending, hu2019_quantum, rosenblum2018_fault-tolerant, reinhold2020_error, ma2020_error}. 

Furthermore, this Ramsey-style parity measurement protocol also allows to perform a direct measurement of the Wigner function of the cavity~\cite{bertet2002_direct, vlastakis2013_deterministically}. The value of the Wigner function at any point of the phase space is equal to the expectation value of the photon number parity after a displacement by $\alpha$: $W(\alpha) = \frac{2}{\pi}\mathrm{Tr}[\bD_\alpha^\dagger\rho\bD_\alpha\mathbf{P}]$. The same technique has been further extended to perform joint photon parity measurements and joint Wigner tomography of two cavities~\cite{wang2015_schrodinger}. 

%%%%%%%%%%%%%%%%%%%%%%%%%%%%%%%%%%%%%%%%%%%%%%%%%%%%%%%%%%%%%%%%%%%%%%%%%%%%%%%%%%%%%%%%%%%%

\section{Towards robust large-scale QC}\label{sec:qc}
In the preceding section, we have introduced the main techniques for the calibration and characterization of a small quantum device consisting of one or qubits and their readout resonators. Building upon this, we can now think about how to incorporate a large number of these physical elements to construct a more complex quantum device that is potentially capable of performing fault-tolerant computation. 

Current efforts towards this goal can be loosely sorted into two main categories. One focuses on scaling up the existing physical elements we have mastered to construct devices on the 100 qubit level and explore the immediate applications such devices could afford. In parallel, another approach is to first tackle the realization of quantum error correction on individual logical elements and then, scale them up in a modular fashion. Both strategies have yielded remarkable progress in recent years. This section serves to review the crucial milestones that have been achieved in the cQED communities and highlight some of the key challenges that still lie ahead as we transition from the current state-of-the-art systems to realizing robust and universal quantum computers.

\subsection{NISQ devices}\label{sec:nisqchip}
As our capabilities to design, fabricate, and manipulate individual physical elements continue to improve, a natural step forward is to construct larger quantum systems with more complexity and computational power. In recent years, we have witnessed a rapid increase in the number of physical qubits, together with their associated readout elements, that are integrated into a single quantum processor in the noisy intermediate-scale quantum (NISQ) regime~\cite{kelly2015_state, takita2016_weight, reagor2018_demonstration, arute2019_quantum, gong2021_quantum}.

We can quantify the development of NISQ devices by considering the number of physical qubits, $n$, present on the quantum processor chip. While this provides an indication of the total Hilbert space available ($2^n$), it does not fully reflect the performance and capabilities of a device. However, quantum bits cannot simply be densely packed on a chip to achieve superior computational power. Doing so typically degrades their individual performance, and more importantly, introduces additional failure modes due to spurious interactions between the physical elements.

Hence, to adequately describe the power of a quantum computer, we must take into account both the number of qubits $n$, and the number of operations that can be performed, which is typically expressed using the circuit depth $d$. The latter corresponds to the number of circuit layers that can be executed before (on average) a single error occurs. Quantitatively, $d\approx 1/ \err_{1\mathrm{step}} = 1/ n\err_{\mathrm{eff}}$, where the effective error rate $\err_{\mathrm{eff}}$ is the average error rate per two-qubit operation. A useful metric that combines these two quantities is the quantum volume $V_{\mathrm{Q}}$~\cite{Moll18,Cross19}. Assuming all-to-all connectivity, where the width of the circuit is equivalent to the number of physical qubits, the quantum volume is defined as $\mathrm{log}_2 (V_{\mathrm{Q}}) = \argmax_n \min (n, d(n))$~\cite{Cross19}. 

However, it is important to note that the quantum volume assumes that the circuits are of equal width and depth. In practice, it is potentially beneficial in the NISQ regime to focus on the short-depth circuits, which could still bring enhancements beyond the reach of classical computing despite the current error rates~\cite{Moll18,Terhal02,Farhi16}.

To illustrate this, let us consider the results described in Ref.~\cite{arute2019_quantum}, in which a $n=53$ qubit device with $\err \sim 10^{-3}$ was used to perform a computation faster than any existing classical machines. This can be understood by looking closely at the quantum volume of a processor, which is designed as a binary metric: can a device run an algorithm? For many algorithms, a single error indicates a catastrophic failure. However, other applications, such as sampling a distribution (as is done in Ref.~\cite{arute2019_quantum}), can tolerate a limited amount of errors simply by averaging the outcomes of the computation. Thus, developing useful applications that could be advantageous to run on a NISQ quantum computer instead of a classical machine has become an active and rapid advancing research frontier.

Moreover, the current NISQ devices also provide a valuable platform to investigate the unique challenges associated with having a large number of quantum elements and test out the various technological solutions that are being rapidly developed the cQED community. 

One of the most pertinent challenges to be address is how to effectively characterize large quantum system. This is crucial as we can only build devices as well as we can measure them. In other words, an effective process to accurately capture the system behavior and gain comprehensive knowledge about its key parameters is indispensable for improving the performance of quantum hardware. This step provides the critical information needed to complete the development cycle of quantum processors, as described in Fig.~\ref{fig:overview}. 

A naive generalization of the various measurement and calibration techniques introduced in Sec.~\ref{sec:interacting_with_quantum_circ} does not suffice to adequately characterize large quantum processors. The reason is two fold. First, if calibration experiments are done sequentially, the number of experiments and the time required become comparable to the time scale over which system parameters drift. While parallelizing these experiments can reduce the total run time, it imposes the additional penalty of cross-talk and unwanted entanglement between individual qubits. If we were to capture these potential correlations between the qubits, we must consider the collective state of the device. This implies the number of parameters increases exponentially with the number of qubits and quickly become intractable. 

Addressing this challenge calls for automated calibration capabilities that integrate control, measurement, and analysis of the measurement outcome in a continuous and efficient workflow. Moreover, new protocols that take into account the complex noise models of a multi-qubit device must also be developed. This is an active field of research, and many novel frameworks have been proposed in recent years~\cite{lilly2020_modeling, gupta2020_adaptive}. These developments highlight the importance of developing NISQ processors, without which we would not be able to verify and improve these calibration protocols. This will, in turn, limit our ability to construct more optimized and robust hardware. 

Another active area of research focuses on improving our ability to reliably design and fabricate devices with larger numbers of physical qubits while ensuring their performance. In a simplified model, each physical qubit should meet three basic conditions: 1.~all control lines are working and have the correct coupling strength to the device; 2.~qubit coherence is above a certain target (e.g.~$50\si{\us}$); and 3.~the relevant system parameters (e.g.~$\omega$, $\chi$, etc.) are within a specified tolerance. 

Based on this, we consider the `yield' of a larger quantum device as the probability that all constituent qubits satisfy the above three requirements. In this simple picture, even if each individual qubit has a high (e.g.~$99\%$) chance of meeting these conditions, the overall yield of a large quantum processor would still deteriorate exponentially to a rather limited level as the number of physical element increases. 

A promising strategy to ameliorate this issue is via a more modular approach, where smaller devices within the same fridge are linked together to form a larger processor~\cite{jiang2007_distributed}. Compared to producing a single monolithic chip, the odds of achieving the same scale by combining multiple smaller patches are much more favorable. Existing flip-chip architectures~\cite{rosenberg17}, in which the readout resonators, Purcell filters, and coupling buses are on a different chip than the qubits, can be seen as a prototype of this technique as they effectively link together different devices. 

As the processing power of the current cQED systems increases, we are also pushing the limits of the associated cryogenic and microwave technologies. For a system up to $\sim 100$ qubits, the heat load imposed by the device, together with the accompanying auxiliary components, can still be managed by using standard techniques as described in Sec.~\ref{sec:environment} and Ref.~\cite{kinner2019_engineering}. However, novel cabling technologies with reduced thermal conductivity and form factor~\cite{Bosman19_MM}, as well as more powerful cryogenic systems will be critical for eventual realization of full scale quantum computers. 

\subsection{Cavity-based logical qubits}\label{sec:qec}
In parallel to the push to construct larger-scale NISQ devices, development of error-corrected logical qubits using continuous variable systems also presents an exciting and highly promising avenue for realizing robust universal quantum computing. 

In particular, recent experimental progress has shown that encoding qubits in multi-photon states of high-coherence superconducting cavities offers a hardware-efficient path towards implementing quantum error correction. There is a growing family of QEC codes, known as the bosonic codes, that employs the cavity's large Hilbert space to store logical information. The construction and implementation of bosonic codes in cQED have become a rapidly-developing area of research in recent years~\cite{gottesman2001_encoding, campagne-ibarcq2020_quantum, puri2020_bias,grimm2020_stabilization, hu2019_quantum, mirrahimi2014_dynamically, michael2016_new}.
This approach has led to the landmark demonstration of the first logical quantum memory with a longer lifetime than any of its physical constituents~\cite{ofek2016_extending}, followed by logical gates between two cavity-based qubits~\cite{rosenblum2018_cnot, gao2019_entangling}, and also error-corrected logical operations~\cite{reinhold2020_error,ma2020_error}.  
For an overview of the progress in this area, readers can refer to some recent review papers~\cite{terhal2020_towards, joshi2021_quantum, cai2021_bosonic}.

\begin{figure}
\includegraphics[scale=1]{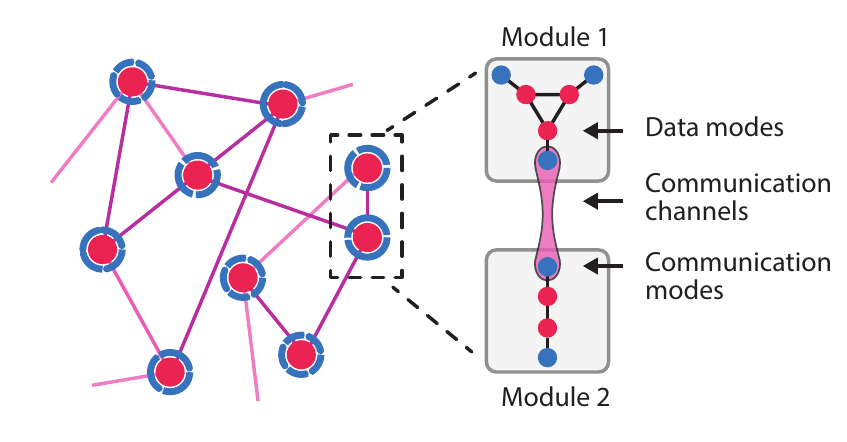}
\caption{\textbf{A conceptual illustration of a modular quantum computer}. Each module consists of a small number of error-corrected logical elements (pink) equipped with robust quantum operations between them. In addition, each module is also designed to have its communication elements (blue) where quantum information can be coherently transferred on the fly across different modules. 
}
\label{fig:modular}
\end{figure}

Furthermore, these cavity-based logical qubits are also motivated by the modular architecture of quantum computing~\cite{monroe2014_large-scale}, where each module is realized with a small number of logical elements equipped with robust first-order-corrected quantum gates (Fig.~\ref{fig:modular}). These modules can potentially be optimized independently and connected with relatively little cross-talk. Communication between modules can be engineered with large on-off ratio and leverage on entanglement distillation protocols to overcome errors~\cite{northup2014_quantum}. This modular architecture is also a natural candidate for implementing distributed computation paradigms in the long run~\cite{kimble2001_conversion, jiang2007_distributed}. 

To realize these high-quality quantum modules endowed with efficient error-correction feature experimentally, robust hardware structures with both good intrinsic coherence and controllability are required. This again highlights one of the key advantages of employing 3D cavity modes as logical elements - they have shown coherence times orders of magnitude longer than the best superconducting qubits to date. While the coherence times of transmons are typically in the 20-100 $\mu$s range~\cite{barends2014_logic} (with a recent demonstration of $>300\,\mu$s~\cite{place2020_new} on a new material platform), standalone 3D superconducting cavities routinely show coherence times on the order of 1\,ms~\cite{reagor2016_quantum, kudra2020_high}, with new preliminary demonstrations up to 2\,s~\cite{romanenko2020_three}. 

Furthermore, it has been shown that the favorable coherence properties of superconducting cavities can be maintained independent of the exact geometry or fabrication processed employed. For example, recent demonstration of lithographically-etched and indium-bonded superconducting cavities achieved a single-photon lifetime approaching 5\,ms~\cite{lei2020_high}. In industrial-scale efforts moving forward, both the planar and the 3D approach to cQED are converging on sophisticated multi-wafer fabrication processes completed with indium bonding, and the vision for such a hardware platform has been nicknamed multilayer microwave integrated quantum circuits (MMIQC)~\cite{brecht2016_multilayer}.

In addition to long coherence times, superconducting cavities also provide an unbounded Hilbert space to allow redundant storage of information, a prerequisite for QEC. Furthermore, expanding the useful Hilbert space by accessing higher excited states does not incur new types of errors (unlike using extra physical qubits), which is crucial for reducing the complexity overhead and enabling near-term benefits of QEC. Broadly speaking, there are three types of first-order errors in a cavity oscillator: single photon loss ($\ba$), single photon gain ($\ba^\dagger$), and dephasing ($\ba^\dagger\ba$). In practice, superconducting cavities can have negligible intrinsic dephasing rates and their effective temperature can be maintained sufficiently low so that the photon gain rate is also negligible. Therefore, if a logical qubit is encoded in the state of a superconducting cavity as a memory unit, there is only one prominent error channel: single photon loss. Performing QEC on this dominant error alone has the prospect of substantially extending the lifetime of quantum information~\cite{ofek2016_extending}. Moreover, it has been demonstrated that this error can be corrected in a passive manner using continuous dissipation~\cite{gertler2020_protecting}.

Quantum error correction against the dominant intrinsic cavity error, single photon loss, represents only the first step towards a fully-functional logical qubit that may outperform physical qubits as quantum building blocks. Another indispensable ingredient is high-quality logical operations between them. Hence, the outstanding challenge at hand is to deliver an error-corrected logic gate with better gate fidelity than native gates on physical qubits, or achieving break-even in terms of gate fidelity. Recent techniques that employ parametric conversion using strong off-resonant pump tones have enabled the demonstration of a coherent swap interaction between two otherwise isolated cavity modes. This interaction forms the basis of bosonic interference and logical entangling gates between two bosonic qubits~\cite{gao2018_programmable, gao2019_entangling, xu2020_demonstration}. In addition, using the idea of path-independence in quantum gates, major developments are underway to improve fault-tolerance of bosonic qubit gates, including recent demonstrations of gates that can tolerate transmon $T_1/T_\phi$ errors~\cite{reinhold2020_error} and cavity photon loss~\cite{ma2020_error}. 

So far, most experiments on bosonic qubits employ a simple cavity-transmon-readout device architecture (Fig.~\ref{fig:cav_tuneup}(d)), which continues to offer exciting opportunities to advance the state of the art. However, the use of transmons as ancillae imposes several important limitations. Firstly, all quantum resources available to the bosonic qubit center around the cavity-transmon dispersive interaction, which inevitably ties the $T_1$ channels of the transmon to cavity dephasing. Secondly, since the transmon is a weakly nonlinear element, parametric processes based on the transmon are relatively slow and may require a lot of microwave power that causes heating or undesirable non-perturbative dynamics~\cite{Lescanne2019_irreversible,zhang2019_engineering, petrescu2020_lifetime}. Looking forward,it is pertinent that future endeavors explore alternative cQED architectures that are more compatible with tunable ancilla-cavity couplings and look beyond simple transmon devices to develop new ancilla circuits with better performance. 

One promising candidate for such new ancilla circuit elements employs three-wave mixing, which has been realized in the Superconducting Nonlinear Asymmetric Inductive eLement (SNAIL)~\cite{frattini2017_three}. Such elements do not produce dispersive shifts to cavity modes and are more effective in modulating bi-modal parametric interactions. It has also been demonstrated that one can control the oscillator quantum states with three-wave mixing processes~\cite{vrajitoarea2019_quantum}. Furthermore, a related ancilla circuit, the asymmetrically threaded SQUID (ATS), has recently demonstrated pristine 2-photon dissipation without introducing any dispersive frequency shifts to the logical qubit~\cite{lescanne2020_exponential}. Finally, it has been proposed that a driven bosonic qubit can be used as an ancilla for error correction and operation of other bosonic qubits~\cite{puri2019_stablized}. 

In order to take advantage of these novel ancilla circuits, which are multi-junction devices, we must be able to provide the necessary magnetic flux to operate them in the appropriate regimes. Therefore, a pressing need in this direction is to integrate flux control lines with high-coherence superconducting cavities~\cite{gargiulo2021_fast,stammeier2018_applying}. 

Another crucial building block for universal quantum computer based on these individually corrected modules is the quantum communication links between them. It has been shown that bosonic qubits can be conveniently released into travelling waves~\cite{pfaff2017_controlled} and recaptured in a separate module~\cite{axline2018_ondemand} using parametric beam-splitter (two-mode swapping) interactions. The catch and release makes use of the temporal mode-matching technique~\cite{cirac1997}, similar to remote state transfer or entanglement experiments with transmons~\cite{kurpiers2018_deterministic}. The added benefit here is that any photon loss in the transmission channel can be first-order corrected~\cite{burkhart2020_error-detected}.

\section{Conclusion and outlook}\label{conclusion}
In the past decade, we have witnessed the remarkable transformation of quantum computing from the realm of scientific curiosity to tangible technological innovations. Rapid progress on the construction of robust quantum hardware has been made in a variety of different physical systems, such as trapped ions~\cite{bruzewicz2019_trapped-ion, brown2016_co-designing}, photons~\cite{wang2019_integrated, slussarenko2019_photonic}, neutral atoms~\cite{saffman2019_quantum}, and spins~\cite{vandersypen2019_quantum, kloeffel2013_prospects}. The cQED platform, in particular, has achieved many major milestones such as demonstration of small-scale QEC~\cite{Riste15, kelly2015_state, Corcoles15,ofek2016_extending, hu2019_quantum, campagne-ibarcq2020_quantum, andersen2020_repeated, bultink2020_protecting} as well as efficient implementations of quantum algorithms on NISQ era devices~\cite{arute2019_quantum, google2020_hartree, Kandala17}.

Together with these achievements, a rich repertoire of knowledge and techniques has been developed in the cQED community, from both the industry and academic teams worldwide. At this pivotal stage of transition from small prototypical devices to robust large-scale quantum computing, a concerted and continual effort in talent-development is indispensable. Here, we have attempted to provide a pr\'{e}cis of some of the useful practices and insights that have been learnt through repeated trial and error in the early days of cQED. Together with the other excellent review articles about the different facets of the cQED technology, we hope that this guide will help accelerate the learning process of a novice experimentalist and allow them to better navigate the process of setting up a successful cQED experiment. 

With the relentless progress and innovation in this field, it suffices to say that the prescriptions we presented here will undoubtedly become obsolete or redundant as the cQED technology continues to mature into more streamlined and robust implementations. Nonetheless, it is our belief that many of these ideas will remain useful and educational for future generations of scientists and engineers going into this exciting and versatile quantum playground. 

\section*{Acknowledgement}
We thank Luigi Frunzio, Shyam Shankar, and Leo DiCarlo for their insightful feedback on the manuscript. We thank E.~Dogan and S.~Shirol for their assistance with example data figures. Y.Y.G.~acknowledges the support of the National Research Foundation Fellowship (Class 2020) and the Ministry of Education, Singapore. C.W.~acknowledges support from the US National Science Foundation (1809114) and Army Research Office (W911NF-17-1-0469).

\bibliographystyle{unsrt}

\bibliography{reference_lib.bib}    %use a bibtex bibliography file

\begin{thebibliography}{100}

\bibitem{vlastakis2013_deterministically}
B.~Vlastakis, G.~Kirchmair, Z.~Leghtas, S.~E. Nigg, L.~Frunzio, S.~M. Girvin,
  M.~Mirrahimi, M.~H. Devoret, and R.~J. Schoelkopf.
\newblock Deterministically encoding quantum information using 100-photon
  schrodinger cat states.
\newblock {\em Science}, 342(6158):607--10, 2013.

\bibitem{wang2015_schrodinger}
C.~Wang, Y.~Y. Gao, P.~Reinhold, R.~W. Heeres, N.~Ofek, K.~Chou, C.~Axline,
  M.~Reagor, J.~Blumoff, K.~M. Sliwa, L.~Frunzio, S.~M. Girvin, L.~Jiang,
  M.~Mirrahimi, M.~H. Devoret, and R.~J. Schoelkopf.
\newblock A schrodinger cat living in two boxes.
\newblock {\em Science}, 352(6289):1087--91, 2016.

\bibitem{asavanant2019_generation}
W.~Asavanant, Y.~Shiozawa, S.~Yokoyama, B.~Charoensombutamon, H.~Emura, R.~N.
  Alexander, S.~Takeda, J.~I. Yoshikawa, N.~C. Menicucci, H.~Yonezawa, and
  A.~Furusawa.
\newblock Generation of time-domain-multiplexed two-dimensional cluster state.
\newblock {\em Science}, 366(6463):373--376, 2019.

\bibitem{larsen2019_deterministic}
M.~V. Larsen, X.~Guo, C.~R. Breum, J.~S. Neergaard-Nielsen, and U.~L. Andersen.
\newblock Deterministic generation of a two-dimensional cluster state.
\newblock {\em Science}, 366(6463):369--372, 2019.

\bibitem{ma2019_dissipatively}
R.~Ma, B.~Saxberg, C.~Owens, N.~Leung, Y.~Lu, J.~Simon, and D.~I. Schuster.
\newblock A dissipatively stabilized mott insulator of photons.
\newblock {\em Nature}, 566(7742):51--57, 2019.

\bibitem{gong2019_genuine}
M.~Gong, M.~C. Chen, Y.~Zheng, S.~Wang, C.~Zha, H.~Deng, Z.~Yan, H.~Rong,
  Y.~Wu, S.~Li, F.~Chen, Y.~Zhao, F.~Liang, J.~Lin, Y.~Xu, C.~Guo, L.~Sun,
  A.~D. Castellano, H.~Wang, C.~Peng, C.~Y. Lu, X.~Zhu, and J.~W. Pan.
\newblock Genuine 12-qubit entanglement on a superconducting quantum processor.
\newblock {\em Phys Rev Lett}, 122(11):110501, 2019.

\bibitem{sagastizabal2020_variational}
R.~Sagastizabal, S.~P. Premaratne, B.~A. Klaver, M.~A. Rol, V.~Negîrneac,
  M.~Moreira, X.~Zou, S.~Johri, N.~Muthusubramanian, M.~Beekman,
  C.~Zachariadis, V.~P. Ostroukh, N.~Haider, A.~Bruno, A.~Y. Matsuura, and
  L.~DiCarlo.
\newblock Variational preparation of finite-temperature states on a quantum
  computer.
\newblock 2020.

\bibitem{ma2020_manipulating}
Y.~Ma, X.~Pan, W.~Cai, X.~Mu, Y.~Xu, L.~Hu, W.~Wang, H.~Wang, Y.~P. Song, Z.~B.
  Yang, S.~B. Zheng, and L.~Sun.
\newblock Manipulating complex hybrid entanglement and testing multipartite
  bell inequalities in a superconducting circuit.
\newblock {\em Phys Rev Lett}, 125(18):180503, 2020.

\bibitem{asaad2020_coherent}
S.~Asaad, V.~Mourik, B.~Joecker, M.~A.~I. Johnson, A.~D. Baczewski, H.~R.
  Firgau, M.~T. Madzik, V.~Schmitt, J.~J. Pla, F.~E. Hudson, K.~M. Itoh, J.~C.
  McCallum, A.~S. Dzurak, A.~Laucht, and A.~Morello.
\newblock Coherent electrical control of a single high-spin nucleus in silicon.
\newblock {\em Nature}, 579(7798):205--209, 2020.

\bibitem{figgatt2019_parallel}
C.~Figgatt, A.~Ostrander, N.~M. Linke, K.~A. Landsman, D.~Zhu, D.~Maslov, and
  C.~Monroe.
\newblock Parallel entangling operations on a universal ion-trap quantum
  computer.
\newblock {\em Nature}, 572(7769):368--372, 2019.

\bibitem{rol2019_fast}
M.~A. Rol, F.~Battistel, F.~K. Malinowski, C.~C. Bultink, B.~M. Tarasinski,
  R.~Vollmer, N.~Haider, N.~Muthusubramanian, A.~Bruno, B.~M. Terhal, and
  L.~DiCarlo.
\newblock Fast, high-fidelity conditional-phase gate exploiting leakage
  interference in weakly anharmonic superconducting qubits.
\newblock {\em Phys Rev Lett}, 123(12):120502, 2019.

\bibitem{paik2016_experimental}
H.~Paik, A.~Mezzacapo, M.~Sandberg, D.~T. McClure, B.~Abdo, A.~D. Corcoles,
  O.~Dial, D.~F. Bogorin, B.~L. Plourde, M.~Steffen, A.~W. Cross, J.~M.
  Gambetta, and J.~M. Chow.
\newblock Experimental demonstration of a resonator-induced phase gate in a
  multiqubit circuit-qed system.
\newblock {\em Phys Rev Lett}, 117(25):250502, 2016.

\bibitem{mckay2016_universal}
David~C. McKay, Stefan Filipp, Antonio Mezzacapo, Easwar Magesan, Jerry~M.
  Chow, and Jay~M. Gambetta.
\newblock Universal gate for fixed-frequency qubits via a tunable bus.
\newblock {\em Physical Review Applied}, 6(6), 2016.

\bibitem{reagor2018_demonstration}
M.~Reagor, C.~B. Osborn, N.~Tezak, A.~Staley, G.~Prawiroatmodjo, M.~Scheer,
  N.~Alidoust, E.~A. Sete, N.~Didier, M.~P. da~Silva, E.~Acala, J.~Angeles,
  A.~Bestwick, M.~Block, B.~Bloom, A.~Bradley, C.~Bui, S.~Caldwell,
  L.~Capelluto, R.~Chilcott, J.~Cordova, G.~Crossman, M.~Curtis, S.~Deshpande,
  T.~El~Bouayadi, D.~Girshovich, S.~Hong, A.~Hudson, P.~Karalekas, K.~Kuang,
  M.~Lenihan, R.~Manenti, T.~Manning, J.~Marshall, Y.~Mohan, W.~O'Brien,
  J.~Otterbach, A.~Papageorge, J.~P. Paquette, M.~Pelstring, A.~Polloreno,
  V.~Rawat, C.~A. Ryan, R.~Renzas, N.~Rubin, D.~Russel, M.~Rust, D.~Scarabelli,
  M.~Selvanayagam, R.~Sinclair, R.~Smith, M.~Suska, T.~W. To, M.~Vahidpour,
  N.~Vodrahalli, T.~Whyland, K.~Yadav, W.~Zeng, and C.~T. Rigetti.
\newblock Demonstration of universal parametric entangling gates on a
  multi-qubit lattice.
\newblock {\em Sci Adv}, 4(2):eaao3603, 2018.

\bibitem{touzard2019_gated}
S.~Touzard, A.~Kou, N.~E. Frattini, V.~V. Sivak, S.~Puri, A.~Grimm, L.~Frunzio,
  S.~Shankar, and M.~H. Devoret.
\newblock Gated conditional displacement readout of superconducting qubits.
\newblock {\em Phys Rev Lett}, 122(8):080502, 2019.

\bibitem{heinsoo2018_rapid}
Johannes Heinsoo, Christian~Kraglund Andersen, Ants Remm, Sebastian Krinner,
  Theodore Walter, Yves Salathé, Simone Gasparinetti, Jean-Claude Besse, Anton
  Potočnik, Andreas Wallraff, and Christopher Eichler.
\newblock Rapid high-fidelity multiplexed readout of superconducting qubits.
\newblock {\em Physical Review Applied}, 10(3), 2018.

\bibitem{mckay2019_three}
D.~C. McKay, S.~Sheldon, J.~A. Smolin, J.~M. Chow, and J.~M. Gambetta.
\newblock Three-qubit randomized benchmarking.
\newblock {\em Phys Rev Lett}, 122(20):200502, 2019.

\bibitem{erhard2019_characterizing}
A.~Erhard, J.~J. Wallman, L.~Postler, M.~Meth, R.~Stricker, E.~A. Martinez,
  P.~Schindler, T.~Monz, J.~Emerson, and R.~Blatt.
\newblock Characterizing large-scale quantum computers via cycle benchmarking.
\newblock {\em Nat Commun}, 10(1):5347, 2019.

\bibitem{wright2019_benchmarking}
K.~Wright, K.~M. Beck, S.~Debnath, J.~M. Amini, Y.~Nam, N.~Grzesiak, J.~S.
  Chen, N.~C. Pisenti, M.~Chmielewski, C.~Collins, K.~M. Hudek, J.~Mizrahi,
  J.~D. Wong-Campos, S.~Allen, J.~Apisdorf, P.~Solomon, M.~Williams, A.~M.
  Ducore, A.~Blinov, S.~M. Kreikemeier, V.~Chaplin, M.~Keesan, C.~Monroe, and
  J.~Kim.
\newblock Benchmarking an 11-qubit quantum computer.
\newblock {\em Nat Commun}, 10(1):5464, 2019.

\bibitem{blais2004_cavity}
Alexandre Blais, Ren-Shou Huang, Andreas Wallraff, S.~M. Girvin, and R.~J.
  Schoelkopf.
\newblock Cavity quantum electrodynamics for superconducting electrical
  circuits: An architecture for quantum computation.
\newblock {\em Physical Review A}, 69(6), 2004.

\bibitem{wallraff2004}
A.~Wallraff, D.~I. Schuster, A.~Blais, L.~Frunzio, R.~Huang, J.~Majer,
  S.~Kumar, S.~M. Girvin, and R.~J. Schoelkopf.
\newblock Strong coupling of a single photon to a superconducting qubit using
  circuit quantum electrodynamics.
\newblock {\em Nature}, 431(7005):162--7, 2004.

\bibitem{reed2012_realization}
M.~D. Reed, L.~DiCarlo, S.~E. Nigg, L.~Sun, L.~Frunzio, S.~M. Girvin, and R.~J.
  Schoelkopf.
\newblock Realization of three-qubit quantum error correction with
  superconducting circuits.
\newblock {\em Nature}, 482(7385):382--5, 2012.

\bibitem{Riste15}
D.~Rist\`{e}, S.~Poletto, M.~Z. Huang, A.~Bruno, V.~Vesterinen, O.~P. Saira,
  and L.~DiCarlo.
\newblock {Detecting bit-flip errors in a logical qubit using stabilizer
  measurements}.
\newblock {\em Nat.\ Commun.}, {6}:6983, {2015}.

\bibitem{kelly2015_state}
J.~Kelly, R.~Barends, A.~G. Fowler, A.~Megrant, E.~Jeffrey, T.~C. White,
  D.~Sank, J.~Y. Mutus, B.~Campbell, Y.~Chen, Z.~Chen, B.~Chiaro, A.~Dunsworth,
  I.~C. Hoi, C.~Neill, P.~J. O'Malley, C.~Quintana, P.~Roushan, A.~Vainsencher,
  J.~Wenner, A.~N. Cleland, and J.~M. Martinis.
\newblock State preservation by repetitive error detection in a superconducting
  quantum circuit.
\newblock {\em Nature}, 519(7541):66--9, 2015.

\bibitem{Corcoles15}
A.~D. C\'orcoles, Easwar Magesan, Srikanth~J. Srinivasan, Andrew~W. Cross,
  M.~Steffen, Jay~M. Gambetta, and Jerry~M. Chow.
\newblock {Demonstration of a quantum error detection code using a square
  lattice of four superconducting qubits}.
\newblock {\em Nat.\ Commun.}, {6}:6979, {2015}.

\bibitem{ofek2016_extending}
N.~Ofek, A.~Petrenko, R.~Heeres, P.~Reinhold, Z.~Leghtas, B.~Vlastakis, Y.~Liu,
  L.~Frunzio, S.~M. Girvin, L.~Jiang, M.~Mirrahimi, M.~H. Devoret, and R.~J.
  Schoelkopf.
\newblock Extending the lifetime of a quantum bit with error correction in
  superconducting circuits.
\newblock {\em Nature}, 536(7617):441--5, 2016.

\bibitem{hu2019_quantum}
L.~Hu, Y.~Ma, W.~Cai, X.~Mu, Y.~Xu, W.~Wang, Y.~Wu, H.~Wang, Y.~P. Song, C.~L.
  Zou, S.~M. Girvin, L.~M. Duan, and L.~Sun.
\newblock Quantum error correction and universal gate set operation on a
  binomial bosonic logical qubit.
\newblock {\em Nature Physics}, 15(5):503--508, 2019.

\bibitem{campagne-ibarcq2020_quantum}
P.~Campagne-Ibarcq, A.~Eickbusch, S.~Touzard, E.~Zalys-Geller, N.~E. Frattini,
  V.~V. Sivak, P.~Reinhold, S.~Puri, S.~Shankar, R.~J. Schoelkopf, L.~Frunzio,
  M.~Mirrahimi, and M.~H. Devoret.
\newblock Quantum error correction of a qubit encoded in grid states of an
  oscillator.
\newblock {\em Nature}, 584(7821):368--372, 2020.

\bibitem{andersen2020_repeated}
Christian~Kraglund Andersen, Ants Remm, Stefania Lazar, Sebastian Krinner,
  Nathan Lacroix, Graham~J. Norris, Mihai Gabureac, Christopher Eichler, and
  Andreas Wallraff.
\newblock Repeated quantum error detection in a surface code.
\newblock {\em Nature Physics}, 2020.

\bibitem{bultink2020_protecting}
C.~C. Bultink, T.~E. O'Brien, R.~Vollmer, N.~Muthusubramanian, M.~W. Beekman,
  M.~A. Rol, X.~Fu, B.~Tarasinski, V.~Ostroukh, B.~Varbanov, A.~Bruno, and
  L.~DiCarlo.
\newblock Protecting quantum entanglement from leakage and qubit errors via
  repetitive parity measurements.
\newblock {\em Sci Adv}, 6(12):eaay3050, 2020.

\bibitem{marques2021logicalqubit}
J.~F. Marques, B.~M. Varbanov, M.~S. Moreira, H.~Ali, N.~Muthusubramanian,
  C.~Zachariadis, F.~Battistel, M.~Beekman, N.~Haider, W.~Vlothuizen, A.~Bruno,
  B.~M. Terhal, and L.~DiCarlo.
\newblock Logical-qubit operations in an error-detecting surface code.
\newblock 2021.

\bibitem{google_quantum_ai_exponential_2021}
{Google Quantum AI}, Zijun Chen, Kevin~J. Satzinger, Juan Atalaya, Alexander~N.
  Korotkov, Andrew Dunsworth, Daniel Sank, Chris Quintana, Matt McEwen, Rami
  Barends, Paul~V. Klimov, Sabrina Hong, Cody Jones, Andre Petukhov, Dvir
  Kafri, Sean Demura, Brian Burkett, Craig Gidney, Austin~G. Fowler, Alexandru
  Paler, Harald Putterman, Igor Aleiner, Frank Arute, Kunal Arya, Ryan Babbush,
  Joseph~C. Bardin, Andreas Bengtsson, Alexandre Bourassa, Michael Broughton,
  Bob~B. Buckley, David~A. Buell, Nicholas Bushnell, Benjamin Chiaro, Roberto
  Collins, William Courtney, Alan~R. Derk, Daniel Eppens, Catherine Erickson,
  Edward Farhi, Brooks Foxen, Marissa Giustina, Ami Greene, Jonathan~A. Gross,
  Matthew~P. Harrigan, Sean~D. Harrington, Jeremy Hilton, Alan Ho, Trent Huang,
  William~J. Huggins, L.~B. Ioffe, Sergei~V. Isakov, Evan Jeffrey, Zhang Jiang,
  Kostyantyn Kechedzhi, Seon Kim, Alexei Kitaev, Fedor Kostritsa, David
  Landhuis, Pavel Laptev, Erik Lucero, Orion Martin, Jarrod~R. McClean, Trevor
  McCourt, Xiao Mi, Kevin~C. Miao, Masoud Mohseni, Shirin Montazeri, Wojciech
  Mruczkiewicz, Josh Mutus, Ofer Naaman, Matthew Neeley, Charles Neill, Michael
  Newman, Murphy~Yuezhen Niu, Thomas~E. O’Brien, Alex Opremcak, Eric Ostby,
  Bálint Pató, Nicholas Redd, Pedram Roushan, Nicholas~C. Rubin, Vladimir
  Shvarts, Doug Strain, Marco Szalay, Matthew~D. Trevithick, Benjamin
  Villalonga, Theodore White, Z.~Jamie Yao, Ping Yeh, Juhwan Yoo, Adam Zalcman,
  Hartmut Neven, Sergio Boixo, Vadim Smelyanskiy, Yu~Chen, Anthony Megrant, and
  Julian Kelly.
\newblock Exponential suppression of bit or phase errors with cyclic error
  correction.
\newblock {\em Nature}, 595(7867):383--387, July 2021.

\bibitem{arute2019_quantum}
F.~Arute, K.~Arya, R.~Babbush, D.~Bacon, J.~C. Bardin, R.~Barends, R.~Biswas,
  S.~Boixo, Fgsl Brandao, D.~A. Buell, B.~Burkett, Y.~Chen, Z.~Chen, B.~Chiaro,
  R.~Collins, W.~Courtney, A.~Dunsworth, E.~Farhi, B.~Foxen, A.~Fowler,
  C.~Gidney, M.~Giustina, R.~Graff, K.~Guerin, S.~Habegger, M.~P. Harrigan,
  M.~J. Hartmann, A.~Ho, M.~Hoffmann, T.~Huang, T.~S. Humble, S.~V. Isakov,
  E.~Jeffrey, Z.~Jiang, D.~Kafri, K.~Kechedzhi, J.~Kelly, P.~V. Klimov,
  S.~Knysh, A.~Korotkov, F.~Kostritsa, D.~Landhuis, M.~Lindmark, E.~Lucero,
  D.~Lyakh, S.~Mandra, J.~R. McClean, M.~McEwen, A.~Megrant, X.~Mi,
  K.~Michielsen, M.~Mohseni, J.~Mutus, O.~Naaman, M.~Neeley, C.~Neill, M.~Y.
  Niu, E.~Ostby, A.~Petukhov, J.~C. Platt, C.~Quintana, E.~G. Rieffel,
  P.~Roushan, N.~C. Rubin, D.~Sank, K.~J. Satzinger, V.~Smelyanskiy, K.~J.
  Sung, M.~D. Trevithick, A.~Vainsencher, B.~Villalonga, T.~White, Z.~J. Yao,
  P.~Yeh, A.~Zalcman, H.~Neven, and J.~M. Martinis.
\newblock Quantum supremacy using a programmable superconducting processor.
\newblock {\em Nature}, 574(7779):505--510, 2019.

\bibitem{devoret2013_superconducting}
M.~H. Devoret and R.~J. Schoelkopf.
\newblock Superconducting circuits for quantum information: an outlook.
\newblock {\em Science}, 339(6124):1169--74, 2013.

\bibitem{wendin2017_quantum}
G.~Wendin.
\newblock Quantum information processing with superconducting circuits: a
  review.
\newblock {\em Rep Prog Phys}, 80(10):106001, 2017.

\bibitem{kjaergaard2020_superconducting}
Morten Kjaergaard, Mollie~E. Schwartz, Jochen Braumüller, Philip Krantz, Joel
  I.~J. Wang, Simon Gustavsson, and William~D. Oliver.
\newblock Superconducting qubits: Current state of play.
\newblock {\em Annual Review of Condensed Matter Physics}, 11(1):369--395,
  2020.

\bibitem{blais2020_quantum}
Alexandre Blais, Steven~M. Girvin, and William~D. Oliver.
\newblock Quantum information processing and quantum optics with circuit
  quantum electrodynamics.
\newblock {\em Nature Physics}, 16(3):247--256, 2020.

\bibitem{blais2020_circuit}
Alexandre {Blais}, Arne~L. {Grimsmo}, S.~M. {Girvin}, and Andreas {Wallraff}.
\newblock {Circuit Quantum Electrodynamics}.
\newblock {\em arXiv e-prints}, page arXiv:2005.12667, May 2020.

\bibitem{krantz2019_engineer}
P.~Krantz, M.~Kjaergaard, F.~Yan, T.~P. Orlando, S.~Gustavsson, and W.~D.
  Oliver.
\newblock A quantum engineer's guide to superconducting qubits.
\newblock {\em Applied Physics Reviews}, 6(2), 2019.

\bibitem{koch2007_charge}
Jens Koch, Terri~M. Yu, Jay Gambetta, A.~A. Houck, D.~I. Schuster, J.~Majer,
  Alexandre Blais, M.~H. Devoret, S.~M. Girvin, and R.~J. Schoelkopf.
\newblock Charge-insensitive qubit design derived from the cooper pair box.
\newblock {\em Physical Review A}, 76(4), 2007.

\bibitem{Girvin12}
S.~M. Girvin.
\newblock Superconducting qubits coupled to microwave photons.
\newblock In {\em Les Houches 2011 Notes}. Oxford University Press, 2012.

\bibitem{thesisSchuster07}
D.~I. Schuster.
\newblock {\em Circuit Quantum Electrodynamics}.
\newblock Ph{D} {D}issertation, Yale University, 2007.

\bibitem{devoret2004_superconducting}
M.~H. {Devoret}, A.~{Wallraff}, and J.~M. {Martinis}.
\newblock {Superconducting Qubits: A Short Review}.
\newblock {\em arXiv e-prints}, November 2004.

\bibitem{devoret2004_implementing}
Michel~H. Devoret and John~M. Martinis.
\newblock Implementing qubits with superconducting integrated circuits.
\newblock {\em Quantum Information Processing}, 3(1-5):163--203, 2004.

\bibitem{langford2013_circuit}
Nathan~K. {Langford}.
\newblock {Circuit QED - Lecture Notes}.
\newblock {\em arXiv e-prints}, page arXiv:1310.1897, October 2013.

\bibitem{Vool2017_introduction}
Uri Vool and Michel Devoret.
\newblock {Introduction to quantum electromagnetic circuits}.
\newblock {\em International Journal of Circuit Theory and Applications},
  45(7):897--934, jul 2017.

\bibitem{manucharyan2009_fluxonium}
V.~E. Manucharyan, J.~Koch, L.~I. Glazman, and M.~H. Devoret.
\newblock Fluxonium: single cooper-pair circuit free of charge offsets.
\newblock {\em Science}, 326(5949):113--6, 2009.

\bibitem{steck2020_quantum}
Daniel~A. Steck.
\newblock Quantum and atom optics, 2020.

\bibitem{haroche2006_exploring}
Serge Haroche and Jean-Michel Raimond.
\newblock Exploring the quantum: Atoms, cavities, and photons.
\newblock {\em Oxford Graduate Texts}, 2006.

\bibitem{nielsen2010_quantum}
Michael~A. Nielsen and Isaac~L. Chuang.
\newblock {\em Quantum Computation and Quantum Information}.
\newblock 2010.

\bibitem{clarke1970_josephson}
John Clarke.
\newblock The josephson effect and e/h.
\newblock {\em American Journal of Physics}, 38(9):1071--1095, 1970.

\bibitem{josephson1962_possible}
B.~D. Josephson.
\newblock Possible new effects in superconductive tunnelling.
\newblock {\em Physics Letters}, 1(7):251--253, 1962.

\bibitem{werninghaus2021_leakage}
M.~Werninghaus, D.~J. Egger, F.~Roy, S.~Machnes, F.~K. Wilhelm, and S.~Filipp.
\newblock Leakage reduction in fast superconducting qubit gates via optimal
  control.
\newblock {\em npj Quantum Information}, 7(1), 2021.

\bibitem{grimsmo2020_quantum}
Arne~L. Grimsmo, Joshua Combes, and Ben~Q. Baragiola.
\newblock Quantum computing with rotation-symmetric bosonic codes.
\newblock {\em Physical Review X}, 10(1), 2020.

\bibitem{jaklevic1964_quantum}
R.~C. Jaklevic, John Lambe, A.~H. Silver, and J.~E. Mercereau.
\newblock Quantum interference effects in josephson tunneling.
\newblock {\em Physical Review Letters}, 12(7):159--160, 1964.

\bibitem{dicarlo2009_demonstration}
L.~DiCarlo, J.~M. Chow, J.~M. Gambetta, L.~S. Bishop, B.~R. Johnson, D.~I.
  Schuster, J.~Majer, A.~Blais, L.~Frunzio, S.~M. Girvin, and R.~J. Schoelkopf.
\newblock Demonstration of two-qubit algorithms with a superconducting quantum
  processor.
\newblock {\em Nature}, 460(7252):240--4, 2009.

\bibitem{Barends2019}
R.~Barends, C.~M. Quintana, A.~G. Petukhov, Yu~Chen, D.~Kafri, K.~Kechedzhi,
  R.~Collins, O.~Naaman, S.~Boixo, F.~Arute, K.~Arya, D.~Buell, B.~Burkett,
  Z.~Chen, B.~Chiaro, A.~Dunsworth, B.~Foxen, A.~Fowler, C.~Gidney,
  M.~Giustina, R.~Graff, T.~Huang, E.~Jeffrey, J.~Kelly, P.~V. Klimov,
  F.~Kostritsa, D.~Landhuis, E.~Lucero, M.~McEwen, A.~Megrant, X.~Mi, J.~Mutus,
  M.~Neeley, C.~Neill, E.~Ostby, P.~Roushan, D.~Sank, K.~J. Satzinger,
  A.~Vainsencher, T.~White, J.~Yao, P.~Yeh, A.~Zalcman, H.~Neven, V.~N.
  Smelyanskiy, and John~M. Martinis.
\newblock {Diabatic Gates for Frequency-Tunable Superconducting Qubits}.
\newblock {\em Physical Review Letters}, 123(21):210501, 2019.

\bibitem{Foxen2020}
B.~Foxen, C.~Neill, A.~Dunsworth, P.~Roushan, B.~Chiaro, A.~Megrant, J.~Kelly,
  Zijun Chen, K.~Satzinger, R.~Barends, F.~Arute, K.~Arya, R.~Babbush,
  D.~Bacon, J.~C. Bardin, S.~Boixo, D.~Buell, B.~Burkett, Yu~Chen, R.~Collins,
  E.~Farhi, A.~Fowler, C.~Gidney, M.~Giustina, R.~Graff, M.~Harrigan, T.~Huang,
  S.~V. Isakov, E.~Jeffrey, Z.~Jiang, D.~Kafri, K.~Kechedzhi, P.~Klimov,
  A.~Korotkov, F.~Kostritsa, D.~Landhuis, E.~Lucero, J.~McClean, M.~McEwen,
  X.~Mi, M.~Mohseni, J.~Y. Mutus, O.~Naaman, M.~Neeley, M.~Niu, A.~Petukhov,
  C.~Quintana, N.~Rubin, D.~Sank, V.~Smelyanskiy, A.~Vainsencher, T.~C. White,
  Z.~Yao, P.~Yeh, A.~Zalcman, H.~Neven, and John~M. Martinis.
\newblock {Demonstrating a Continuous Set of Two-qubit Gates for Near-term
  Quantum Algorithms}.
\newblock {\em Physical Review Letters}, 125(12):120504, 2020.

\bibitem{Reed10b}
M~D Reed, B~R Johnson, A~A Houck, L~DiCarlo, J~M Chow, D~I Schuster, L~Frunzio,
  and R~J Schoelkopf.
\newblock {Fast reset and suppressing spontaneous emission of a superconducting
  qubit}.
\newblock {\em Appl. Phys. Lett.}, 96(20):203110, 2010.

\bibitem{Klimov2018}
P.~V. Klimov, J.~Kelly, Z.~Chen, M.~Neeley, A.~Megrant, B.~Burkett, R.~Barends,
  K.~Arya, B.~Chiaro, Yu~Chen, A.~Dunsworth, A.~Fowler, B.~Foxen, C.~Gidney,
  M.~Giustina, R.~Graff, T.~Huang, E.~Jeffrey, Erik Lucero, J.~Y. Mutus,
  O.~Naaman, C.~Neill, C.~Quintana, P.~Roushan, Daniel Sank, A.~Vainsencher,
  J.~Wenner, T.~C. White, S.~Boixo, R.~Babbush, V.~N. Smelyanskiy, H.~Neven,
  and John~M. Martinis.
\newblock {Fluctuations of Energy-Relaxation Times in Superconducting Qubits}.
\newblock {\em Physical Review Letters}, 121(9):90502, 2018.

\bibitem{Luthi18}
Florian Luthi, Thijs Stavenga, OW~Enzing, Alessandro Bruno, Christian Dickel,
  NK~Langford, Michiel~Adriaan Rol, Thomas~Sand Jespersen, Jesper Nyg{\aa}rd,
  P~Krogstrup, and Leo DiCarlo.
\newblock Evolution of nanowire transmon qubits and their coherence in a
  magnetic field.
\newblock {\em Phys. Rev. Lett.}, 120(10):100502, 2018.

\bibitem{Yamamoto2008_fluxpumped}
T.~Yamamoto, K.~Inomata, M.~Watanabe, K.~Matsuba, T.~Miyazaki, W.~D. Oliver,
  Y.~Nakamura, and J.~S. Tsai.
\newblock {Flux-driven Josephson parametric amplifier}.
\newblock {\em Applied Physics Letters}, 93(4):1--4, 2008.

\bibitem{lescanne2020_exponential}
Raphaël Lescanne, Marius Villiers, Théau Peronnin, Alain Sarlette, Matthieu
  Delbecq, Benjamin Huard, Takis Kontos, Mazyar Mirrahimi, and Zaki Leghtas.
\newblock Exponential suppression of bit-flips in a qubit encoded in an
  oscillator.
\newblock {\em Nature Physics}, 16(5):509--513, 2020.

\bibitem{Mooij1999}
J~E Mooij.
\newblock {Josephson Persistent-Current Qubit}.
\newblock {\em Science}, 285(5430):1036--1039, aug 1999.

\bibitem{yan2016_theflux}
F.~Yan, S.~Gustavsson, A.~Kamal, J.~Birenbaum, A.~P. Sears, D.~Hover, T.~J.
  Gudmundsen, D.~Rosenberg, G.~Samach, S.~Weber, J.~L. Yoder, T.~P. Orlando,
  J.~Clarke, A.~J. Kerman, and W.~D. Oliver.
\newblock The flux qubit revisited to enhance coherence and reproducibility.
\newblock {\em Nat Commun}, 7:12964, 2016.

\bibitem{Nguyen2019}
Long~B. Nguyen, Yen~Hsiang Lin, Aaron Somoroff, Raymond Mencia, Nicholas
  Grabon, and Vladimir~E. Manucharyan.
\newblock {High-Coherence Fluxonium Qubit}.
\newblock {\em Physical Review X}, 9(4):41041, 2019.

\bibitem{gottesman2001_encoding}
Daniel Gottesman, Alexei Kitaev, and John Preskill.
\newblock Encoding a qubit in an oscillator.
\newblock {\em Physical Review A}, 64(1), 2001.

\bibitem{Gladchenko2009_cos2phi}
Sergey Gladchenko, David Olaya, Eva Dupont-Ferrier, Benoit Dou{\c{c}}ot, Lev~B.
  Ioffe, and Michael~E. Gershenson.
\newblock {Superconducting nanocircuits for topologically protected qubits}.
\newblock {\em Nature Physics}, 5(1):48--53, 2009.

\bibitem{smith2020_superconducting}
W.~C. Smith, A.~Kou, X.~Xiao, U.~Vool, and M.~H. Devoret.
\newblock Superconducting circuit protected by two-cooper-pair tunneling.
\newblock {\em npj Quantum Information}, 6(1), 2020.

\bibitem{brooks2013_protected}
Peter Brooks, Alexei Kitaev, and John Preskill.
\newblock Protected gates for superconducting qubits.
\newblock {\em Physical Review A}, 87(5), 2013.

\bibitem{gyenis2020_experimental}
András Gyenis, Pranav~S. Mundada, Agustin Di~Paolo, Thomas~M. Hazard, Xinyuan
  You, David~I. Schuster, Jens Koch, Alexandre Blais, and Andrew~A. Houck.
\newblock Experimental realization of a protected superconducting circuit
  derived from the $0-\pi$ qubit.
\newblock {\em PRX Quantum}, 2(1), 2021.

\bibitem{grynberg2010_introduction}
Gilbert Grynberg, Alain Aspect, Claude Fabre, and Claude Cohen-Tannoudji.
\newblock {\em Introduction to Quantum Optics}.
\newblock 2010.

\bibitem{mirrahimi2014_dynamically}
Mazyar Mirrahimi, Zaki Leghtas, Victor~V. Albert, Steven Touzard, Robert~J.
  Schoelkopf, Liang Jiang, and Michel~H. Devoret.
\newblock Dynamically protected cat-qubits: a new paradigm for universal
  quantum computation.
\newblock {\em New Journal of Physics}, 16(4):045014, 2014.

\bibitem{leghtas2015_confining}
Z.~Leghtas, S.~Touzard, I.~M. Pop, A.~Kou, B.~Vlastakis, A.~Petrenko, K.~M.
  Sliwa, A.~Narla, S.~Shankar, M.~J. Hatridge, M.~Reagor, L.~Frunzio, R.~J.
  Schoelkopf, M.~Mirrahimi, and M.~H. Devoret.
\newblock Confining the state of light to a quantum manifold by engineered
  two-photon loss.
\newblock {\em Science}, 347(6224):853--7, 2015.

\bibitem{grimm2020_stabilization}
A~Grimm, N~E Frattini, S~Puri, S~O Mundhada, S~Touzard, M~Mirrahimi, S~M
  Girvin, S~Shankar, and M~H Devoret.
\newblock {Stabilization and operation of a Kerr-cat qubit}.
\newblock {\em Nature}, 584(7820):205--209, 2020.

\bibitem{Roch14}
N.~Roch, M.~E. Schwartz, F.~Motzoi, C.~Macklin, R.~Vijay, A.~W. Eddins, A.~N.
  Korotkov, K.~B. Whaley, M.~Sarovar, and I.~Siddiqi.
\newblock Observation of measurement-induced entanglement and quantum
  trajectories of remote superconducting qubits.
\newblock {\em Phys. Rev. Lett.}, 112:170501, Apr 2014.

\bibitem{Narla16}
A.~Narla, S.~Shankar, M.~Hatridge, Z.~Leghtas, K.~M. Sliwa, E.~Zalys-Geller,
  S.~O. Mundhada, W.~Pfaff, L.~Frunzio, R.~J. Schoelkopf, and M.~H. Devoret.
\newblock Robust concurrent remote entanglement between two superconducting
  qubits.
\newblock {\em Phys. Rev. X}, 6:031036, Sep 2016.

\bibitem{campagne-ibarcq2018_deterministic}
P.~Campagne-Ibarcq, E.~Zalys-Geller, A.~Narla, S.~Shankar, P.~Reinhold,
  L.~Burkhart, C.~Axline, W.~Pfaff, L.~Frunzio, R.~J. Schoelkopf, and M.~H.
  Devoret.
\newblock Deterministic remote entanglement of superconducting circuits through
  microwave two-photon transitions.
\newblock {\em Phys Rev Lett}, 120(20):200501, 2018.

\bibitem{axline2018_ondemand}
Christopher~J. Axline, Luke~D. Burkhart, Wolfgang Pfaff, Mengzhen Zhang, Kevin
  Chou, Philippe Campagne-Ibarcq, Philip Reinhold, Luigi Frunzio, S.~M. Girvin,
  Liang Jiang, M.~H. Devoret, and R.~J. Schoelkopf.
\newblock On-demand quantum state transfer and entanglement between remote
  microwave cavity memories.
\newblock {\em Nature Physics}, 14(7):705--710, 2018.

\bibitem{Teufel2011_sideband}
J.~D. Teufel, T.~Donner, Dale Li, J.~W. Harlow, M.~S. Allman, K.~Cicak, A.~J.
  Sirois, J.~D. Whittaker, K.~W. Lehnert, and R.~W. Simmonds.
\newblock {Sideband cooling of micromechanical motion to the quantum ground
  state}.
\newblock {\em Nature}, 475(7356):359--363, 2011.

\bibitem{rosenblum2018_cnot}
S.~Rosenblum, Y.~Y. Gao, P.~Reinhold, C.~Wang, C.~J. Axline, L.~Frunzio, S.~M.
  Girvin, L.~Jiang, M.~Mirrahimi, M.~H. Devoret, and R.~J. Schoelkopf.
\newblock A cnot gate between multiphoton qubits encoded in two cavities.
\newblock {\em Nat Commun}, 9(1):652, 2018.

\bibitem{gao2018_programmable}
Yvonne~Y. Gao, Brian~J. Lester, Yaxing Zhang, Chen Wang, Serge Rosenblum, Luigi
  Frunzio, Liang Jiang, S.~ M Girvin, and Robert~J. Schoelkopf.
\newblock Programmable interference between two microwave quantum memories.
\newblock {\em Physical Review X}, 8(2), 2018.

\bibitem{gao2019_entangling}
Y.~Y. Gao, B.~J. Lester, K.~S. Chou, L.~Frunzio, M.~H. Devoret, L.~Jiang, S.~M.
  Girvin, and R.~J. Schoelkopf.
\newblock Entanglement of bosonic modes through an engineered exchange
  interaction.
\newblock {\em Nature}, 566(7745):509--512, 2019.

\bibitem{Lescanne2019_irreversible}
Rapha{\"{e}}l Lescanne, Samuel Del{\'{e}}glise, Emanuele Albertinale, Ulysse
  R{\'{e}}glade, Thibault Capelle, Edouard Ivanov, Thibaut Jacqmin, Zaki
  Leghtas, and Emmanuel Flurin.
\newblock {Irreversible Qubit-Photon Coupling for the Detection of Itinerant
  Microwave Photons}.
\newblock {\em Physical Review X}, 10(2):021038, 2020.

\bibitem{joshi2021_quantum}
Atharv Joshi, Kyungjoo Noh, and Yvonne~Y. Gao.
\newblock Quantum information processing with bosonic qubits in circuit qed.
\newblock {\em Quantum Science and Technology}, 6(3), 2021.

\bibitem{hu2018_simulation}
Ling Hu, Yue-Chi Ma, Yuan Xu, Wei-Ting Wang, Yu-Wei Ma, Ke~Liu, Hai-Yan Wang,
  Yi-Pu Song, Man-Hong Yung, and Lu-Yan Sun.
\newblock Simulation of molecular spectroscopy with circuit quantum
  electrodynamics.
\newblock {\em Science Bulletin}, 63(5):293--299, 2018.

\bibitem{wang2020_efficient}
Christopher~S. Wang, Jacob~C. Curtis, Brian~J. Lester, Yaxing Zhang, Yvonne~Y.
  Gao, Jessica Freeze, Victor~S. Batista, Patrick~H. Vaccaro, Isaac~L. Chuang,
  Luigi Frunzio, Liang Jiang, S.~ M Girvin, and Robert~J. Schoelkopf.
\newblock Efficient multiphoton sampling of molecular vibronic spectra on a
  superconducting bosonic processor.
\newblock {\em Physical Review X}, 10(2), 2020.

\bibitem{gargiulo2021_fast}
O.~Gargiulo, S.~Oleschko, J.~Prat-Camps, M.~Zanner, and G.~Kirchmair.
\newblock Fast flux control of 3d transmon qubits using a magnetic hose.
\newblock {\em Applied Physics Letters}, 118(1), 2021.

\bibitem{Dickel18_thesis}
Christian Dickel.
\newblock {\em Scalability and modularity for transmon-based quantum
  processors}.
\newblock Ph{D} {D}issertation, Delft University of Technology, 2018.

\bibitem{huang2021_microwave}
Sihao Huang, Benjamin Lienhard, Greg Calusine, Antti Vepsäläinen, Jochen
  Braumüller, David~K. Kim, Alexander~J. Melville, Bethany~M. Niedzielski,
  Jonilyn~L. Yoder, Bharath Kannan, Terry~P. Orlando, Simon Gustavsson, and
  William~D. Oliver.
\newblock Microwave package design for superconducting quantum processors.
\newblock {\em PRX Quantum}, 2(2), 2021.

\bibitem{reagor2016_quantum}
M.~Reagor, W.~Pfaff, C.~Axline, R.~W. Heeres, N.~Ofek, K.~Sliwa, E.~Holland,
  C.~Wang, J.~Blumoff, K.~Chou, M.~J. Hatridge, L.~Frunzio, M.~H. Devoret,
  L.~Jiang, and R.~J. Schoelkopf.
\newblock Quantum memory with millisecond coherence in circuit qed.
\newblock {\em Physical Review B}, 94(1), 2016.

\bibitem{axline2016_architecture}
C.~Axline, M.~Reagor, R.~Heeres, P.~Reinhold, C.~Wang, K.~Shain, W.~Pfaff,
  Y.~Chu, L.~Frunzio, and R.~J. Schoelkopf.
\newblock An architecture for integrating planar and 3d cqed devices.
\newblock {\em Applied Physics Letters}, 109(4):042601, 2016.

\bibitem{Sears12}
A.~P. Sears, A.~Petrenko, G.~Catelani, L.~Sun, Hanhee Paik, G.~Kirchmair,
  L.~Frunzio, L.~I. Glazman, S.~M. Girvin, and R.~J. Schoelkopf.
\newblock Photon shot noise dephasing in the strong-dispersive limit of circuit
  {QED}.
\newblock {\em Phys. Rev. B}, 86:180504, Nov 2012.

\bibitem{brecht2016_multilayer}
Teresa Brecht, Wolfgang Pfaff, Chen Wang, Yiwen Chu, Luigi Frunzio, Michel~H.
  Devoret, and Robert~J. Schoelkopf.
\newblock Multilayer microwave integrated quantum circuits for scalable quantum
  computing.
\newblock {\em npj Quantum Information}, 2(1), 2016.

\bibitem{goano2001_general}
M.~Goano, F.~Bertazzi, P.~Caravelli, G.~Ghione, and T.~A. Driscoll.
\newblock A general conformal-mapping approach to the optimum electrode design
  of coplanar waveguides with arbitrary cross section.
\newblock {\em IEEE Transactions on Microwave Theory and Techniques},
  49(9):1573--1580, 2001.

\bibitem{martinis2014_calculation}
John~M. {Martinis}, Rami {Barends}, and Alexander~N. {Korotkov}.
\newblock {Calculation of Coupling Capacitance in Planar Electrodes}.
\newblock {\em arXiv e-prints}, page arXiv:1410.3458, October 2014.

\bibitem{sears2012_photon}
A.~P. Sears, A.~Petrenko, G.~Catelani, L.~Sun, Hanhee Paik, G.~Kirchmair,
  L.~Frunzio, L.~I. Glazman, S.~M. Girvin, and R.~J. Schoelkopf.
\newblock Photon shot noise dephasing in the strong-dispersive limit of circuit
  qed.
\newblock {\em Physical Review B}, 86(18), 2012.

\bibitem{stammeier2018_applying}
M.~Stammeier, S.~Garcia, and A.~Wallraff.
\newblock Applying electric and magnetic field bias in a 3d superconducting
  waveguide cavity with high quality factor.
\newblock {\em Quantum Science and Technology}, 3(4), 2018.

\bibitem{Houck08}
A.~A. Houck, J.~A. Schreier, B.~R. Johnson, J.~M. Chow, Jens Koch, J.~M.
  Gambetta, D.~I. Schuster, L.~Frunzio, M.~H. Devoret, S.~M. Girvin, and R.~J.
  Schoelkopf.
\newblock Controlling the spontaneous emission of a superconducting transmon
  qubit.
\newblock {\em Phys. Rev. Lett.}, 101:080502, 2008.

\bibitem{thesisChow10}
J.~M. Chow.
\newblock {\em Quantum Information Processing with Superconducting Qubits}.
\newblock Ph{D} {D}issertation, Yale University, 2010.

\bibitem{Gely_2020}
Mario~F Gely and Gary~A Steele.
\newblock Qucat: quantum circuit analyzer tool in python.
\newblock {\em New Journal of Physics}, 22(1):013025, Jan 2020.

\bibitem{groszkowski2021scqubits}
Peter Groszkowski and Jens Koch.
\newblock Scqubits: a python package for superconducting qubits, 2021.

\bibitem{minev2020_energy}
Zlatko~K. {Minev}, Zaki {Leghtas}, Shantanu~O. {Mundhada}, Lysander
  {Christakis}, Ioan~M. {Pop}, and Michel~H. {Devoret}.
\newblock {Energy-participation quantization of Josephson circuits}.
\newblock arXiv:2010.00620.

\bibitem{Gao08_thesis}
J.~Gao.
\newblock {\em The Physics of Superconducting Microwave Resonators,}.
\newblock Ph{D} {D}issertation, California Institute of Technology, 2008.

\bibitem{nigg2014_blackbox}
S.~E. Nigg, H.~Paik, B.~Vlastakis, G.~Kirchmair, S.~Shankar, L.~Frunzio, M.~H.
  Devoret, R.~J. Schoelkopf, and S.~M. Girvin.
\newblock Black-box superconducting circuit quantization.
\newblock {\em Phys Rev Lett}, 108(24):240502, 2012.

\bibitem{Minev18_thesis}
Z.K. Minev.
\newblock {\em Catching and Reversing a Quantum Jump Mid-Flight}.
\newblock Ph{D} {D}issertation, Yale University, 2018.

\bibitem{solgun2015_multiport}
Firat Solgun and David~P. DiVincenzo.
\newblock Multiport impedance quantization.
\newblock {\em Annals of Physics}, 361:605--669, 2015.

\bibitem{minev2020_git}
Zlatko~K. {Minev} and Zaki {Leghtas}.
\newblock Automated python module for the design and quantization of josephson
  quantum circuits.
\newblock \url{https://github.com/zlatko-minev/pyEPR}, 2020.

\bibitem{wang2015_surface}
C.~Wang, C.~Axline, Y.~Y. Gao, T.~Brecht, Y.~Chu, L.~Frunzio, M.~H. Devoret,
  and R.~J. Schoelkopf.
\newblock Surface participation and dielectric loss in superconducting qubits.
\newblock {\em Applied Physics Letters}, 107(16):162601, 2015.

\bibitem{murray2018_analytical}
Conal~E. Murray, Jay~M. Gambetta, Douglas~T. McClure, and Matthias Steffen.
\newblock Analytical determination of participation in superconducting coplanar
  architectures.
\newblock {\em IEEE Transactions on Microwave Theory and Techniques},
  66(8):3724--3733, 2018.

\bibitem{wenner2011_surface}
J.~Wenner, R.~Barends, R.~C. Bialczak, Yu~Chen, J.~Kelly, Erik Lucero, Matteo
  Mariantoni, A.~Megrant, P.~J.~J. O’Malley, D.~Sank, A.~Vainsencher,
  H.~Wang, T.~C. White, Y.~Yin, J.~Zhao, A.~N. Cleland, and John~M. Martinis.
\newblock Surface loss simulations of superconducting coplanar waveguide
  resonators.
\newblock {\em Applied Physics Letters}, 99(11):113513, 2011.

\bibitem{Bruno15Reduced}
A.~Bruno, G.~de~Lange, S.~Asaad, K.~L. van~der Enden, N.~K. Langford, and
  L.~DiCarlo.
\newblock Reducing intrinsic loss in superconducting resonators by surface
  treatment and deep etching of silicon substrates.
\newblock {\em Applied Physics Letters}, 106(18):182601, 2015.

\bibitem{vissers2012_reduced}
Michael~R. Vissers, Jeffrey~S. Kline, Jiansong Gao, David~S. Wisbey, and
  David~P. Pappas.
\newblock Reduced microwave loss in trenched superconducting coplanar
  waveguides.
\newblock {\em Applied Physics Letters}, 100(8), 2012.

\bibitem{calusine2018_analysis}
G.~Calusine, A.~Melville, W.~Woods, R.~Das, C.~Stull, V.~Bolkhovsky, D.~Braje,
  D.~Hover, D.~K. Kim, X.~Miloshi, D.~Rosenberg, A.~Sevi, J.~L. Yoder,
  E.~Dauler, and W.~D. Oliver.
\newblock Analysis and mitigation of interface losses in trenched
  superconducting coplanar waveguide resonators.
\newblock {\em Applied Physics Letters}, 112(6), 2018.

\bibitem{paik2011_observation}
H.~Paik, D.~I. Schuster, L.~S. Bishop, G.~Kirchmair, G.~Catelani, A.~P. Sears,
  B.~R. Johnson, M.~J. Reagor, L.~Frunzio, L.~I. Glazman, S.~M. Girvin, M.~H.
  Devoret, and R.~J. Schoelkopf.
\newblock Observation of high coherence in josephson junction qubits measured
  in a three-dimensional circuit qed architecture.
\newblock {\em Phys Rev Lett}, 107(24):240501, 2011.

\bibitem{maloney1972_superconducting}
M.~David Maloney, Francisco de~la Cruz, and Manuel Cardona.
\newblock Superconducting parameters and size effects of aluminum films and
  foils.
\newblock {\em Physical Review B}, 5(9):3558--3572, 1972.

\bibitem{nazaretski2009_direct}
E.~Nazaretski, J.~P. Thibodaux, I.~Vekhter, L.~Civale, J.~D. Thompson, and
  R.~Movshovich.
\newblock Direct measurements of the penetration depth in a superconducting
  film using magnetic force microscopy.
\newblock {\em Applied Physics Letters}, 95(26), 2009.

\bibitem{reagor2013_reaching}
Matthew Reagor, Hanhee Paik, Gianluigi Catelani, Luyan Sun, Christopher Axline,
  Eric Holland, Ioan~M. Pop, Nicholas~A. Masluk, Teresa Brecht, Luigi Frunzio,
  Michel~H. Devoret, Leonid Glazman, and Robert~J. Schoelkopf.
\newblock Reaching 10 ms single photon lifetimes for superconducting aluminum
  cavities.
\newblock {\em Applied Physics Letters}, 102(19):192604, 2013.

\bibitem{catelani2011_quasiparticle}
G.~Catelani, J.~Koch, L.~Frunzio, R.~J. Schoelkopf, M.~H. Devoret, and L.~I.
  Glazman.
\newblock Quasiparticle relaxation of superconducting qubits in the presence of
  flux.
\newblock {\em Physical Review Letters}, 106(7), 2011.

\bibitem{nsanzinesa2014_trapping}
I.~Nsanzineza and B.~L. Plourde.
\newblock Trapping a single vortex and reducing quasiparticles in a
  superconducting resonator.
\newblock {\em Phys Rev Lett}, 113(11):117002, 2014.

\bibitem{dunsworth2017_characterization}
A.~Dunsworth, A.~Megrant, C.~Quintana, Zijun Chen, R.~Barends, B.~Burkett,
  B.~Foxen, Yu~Chen, B.~Chiaro, A.~Fowler, R.~Graff, E.~Jeffrey, J.~Kelly,
  E.~Lucero, J.~Y. Mutus, M.~Neeley, C.~Neill, P.~Roushan, D.~Sank,
  A.~Vainsencher, J.~Wenner, T.~C. White, and John~M. Martinis.
\newblock Characterization and reduction of capacitive loss induced by
  sub-micron josephson junction fabrication in superconducting qubits.
\newblock {\em Applied Physics Letters}, 111(2), 2017.

\bibitem{brecht2015_demonstration}
T.~Brecht, M.~Reagor, Y.~Chu, W.~Pfaff, C.~Wang, L.~Frunzio, M.~H. Devoret, and
  R.~J. Schoelkopf.
\newblock Demonstration of superconducting micromachined cavities.
\newblock {\em Applied Physics Letters}, 107(19), 2015.

\bibitem{Tinkham96}
M.~Tinkham.
\newblock {\em Introduction to Superconductivity}.
\newblock Dover Publications, New York, 2nd edition, 2004.

\bibitem{serniak2019_direct}
K.~Serniak, S.~Diamond, M.~Hays, V.~Fatemi, S.~Shankar, L.~Frunzio, R.~J.
  Schoelkopf, and M.~H. Devoret.
\newblock Direct dispersive monitoring of charge parity in
  offset-charge-sensitive transmons.
\newblock {\em Physical Review Applied}, 12(1), 2019.

\bibitem{wang2014_measurement}
C.~Wang, Y.~Y. Gao, I.~M. Pop, U.~Vool, C.~Axline, T.~Brecht, R.~W. Heeres,
  L.~Frunzio, M.~H. Devoret, G.~Catelani, L.~I. Glazman, and R.~J. Schoelkopf.
\newblock Measurement and control of quasiparticle dynamics in a
  superconducting qubit.
\newblock {\em Nat Commun}, 5:5836, 2014.

\bibitem{hosseinkhani2017_optimal}
A.~Hosseinkhani, R.~P. Riwar, R.~ J Schoelkopf, L.~ I Glazman, and
  G.~Catelani.
\newblock Optimal configurations for normal-metal traps in transmon qubits.
\newblock {\em Physical Review Applied}, 8(6), 2017.

\bibitem{dolan1977_offset}
G.~J. Dolan.
\newblock Offset masks for lift‐off photoprocessing.
\newblock {\em Applied Physics Letters}, 31(5):337--339, 1977.

\bibitem{lecocq2011_junction}
F.~Lecocq, I.~M. Pop, Z.~Peng, I.~Matei, T.~Crozes, T.~Fournier, C.~Naud,
  W.~Guichard, and O.~Buisson.
\newblock Junction fabrication by shadow evaporation without a suspended
  bridge.
\newblock {\em Nanotechnology}, 22(31):315302, 2011.

\bibitem{mamin2021_merged}
H.~J. {Mamin}, E.~{Huang}, S.~{Carnevale}, C.~T. {Rettner}, N.~{Arellano},
  M.~H. {Sherwood}, C.~{Kurter}, B.~{Trimm}, M.~{Sandberg}, R.~M. {Shelby},
  M.~A. {Mueed}, B.~A. {Madon}, A.~{Pushp}, M.~{Steffen}, and D.~{Rugar}.
\newblock {Merged-Element Transmons: Design and Qubit Performance}.
\newblock {\em arXiv e-prints}, page arXiv:2103.09163, March 2021.

\bibitem{kreikebaum2020_improving}
J.~M. Kreikebaum, K.~P. O’Brien, A.~Morvan, and I.~Siddiqi.
\newblock Improving wafer-scale josephson junction resistance variation in
  superconducting quantum coherent circuits.
\newblock {\em Superconductor Science and Technology}, 33(6), 2020.

\bibitem{costache2012_lateral}
Marius~V. Costache, Germàn Bridoux, Ingmar Neumann, and Sergio~O. Valenzuela.
\newblock Lateral metallic devices made by a multiangle shadow evaporation
  technique.
\newblock {\em Journal of Vacuum Science \& Technology B, Nanotechnology and
  Microelectronics: Materials, Processing, Measurement, and Phenomena}, 30(4),
  2012.

\bibitem{wu2017_overlap}
X.~Wu, J.~L. Long, H.~S. Ku, R.~E. Lake, M.~Bal, and D.~P. Pappas.
\newblock Overlap junctions for high coherence superconducting qubits.
\newblock {\em Applied Physics Letters}, 111(3), 2017.

\bibitem{stehli2020_coherent}
Alexander Stehli, Jan~David Brehm, Tim Wolz, Paul Baity, Sergey Danilin,
  Valentino Seferai, Hannes Rotzinger, Alexey~V. Ustinov, and Martin Weides.
\newblock Coherent superconducting qubits from a subtractive junction
  fabrication process.
\newblock {\em Applied Physics Letters}, 117(12), 2020.

\bibitem{bilmes2021_in-situ}
Alexander {Bilmes}, Alexander~K. {Neumann}, Serhii {Volosheniuk}, Alexey~V.
  {Ustinov}, and J{\"u}rgen {Lisenfeld}.
\newblock {In-situ bandaged Josephson junctions for superconducting quantum
  processors}.
\newblock {\em arXiv e-prints}, page arXiv:2101.01453, January 2021.

\bibitem{steffen2017_recent}
M.~Steffen, M.~Sandberg, and S.~Srinivasan.
\newblock Recent research trends for high coherence quantum circuits.
\newblock {\em Superconductor Science and Technology}, 30(3), 2017.

\bibitem{pop2012_fabrication}
I.~M. Pop, T.~Fournier, T.~Crozes, F.~Lecocq, I.~Matei, B.~Pannetier,
  O.~Buisson, and W.~Guichard.
\newblock Fabrication of stable and reproducible submicron tunnel junctions.
\newblock {\em Journal of Vacuum Science $\&$ Technology B, Nanotechnology and
  Microelectronics: Materials, Processing, Measurement, and Phenomena}, 30(1),
  2012.

\bibitem{muller2019_towards}
C.~Muller, J.~H. Cole, and J.~Lisenfeld.
\newblock Towards understanding two-level-systems in amorphous solids: insights
  from quantum circuits.
\newblock {\em Rep Prog Phys}, 82(12):124501, 2019.

\bibitem{ambegaokar1963_tunneling}
Vinay Ambegaokar and Alexis Baratoff.
\newblock Tunneling between superconductors.
\newblock {\em Physical Review Letters}, 10(11):486--489, 1963.

\bibitem{gloos2020_wide}
K.~Gloos, R.~S. Poikolainen, and J.~P. Pekola.
\newblock Wide-range thermometer based on the temperature-dependent conductance
  of planar tunnel junctions.
\newblock {\em Applied Physics Letters}, 77(18):2915--2917, 2000.

\bibitem{kudra2020_high}
M.~Kudra, J.~Biznárová, A.~Fadavi~Roudsari, J.~J. Burnett, D.~Niepce,
  S.~Gasparinetti, B.~Wickman, and P.~Delsing.
\newblock High quality three-dimensional aluminum microwave cavities.
\newblock {\em Applied Physics Letters}, 117(7), 2020.

\bibitem{lei2020_high}
Chan~U. Lei, Lev Krayzman, Suhas Ganjam, Luigi Frunzio, and Robert~J.
  Schoelkopf.
\newblock High coherence superconducting microwave cavities with indium bump
  bonding.
\newblock {\em Applied Physics Letters}, 116(15), 2020.

\bibitem{brecht2017_micromachined}
T.~Brecht, Y.~Chu, C.~Axline, W.~Pfaff, J.~ Z Blumoff, K.~Chou, L.~Krayzman,
  L.~Frunzio, and R.~ J Schoelkopf.
\newblock Micromachined integrated quantum circuit containing a superconducting
  qubit.
\newblock {\em Physical Review Applied}, 7(4), 2017.

\bibitem{kinner2019_engineering}
S.~Krinner, S.~Storz, P.~Kurpiers, P.~Magnard, J.~Heinsoo, R.~Keller,
  J.~Lütolf, C.~Eichler, and A.~Wallraff.
\newblock Engineering cryogenic setups for 100-qubit scale superconducting
  circuit systems.
\newblock {\em EPJ Quantum Technology}, 6(1), 2019.

\bibitem{Santavicca08}
D~F Santavicca and D~E Prober.
\newblock Impedance-matched low-pass stripline filters.
\newblock {\em Meas. Sci. Technol.}, 19:087001, 2008.

\bibitem{Slichter2012_measurement}
D~H Slichter, R~Vijay, S~J Weber, S~Boutin, M~Boissonneault, J~M Gambetta,
  A~Blais, and I~Siddiqi.
\newblock {Measurement-Induced Qubit State Mixing in Circuit QED from
  Up-Converted Dephasing Noise}.
\newblock {\em Physical Review Letters}, 109(15):153601, oct 2012.

\bibitem{Gambetta06}
Jay Gambetta, Alexandre Blais, D.~I. Schuster, A.~Wallraff, L.~Frunzio,
  J.~Majer, M.~H. Devoret, S.~M. Girvin, and R.~J. Schoelkopf.
\newblock Qubit-photon interactions in a cavity: Measurement-induced dephasing
  and number splitting.
\newblock {\em Phys. Rev. A}, 74:042318, Oct 2006.

\bibitem{schoelkopf2003_qubits}
R.J. Schoelkopf, A.A. Clerk, S.M. Girvin, K.W. Lehnert, and M.H. Devoret.
\newblock Qubits as spectrometers of quantum noise.
\newblock {\em Quantum Noise in Mesoscopic Physics, Springer, Dordrecht}, 2003.

\bibitem{wang2019_cavity}
Z.~Wang, S.~Shankar, Z.~K. Minev, P.~Campagne-Ibarcq, A.~Narla, and M.~H.
  Devoret.
\newblock Cavity attenuators for superconducting qubits.
\newblock {\em Physical Review Applied}, 11(1), 2019.

\bibitem{Caves1982_quantum}
Carlton~M. Caves.
\newblock {Quantum limits on noise in linear amplifiers}.
\newblock {\em Physical Review D}, 26(8):1817--1839, 1982.

\bibitem{Clerk10}
A.~A. Clerk, M.~H. Devoret, S.~M. Girvin, Florian Marquardt, and R.~J.
  Schoelkopf.
\newblock {Introduction to quantum noise, measurement, and amplification}.
\newblock {\em Reviews of Modern Physics}, 82(2):1155--1208, 2010.

\bibitem{Macklin15}
C.~Macklin, K.~O{\textquoteright}Brien, D.~Hover, M.~E. Schwartz,
  V.~Bolkhovsky, X.~Zhang, W.~D. Oliver, and I.~Siddiqi.
\newblock A near{\textendash}quantum-limited josephson traveling-wave
  parametric amplifier.
\newblock {\em Science}, 350(6258):307--310, 2015.

\bibitem{Roy2015_broadband}
Tanay Roy, Suman Kundu, Madhavi Chand, A.~M. Vadiraj, A.~Ranadive, N.~Nehra,
  Meghan~P. Patankar, J.~Aumentado, A.~A. Clerk, and R.~Vijay.
\newblock {Broadband parametric amplification with impedance engineering:
  Beyond the gain-bandwidth product}.
\newblock {\em Applied Physics Letters}, 107(26), 2015.

\bibitem{planat2020_photonic}
Luca Planat, Arpit Ranadive, Rémy Dassonneville, Javier Puertas~Martínez,
  Sébastien Léger, Cécile Naud, Olivier Buisson, Wiebke Hasch-Guichard,
  Denis~M. Basko, and Nicolas Roch.
\newblock Photonic-crystal josephson traveling-wave parametric amplifier.
\newblock {\em Physical Review X}, 10(2), 2020.

\bibitem{Kundu2019_multiplexed}
Suman Kundu, Nicolas Gheeraert, Sumeru Hazra, Tanay Roy, Kishor~V. Salunkhe,
  Meghan~P. Patankar, and R.~Vijay.
\newblock {Multiplexed readout of four qubits in 3D circuit QED architecture
  using a broadband Josephson parametric amplifier}.
\newblock {\em Applied Physics Letters}, 114(17), 2019.

\bibitem{aumentado2020_superconducting}
Jose Aumentado.
\newblock Superconducting parametric amplifiers: The state of the art in
  josephson parametric amplifiers.
\newblock {\em IEEE Microwave Magazine}, 21(8):45--59, 2020.

\bibitem{clerk2010_introduction}
A.~A. Clerk, M.~H. Devoret, S.~M. Girvin, Florian Marquardt, and R.~J.
  Schoelkopf.
\newblock Introduction to quantum noise, measurement, and amplification.
\newblock {\em Reviews of Modern Physics}, 82(2):1155--1208, 2010.

\bibitem{vool2014_nonpoissonian}
U.~Vool, I.~M. Pop, K.~Sliwa, B.~Abdo, C.~Wang, T.~Brecht, Y.~Y. Gao,
  S.~Shankar, M.~Hatridge, G.~Catelani, M.~Mirrahimi, L.~Frunzio, R.~J.
  Schoelkopf, L.~I. Glazman, and M.~H. Devoret.
\newblock Non-poissonian quantum jumps of a fluxonium qubit due to
  quasiparticle excitations.
\newblock {\em Physical Review Letters}, 113(24), 2014.

\bibitem{van_Dijk_2019}
J.P.G. van Dijk, E.~Kawakami, R.N. Schouten, M.~Veldhorst, L.M.K. Vandersypen,
  M.~Babaie, E.~Charbon, and F.~Sebastiano.
\newblock Impact of classical control electronics on qubit fidelity.
\newblock {\em Physical Review Applied}, 12(4), Oct 2019.

\bibitem{Chen16b}
Zijun Chen, Julian Kelly, Chris Quintana, R.~Barends, B.~Campbell, Yu~Chen,
  B.~Chiaro, A.~Dunsworth, A.~G. Fowler, E.~Lucero, E.~Jeffrey, A.~Megrant,
  J.~Mutus, M.~Neeley, C.~Neill, P.~J.~J. O'Malley, P.~Roushan, D.~Sank,
  A.~Vainsencher, J.~Wenner, T.~C. White, A.~N. Korotkov, and John~M. Martinis.
\newblock Measuring and suppressing quantum state leakage in a superconducting
  qubit.
\newblock {\em Phys. Rev. Lett.}, 116:020501, 2016.

\bibitem{motzoi2009_simple}
F.~Motzoi, J.~M. Gambetta, P.~Rebentrost, and F.~K. Wilhelm.
\newblock Simple pulses for elimination of leakage in weakly nonlinear qubits.
\newblock {\em Phys Rev Lett}, 103(11):110501, 2009.

\bibitem{Chow10b}
J.~M. Chow, L.~DiCarlo, J.~M. Gambetta, F.~Motzoi, L.~Frunzio, S.~M. Girvin,
  and R.~J. Schoelkopf.
\newblock Optimized driving of superconducting artificial atoms for improved
  single-qubit gates.
\newblock {\em Phys. Rev. A}, 82:040305, 2010.

\bibitem{Krinner19}
S.~Krinner, S.~Storz, P.~Kurpiers, P.~Magnard, J.~Heinsoo, R.~Keller,
  J.~Lütolf, C.~Eichler, and A.~Wallraff.
\newblock Engineering cryogenic setups for 100-qubit scale superconducting
  circuit systems.
\newblock {\em EPJ Quantum Technology}, 6(1), May 2019.

\bibitem{Pozar05}
D.~M. Pozar.
\newblock {\em Microwave Engineering}.
\newblock John Wiley \& Sons, Hoboken, 3 edition, 2005.

\bibitem{jolin2020_calibration}
S.~W. Jolin, R.~Borgani, M.~O. Tholen, D.~Forchheimer, and D.~B. Haviland.
\newblock Calibration of mixer amplitude and phase imbalance in superconducting
  circuits.
\newblock {\em Rev Sci Instrum}, 91(12):124707, 2020.

\bibitem{kelly18physical}
Julian Kelly, Peter O'Malley, Matthew Neeley, Hartmut Neven, and John~M.
  Martinis.
\newblock Physical qubit calibration on a directed acyclic graph.
\newblock {\em ArXiv:1803.03226}, 2018.

\bibitem{Bruno15}
A.~Bruno, G.~de~Lange, S.~Asaad, K.~L. van~der Enden, N.~K. Langford, and
  L.~DiCarlo.
\newblock Reducing intrinsic loss in superconducting resonators by surface
  treatment and deep etching of silicon substrates.
\newblock {\em Appl. Phys. Lett.}, 106:182601, 2015.

\bibitem{Reed10}
M.~D. Reed, L.~DiCarlo, B.~R. Johnson, L.~Sun, D.~I. Schuster, L.~Frunzio, and
  R.~J. Schoelkopf.
\newblock High-fidelity readout in circuit quantum electrodynamics using the
  {J}aynes-{C}ummings nonlinearity.
\newblock {\em Phys. Rev. Lett.}, 105:173601, 2010.

\bibitem{Schuster05}
D.~I. Schuster, A.~Wallraff, A.~Blais, L.~Frunzio, R.-S. Huang, J.~Majer, S.~M.
  Girvin, and R.~J. Schoelkopf.
\newblock ac {S}tark shift and dephasing of a superconducting qubit strongly
  coupled to a cavity field.
\newblock {\em Phys. Rev. Lett.}, 94:123602, 2005.

\bibitem{schuster2007_resolving}
D.~I. Schuster, A.~A. Houck, J.~A. Schreier, A.~Wallraff, J.~M. Gambetta,
  A.~Blais, L.~Frunzio, J.~Majer, B.~Johnson, M.~H. Devoret, S.~M. Girvin, and
  R.~J. Schoelkopf.
\newblock Resolving photon number states in a superconducting circuit.
\newblock {\em Nature}, 445(7127):515--8, 2007.

\bibitem{Rabi36}
II~Rabi.
\newblock On the process of space quantization.
\newblock {\em Phys. Rev.}, 49(4):324, 1936.

\bibitem{Wallraff05}
A.~Wallraff, D.~I. Schuster, A.~Blais, L.~Frunzio, J.~Majer, M.~H. Devoret,
  S.~M. Girvin, and R.~J. Schoelkopf.
\newblock Approaching unit visibility for control of a superconducting qubit
  with dispersive readout.
\newblock {\em Phys. Rev. Lett.}, 95:060501, 2005.

\bibitem{Gambetta11}
J.~M. Gambetta, F.~Motzoi, S.~T. Merkel, and F.~K. Wilhelm.
\newblock Analytic control methods for high-fidelity unitary operations in a
  weakly nonlinear oscillator.
\newblock {\em Phys. Rev. A}, 83:012308, 2011.

\bibitem{Vandersypen05}
L.~M.~K. Vandersypen and I.~L. Chuang.
\newblock {NMR} techniques for quantum control and computation.
\newblock {\em Rev. Mod. Phys.}, 76:1037--1069, Jan 2005.

\bibitem{rosenblum2018_fault-tolerant}
S.~Rosenblum, P.~Reinhold, M.~Mirrahimi, L.~Jiang, L.~Frunzio, and R.~J.
  Schoelkopf.
\newblock Fault-tolerant detection of a quantum error.
\newblock {\em Science}, 361(6399):266--270, 2018.

\bibitem{Martinis03}
John~M. Martinis, S.~Nam, J.~Aumentado, K.~M. Lang, and C.~Urbina.
\newblock Decoherence of a superconducting qubit due to bias noise.
\newblock {\em Phys. Rev. B}, 67(9):094510, 2003.

\bibitem{ReedPhD13}
M.~Reed.
\newblock {\em Entanglement and quantum error correction with superconducting
  qubits}.
\newblock Ph{D} {D}issertation, Yale University, 2013.

\bibitem{Asaad16}
S.~Asaad, C.~Dickel, S.~Poletto, A.~Bruno, N.~K. Langford, M.~A. Rol,
  D.~Deurloo, and L.~DiCarlo.
\newblock Independent, extensible control of same-frequency superconducting
  qubits by selective broadcasting.
\newblock {\em npj Quantum Inf.}, 2:16029, 2016.

\bibitem{Fu17}
X.~Fu, M.~A. Rol, C.~C. Bultink, J.~van Someren, N.~Khammassi, I.~Ashraf,
  R.~F.~L. Vermeulen, J.~C. de~Sterke, W.~J. Vlothuizen, R.~N. Schouten, C.~G.
  Almudever, L.~DiCarlo, and K.~Bertels.
\newblock An experimental microarchitecture for a superconducting quantum
  processor.
\newblock In {\em Proceedings of the 50th Annual IEEE/ACM International
  Symposium on Microarchitecture}, MICRO-50 '17, pages 813--825, New York, NY,
  USA, 2017. ACM.

\bibitem{Bultink16}
C.~C. Bultink, M.~A. Rol, T.~E. O'Brien, X.~Fu, B.~C.~S. Dikken, C.~Dickel,
  R.~F.~L. Vermeulen, J.~C. de~Sterke, A.~Bruno, R.~N. Schouten, and
  L.~DiCarlo.
\newblock Active resonator reset in the nonlinear dispersive regime of circuit
  {Q}{E}{D}.
\newblock {\em Phys. Rev. Appl.}, 6:034008, 2016.

\bibitem{Fu19}
X.~Fu, L.~Riesebos, M.~A. Rol, J.~van Straten, J.~van Someren, N.~Khammassi,
  I.~Ashraf, R.~F.~L. Vermeulen, V.~Newsum, K.~K.~L. Loh, J.~C. de~Sterke,
  W.~J. Vlothuizen, R.~N. Schouten, C.~G. Almudever, L.~DiCarlo, and
  K.~Bertels.
\newblock {eQASM}: An executable quantum instruction set architecture.
\newblock In {\em Proceedings of 25th IEEE International Symposium on
  High-Performance Computer Architecture (HPCA)}, pages 224--237. IEEE, 2019.

\bibitem{Boissonneault2009_dispersive}
Maxime Boissonneault, J~M Gambetta, and Alexandre Blais.
\newblock {Dispersive regime of circuit QED: Photon-dependent qubit dephasing
  and relaxation rates}.
\newblock {\em Physical Review A}, 79(1):13819, 2009.

\bibitem{Sank16}
Daniel Sank, Zijun Chen, Mostafa Khezri, J.~Kelly, R.~Barends, B.~Campbell,
  Y.~Chen, B.~Chiaro, A.~Dunsworth, A.~Fowler, E.~Jeffrey, E.~Lucero,
  A.~Megrant, J.~Mutus, M.~Neeley, C.~Neill, P.~J.~J. O'Malley, C.~Quintana,
  P.~Roushan, A.~Vainsencher, T.~White, J.~Wenner, Alexander~N. Korotkov, and
  John~M. Martinis.
\newblock Measurement-induced state transitions in a superconducting qubit:
  Beyond the rotating wave approximation.
\newblock {\em Phys. Rev. Lett.}, 117:190503, Nov 2016.

\bibitem{Gambetta08}
Jay Gambetta, Alexandre Blais, M.~Boissonneault, A.~A. Houck, D.~I. Schuster,
  and S.~M. Girvin.
\newblock {Quantum trajectory approach to circuit QED: Quantum jumps and the
  Zeno effect}.
\newblock {\em Phys. Rev. A}, 77(1):012112, January 2008.

\bibitem{Jeffrey14}
Evan Jeffrey, Daniel Sank, J.~Y. Mutus, T.~C. White, J.~Kelly, R.~Barends,
  Y.~Chen, Z.~Chen, B.~Chiaro, A.~Dunsworth, A.~Megrant, P.~J.~J. O'Malley,
  C.~Neill, P.~Roushan, A.~Vainsencher, J.~Wenner, A.~N. Cleland, and John~M.
  Martinis.
\newblock Fast accurate state measurement with superconducting qubits.
\newblock {\em Phys. Rev. Lett.}, 112:190504, May 2014.

\bibitem{didier2015_fast}
N.~Didier, J.~Bourassa, and A.~Blais.
\newblock Fast quantum nondemolition readout by parametric modulation of
  longitudinal qubit-oscillator interaction.
\newblock {\em Phys Rev Lett}, 115(20):203601, 2015.

\bibitem{Bultink18}
C~C Bultink, B~Tarasinski, N~Haandbaek, S~Poletto, N~Haider, D~J Michalak,
  A~Bruno, and L~DiCarlo.
\newblock General method for extracting the quantum efficiency of dispersive
  qubit readout in circuit qed.
\newblock {\em Appl. Phys. Lett.}, 112:092601, 2018.

\bibitem{McClure16}
D.~T. McClure, Hanhee Paik, L.~S. Bishop, M.~Steffen, Jerry~M. Chow, and Jay~M.
  Gambetta.
\newblock Rapid driven reset of a qubit readout resonator.
\newblock {\em Phys. Rev. Appl.}, 5:011001, 2016.

\bibitem{Ryan15}
Colm~A. Ryan, Blake~R. Johnson, Jay~M. Gambetta, Jerry~M. Chow, Marcus~P.
  da~Silva, Oliver~E. Dial, and Thomas~A. Ohki.
\newblock Tomography via correlation of noisy measurement records.
\newblock {\em Phys. Rev. A}, 91:022118, 2015.

\bibitem{Hatridge2013_weak}
M~Hatridge, S~Shankar, M~Mirrahimi, F~Schackert, K~Geerlings, T~Brecht, K~M
  Sliwa, B~Abdo, L~Frunzio, S~M Girvin, R~J Schoelkopf, and M~H Devoret.
\newblock {Quantum Back-Action of an Individual Variable-Strength Measurement}.
\newblock {\em Science}, 339(6116):178--181, 2013.

\bibitem{Hacohen-Gourgy2016}
Shay Hacohen-Gourgy, Leigh~S Martin, Emmanuel Flurin, Vinay~V Ramasesh,
  K~Birgitta Whaley, and Irfan Siddiqi.
\newblock {Quantum dynamics of simultaneously measured non-commuting
  observables}.
\newblock {\em Nature}, 538:491, 2016.

\bibitem{Eddins2018}
Andrew Eddins, Sydney Schreppler, D.~M. Toyli, L.~S. Martin, Shay
  Hacohen-Gourgy, L.~C.~G. Govia, Hugo Ribeiro, A.~A. Clerk, and Irfan Siddiqi.
\newblock {Stroboscopic Qubit Measurement with Squeezed Illumination}.
\newblock {\em Physical Review Letters}, 120(4):040505, jan 2018.

\bibitem{Strauch03}
Frederick~W. Strauch, Philip~R. Johnson, Alex~J. Dragt, C.~J. Lobb, J.~R.
  Anderson, and F.~C. Wellstood.
\newblock Quantum logic gates for coupled superconducting phase qubits.
\newblock {\em Phys. Rev. Lett.}, 91:167005, 2003.

\bibitem{Neeley10}
Matthew Neeley, Radoslaw~C Bialczak, M~Lenander, E~Lucero, Matteo Mariantoni,
  A~D O'Connell, D~Sank, H~Wang, M~Weides, J~Wenner, Y~Yin, T~Yamamoto, A~N
  Cleland, and John~M Martinis.
\newblock {Generation of three-qubit entangled states using superconducting
  phase qubits.}
\newblock {\em Nature}, 467(7315):570, September 2010.

\bibitem{barends2014_logic}
R.~Barends, J.~Kelly, A.~Megrant, A.~Veitia, D.~Sank, E.~Jeffrey, T.~C. White,
  J.~Mutus, A.~G. Fowler, B.~Campbell, Y.~Chen, Z.~Chen, B.~Chiaro,
  A.~Dunsworth, C.~Neill, P.~O'Malley, P.~Roushan, A.~Vainsencher, J.~Wenner,
  A.~N. Korotkov, A.~N. Cleland, and J.~M. Martinis.
\newblock Superconducting quantum circuits at the surface code threshold for
  fault tolerance.
\newblock {\em Nature}, 508(7497):500--3, 2014.

\bibitem{Chen14}
Yu~Chen, C.~Neill, P.~Roushan, N.~Leung, M.~Fang, R.~Barends, J.~Kelly,
  B.~Campbell, Z.~Chen, B.~Chiaro, A.~Dunsworth, E.~Jeffrey, A.~Megrant, J.~Y.
  Mutus, P.~J.~J. O'Malley, C.~M. Quintana, D.~Sank, A.~Vainsencher, J.~Wenner,
  T.~C. White, Michael~R. Geller, A.~N. Cleland, and John~M. Martinis.
\newblock Qubit architecture with high coherence and fast tunable coupling.
\newblock {\em Phys. Rev. Lett.}, 113:220502, Nov 2014.

\bibitem{sung2020_realization}
Youngkyu {Sung}, Leon {Ding}, Jochen {Braum{\"u}ller}, Antti
  {Veps{\"a}l{\"a}inen}, Bharath {Kannan}, Morten {Kjaergaard}, Ami {Greene},
  Gabriel~O. {Samach}, Chris {McNally}, David {Kim}, Alexander {Melville},
  Bethany~M. {Niedzielski}, Mollie~E. {Schwartz}, Jonilyn~L. {Yoder}, Terry~P.
  {Orlando}, Simon {Gustavsson}, and William~D. {Oliver}.
\newblock {Realization of high-fidelity CZ and ZZ-free iSWAP gates with a
  tunable coupler}.
\newblock {\em arXiv e-prints}, page arXiv:2011.01261, November 2020.

\bibitem{Rigetti10}
Chad Rigetti and Michel Devoret.
\newblock Fully microwave-tunable universal gates in superconducting qubits
  with linear couplings and fixed transition frequencies.
\newblock {\em Phys. Rev. B}, 81:134507, Apr 2010.

\bibitem{Poletto12}
S.~Poletto, Jay~M. Gambetta, Seth~T. Merkel, John~A. Smolin, Jerry~M. Chow,
  A.~D. C\'orcoles, George~A. Keefe, Mary~B. Rothwell, J.~R. Rozen, D.~W.
  Abraham, Chad Rigetti, and M.~Steffen.
\newblock Entanglement of two superconducting qubits in a waveguide cavity via
  monochromatic two-photon excitation.
\newblock {\em Phys. Rev. Lett.}, 109:240505, Dec 2012.

\bibitem{chow2013_microwave}
Jerry~M. Chow, Jay~M. Gambetta, Andrew~W. Cross, Seth~T. Merkel, Chad Rigetti,
  and M.~Steffen.
\newblock Microwave-activated conditional-phase gate for superconducting
  qubits.
\newblock {\em New Journal of Physics}, 15(11), 2013.

\bibitem{sheldon2016_procedure}
Sarah Sheldon, Easwar Magesan, Jerry~M. Chow, and Jay~M. Gambetta.
\newblock Procedure for systematically tuning up cross-talk in the
  cross-resonance gate.
\newblock {\em Physical Review A}, 93(6), 2016.

\bibitem{Schutjens13}
R.~Schutjens, F.~Abu Dagga, D.~J. Egger, and F.~K. Wilhelm.
\newblock Single-qubit gates in frequency-crowded transmon systems.
\newblock {\em Phys. Rev. A}, 88:052330, Nov 2013.

\bibitem{Brink18}
M.~{Brink}, J.~M. {Chow}, J.~{Hertzberg}, E.~{Magesan}, and S.~{Rosenblatt}.
\newblock Device challenges for near term superconducting quantum processors:
  frequency collisions.
\newblock In {\em 2018 IEEE International Electron Devices Meeting (IEDM)},
  pages 6.1.1--6.1.3, Dec 2018.

\bibitem{Caldwell18}
S.~A. Caldwell, N.~Didier, C.~A. Ryan, E.~A. Sete, A.~Hudson, P.~Karalekas,
  R.~Manenti, M.~P. da~Silva, R.~Sinclair, E.~Acala, N.~Alidoust, J.~Angeles,
  A.~Bestwick, M.~Block, B.~Bloom, A.~Bradley, C.~Bui, L.~Capelluto,
  R.~Chilcott, J.~Cordova, G.~Crossman, M.~Curtis, S.~Deshpande, T.~El
  Bouayadi, D.~Girshovich, S.~Hong, K.~Kuang, M.~Lenihan, T.~Manning,
  A.~Marchenkov, J.~Marshall, R.~Maydra, Y.~Mohan, W.~O'Brien, C.~Osborn,
  J.~Otterbach, A.~Papageorge, J.-P. Paquette, M.~Pelstring, A.~Polloreno,
  G.~Prawiroatmodjo, V.~Rawat, M.~Reagor, R.~Renzas, N.~Rubin, D.~Russell,
  M.~Rust, D.~Scarabelli, M.~Scheer, M.~Selvanayagam, R.~Smith, A.~Staley,
  M.~Suska, N.~Tezak, D.~C. Thompson, T.-W. To, M.~Vahidpour, N.~Vodrahalli,
  T.~Whyland, K.~Yadav, W.~Zeng, and C.~Rigetti.
\newblock Parametrically activated entangling gates using transmon qubits.
\newblock {\em Phys. Rev. Applied}, 10:034050, Sep 2018.

\bibitem{Hong19}
Sabrina~S. Hong, Alexander~T. Papageorge, Prasahnt Sivarajah, Genya Crossman,
  Nicolas Didier, Anthony~M. Polloreno, Eyob~A. Sete, Stefan~W. Turkowski,
  Marcus~P. da~Silva, and Blake~R. Johnson.
\newblock Demonstration of a parametrically activated entangling gate protected
  from flux noise.
\newblock {\em Phys. Rev. A}, 101:012302, Jan 2020.

\bibitem{Versluis17}
R.~Versluis, S.~Poletto, N.~Khammassi, B.~M. Tarasinski, N.~Haider, D.~J.
  Michalak, A.~Bruno, K.~Bertels, and L.~DiCarlo.
\newblock Scalable quantum circuit and control for a superconducting surface
  code.
\newblock {\em Phys. Rev. Appl.}, 8:034021, Sep 2017.

\bibitem{kafri2017_tunable}
Dvir Kafri, Chris Quintana, Yu~Chen, Alireza Shabani, John~M. Martinis, and
  Hartmut Neven.
\newblock Tunable inductive coupling of superconducting qubits in the strongly
  nonlinear regime.
\newblock {\em Physical Review A}, 95(5), 2017.

\bibitem{yan2018_tunable}
Fei Yan, Philip Krantz, Youngkyu Sung, Morten Kjaergaard, Daniel~L. Campbell,
  Terry~P. Orlando, Simon Gustavsson, and William~D. Oliver.
\newblock Tunable coupling scheme for implementing high-fidelity two-qubit
  gates.
\newblock {\em Physical Review Applied}, 10(5), 2018.

\bibitem{Nielsen_2020}
Erik Nielsen, Kenneth Rudinger, Timothy Proctor, Antonio Russo, Kevin Young,
  and Robin Blume-Kohout.
\newblock Probing quantum processor performance with pygsti.
\newblock {\em Quantum Science and Technology}, 5(4):044002, Jul 2020.

\bibitem{chow2012_universal}
J.~M. Chow, J.~M. Gambetta, A.~D. Corcoles, S.~T. Merkel, J.~A. Smolin,
  C.~Rigetti, S.~Poletto, G.~A. Keefe, M.~B. Rothwell, J.~R. Rozen, M.~B.
  Ketchen, and M.~Steffen.
\newblock Universal quantum gate set approaching fault-tolerant thresholds with
  superconducting qubits.
\newblock {\em Phys Rev Lett}, 109(6):060501, 2012.

\bibitem{Kitaev02}
A.Y. Kitaev, A.H. Shen, and M.N. Vyalyi.
\newblock {\em Classical and Quantum Computation}.
\newblock American Mathematical Society, 2002.

\bibitem{greenbaum2015_introduction}
Daniel {Greenbaum}.
\newblock {Introduction to Quantum Gate Set Tomography}.
\newblock arXiv:1509.02921, 2015.

\bibitem{Kirchhoff18}
Susanna {Kirchhoff}, Torsten {Ke{\ss}ler}, Per~J. {Liebermann}, Elie
  {Ass{\'e}mat}, Shai {Machnes}, Felix {Motzoi}, and Frank~K. {Wilhelm}.
\newblock {Optimized cross-resonance gate for coupled transmon systems}.
\newblock {\em Physical Review A}, 97:042348, Apr 2018.

\bibitem{Sanders15}
Yuval~R Sanders, Joel~J Wallman, and Barry~C Sanders.
\newblock Bounding quantum gate error rate based on reported average fidelity.
\newblock {\em New Journal of Physics}, 18(1):012002, dec 2015.

\bibitem{Wallman15}
Joel Wallman, Chris Granade, Robin Harper, and Steven~T Flammia.
\newblock Estimating the coherence of noise.
\newblock {\em New Journal of Physics}, 17(11):113020, nov 2015.

\bibitem{Feng16}
Guanru Feng, Joel~J. Wallman, Brandon Buonacorsi, Franklin~H. Cho, Daniel~K.
  Park, Tao Xin, Dawei Lu, Jonathan Baugh, and Raymond Laflamme.
\newblock Estimating the coherence of noise in quantum control of a solid-state
  qubit.
\newblock {\em Phys. Rev. Lett.}, 117(26), Dec 2016.

\bibitem{Dirkse19}
Bas Dirkse, Jonas Helsen, and Stephanie Wehner.
\newblock Efficient unitarity randomized benchmarking of few-qubit clifford
  gates.
\newblock {\em Phys. Rev. A}, 99:012315, Jan 2019.

\bibitem{Blume13}
Robin Blume-Kohout, John~King Gamble, Erik Nielsen, Jonathan Mizrahi,
  Jonathan~D. Sterk, and Peter Maunz.
\newblock Robust, self-consistent, closed-form tomography of quantum logic
  gates on a trapped ion qubit.
\newblock arXiv:1310.4492, 2013.

\bibitem{Blume-Kohout17}
Robin Blume-Kohout, John~King Gamble, Erik Nielsen, Kenneth Rudinger, Jonathan
  Mizrahi, Kevin Fortier, and Peter Maunz.
\newblock Demonstration of qubit operations below a rigorous fault tolerance
  threshold with gate set tomography.
\newblock {\em Nat.\ Commun.}, (14485), 2017.

\bibitem{Magesan11}
Easwar Magesan, J.~M. Gambetta, and Joseph Emerson.
\newblock Scalable and robust randomized benchmarking of quantum processes.
\newblock {\em Phys. Rev. Lett.}, 106:180504, 2011.

\bibitem{Magesan12}
Easwar Magesan, Jay~M. Gambetta, and Joseph Emerson.
\newblock Characterizing quantum gates via randomized benchmarking.
\newblock {\em Phys. Rev. A}, 85:042311, 2012.

\bibitem{Magesan12b}
Easwar Magesan, Jay~M. Gambetta, B.~R. Johnson, Colm~A. Ryan, Jerry~M. Chow,
  Seth~T. Merkel, Marcus~P. da~Silva, George~A. Keefe, Mary~B. Rothwell,
  Thomas~A. Ohki, Mark~B. Ketchen, and M.~Steffen.
\newblock Efficient measurement of quantum gate error by interleaved randomized
  benchmarking.
\newblock {\em Phys. Rev. Lett.}, 109:080505, 2012.

\bibitem{kimmel2014_robust}
Shelby Kimmel, Marcus~P. da~Silva, Colm~A. Ryan, Blake~R. Johnson, and Thomas
  Ohki.
\newblock Robust extraction of tomographic information via randomized
  benchmarking.
\newblock {\em Physical Review X}, 4(1), 2014.

\bibitem{Boixo18}
Sergio Boixo, Sergei~V. Isakov, Vadim~N. Smelyanskiy, Ryan Babbush, Nan Ding,
  Zhang Jiang, Michael~J. Bremner, John~M. Martinis, and Hartmut Neven.
\newblock Characterizing quantum supremacy in near-term devices.
\newblock {\em Nat.\ Phys.}, 14(6):595--600, 2018.

\bibitem{sarovar19}
Mohan Sarovar, Timothy Proctor, Kenneth Rudinger, Kevin Young, Erik Nielsen,
  and Robin Blume-Kohout.
\newblock Detecting crosstalk errors in quantum information processors.
\newblock {\em ArXiv:1908.09855}, 2019.

\bibitem{Helsen19_characterbenchmarking}
Jonas Helsen, Xiao Xue, Lieven M.~K. Vandersypen, and Stephanie Wehner.
\newblock A new class of efficient randomized benchmarking protocols.
\newblock {\em npj Quantum Information}, 5(1):71, 2019.

\bibitem{Xue18}
X.~Xue, T.~F. Watson, J.~Helsen, D.~R. Ward, D.~E. Savage, M.~G. Lagally, S.~N.
  Coppersmith, M.~A. Eriksson, S.~Wehner, and L.~M.~K. Vandersypen.
\newblock Benchmarking gate fidelities in a si/sige two-qubit device.
\newblock arXiv:1811.04002, 2018.

\bibitem{Gambetta12}
Jay~M. Gambetta, A.~D. C\'orcoles, S.~T. Merkel, B.~R. Johnson, John~A. Smolin,
  Jerry~M. Chow, Colm~A. Ryan, Chad Rigetti, S.~Poletto, Thomas~A. Ohki,
  Mark~B. Ketchen, and M.~Steffen.
\newblock Characterization of addressability by simultaneous randomized
  benchmarking.
\newblock {\em Phys. Rev. Lett.}, 109:240504, Dec 2012.

\bibitem{Wallman16}
Joel~J Wallman, Marie Barnhill, and Joseph Emerson.
\newblock Robust characterization of leakage errors.
\newblock {\em New Journal of Physics}, 18(4):043021, apr 2016.

\bibitem{Wood18}
Christopher~J. Wood and Jay~M. Gambetta.
\newblock Quantification and characterization of leakage errors.
\newblock {\em Phys. Rev. A}, 97:032306, Mar 2018.

\bibitem{proctor19}
Timothy Proctor, Melissa Revelle, Erik Nielsen, Kenneth Rudinger, Daniel
  Lobser, Peter Maunz, Robin Blume-Kohout, and Kevin Young.
\newblock Detecting, tracking, and eliminating drift in quantum information
  processors.
\newblock {\em ArXiv:1907.13608}, 2019.

\bibitem{Nielsen16}
Erik Nielsen, Travis Scholten, Kenneth Rudinger, and Jonathan Gross.
\newblock py{GST}i: Version 0.9.1 beta, June 2016.

\bibitem{Rol20_thesis}
M.A. Rol.
\newblock {\em Control for programmable superconducting quantum systems}.
\newblock Ph{D} {D}issertation, Delft University of Technology, 2020.

\bibitem{heeres2017_implementing}
R.~W. Heeres, P.~Reinhold, N.~Ofek, L.~Frunzio, L.~Jiang, M.~H. Devoret, and
  R.~J. Schoelkopf.
\newblock Implementing a universal gate set on a logical qubit encoded in an
  oscillator.
\newblock {\em Nat Commun}, 8(1):94, 2017.

\bibitem{kirchmair2013_observation}
G.~Kirchmair, B.~Vlastakis, Z.~Leghtas, S.~E. Nigg, H.~Paik, E.~Ginossar,
  M.~Mirrahimi, L.~Frunzio, S.~M. Girvin, and R.~J. Schoelkopf.
\newblock Observation of quantum state collapse and revival due to the
  single-photon kerr effect.
\newblock {\em Nature}, 495(7440):205--9, 2013.

\bibitem{vlastakis2013_deterministically_supp}
B.~Vlastakis, G.~Kirchmair, Z.~Leghtas, S.~E. Nigg, L.~Frunzio, S.~M. Girvin,
  M.~Mirrahimi, M.~H. Devoret, and R.~J. Schoelkopf.
\newblock Supplementary materials: Deterministically encoding quantum
  information using 100-photon schrodinger cat states.
\newblock {\em Science}, 342(6158):607--10, 2013.

\bibitem{krastanov2015_universal}
Stefan Krastanov, Victor~V. Albert, Chao Shen, Chang-Ling Zou, Reinier~W.
  Heeres, Brian Vlastakis, Robert~J. Schoelkopf, and Liang Jiang.
\newblock Universal control of an oscillator with dispersive coupling to a
  qubit.
\newblock {\em Physical Review A}, 92(4), 2015.

\bibitem{heeres2015_cavity}
R.~W. Heeres, B.~Vlastakis, E.~Holland, S.~Krastanov, V.~V. Albert, L.~Frunzio,
  L.~Jiang, and R.~J. Schoelkopf.
\newblock Cavity state manipulation using photon-number selective phase gates.
\newblock {\em Phys Rev Lett}, 115(13):137002, 2015.

\bibitem{reagor2015_thesis}
M.~Reagor.
\newblock {\em Superconducting Cavities for Circuit Quantum Electrodynamics}.
\newblock Ph{D} {D}issertation, Yale University, 2015.

\bibitem{sun2014_tracking}
L.~Sun, A.~Petrenko, Z.~Leghtas, B.~Vlastakis, G.~Kirchmair, K.~M. Sliwa,
  A.~Narla, M.~Hatridge, S.~Shankar, J.~Blumoff, L.~Frunzio, M.~Mirrahimi,
  M.~H. Devoret, and R.~J. Schoelkopf.
\newblock Tracking photon jumps with repeated quantum non-demolition parity
  measurements.
\newblock {\em Nature}, 511(7510):444--8, 2014.

\bibitem{reinhold2020_error}
Philip Reinhold, Serge Rosenblum, Wen-Long Ma, Luigi Frunzio, Liang Jiang, and
  Robert~J. Schoelkopf.
\newblock Error-corrected gates on an encoded qubit.
\newblock {\em Nature Physics}, 2020.

\bibitem{ma2020_error}
Y.~Ma, Y.~Xu, X.~Mu, W.~Cai, L.~Hu, W.~Wang, X.~Pan, H.~Wang, Y.~P. Song, C.~L.
  Zou, and L.~Sun.
\newblock Error-transparent operations on a logical qubit protected by quantum
  error correction.
\newblock {\em Nature Physics}, 2020.

\bibitem{bertet2002_direct}
P.~Bertet, A.~Auffeves, P.~Maioli, S.~Osnaghi, T.~Meunier, M.~Brune, J.~M.
  Raimond, and S.~Haroche.
\newblock Direct measurement of the wigner function of a one-photon fock state
  in a cavity.
\newblock {\em Physical Review Letters}, 89(20), 2002.

\bibitem{takita2016_weight}
M.~Takita, A.~D. Corcoles, E.~Magesan, B.~Abdo, M.~Brink, A.~Cross, J.~M. Chow,
  and J.~M. Gambetta.
\newblock Demonstration of weight-four parity measurements in the surface code
  architecture.
\newblock {\em Phys Rev Lett}, 117(21):210505, 2016.

\bibitem{gong2021_quantum}
M.~Gong, S.~Wang, C.~Zha, M.~C. Chen, H.~L. Huang, Y.~Wu, Q.~Zhu, Y.~Zhao,
  S.~Li, S.~Guo, H.~Qian, Y.~Ye, F.~Chen, C.~Ying, J.~Yu, D.~Fan, D.~Wu, H.~Su,
  H.~Deng, H.~Rong, K.~Zhang, S.~Cao, J.~Lin, Y.~Xu, L.~Sun, C.~Guo, N.~Li,
  F.~Liang, V.~M. Bastidas, K.~Nemoto, W.~J. Munro, Y.~H. Huo, C.~Y. Lu, C.~Z.
  Peng, X.~Zhu, and J.~W. Pan.
\newblock Quantum walks on a programmable two-dimensional 62-qubit
  superconducting processor.
\newblock {\em Science}, 2021.

\bibitem{Moll18}
Nikolaj Moll, Panagiotis Barkoutsos, Lev~S Bishop, Jerry~M Chow, Andrew Cross,
  Daniel~J Egger, Stefan Filipp, Andreas Fuhrer, Jay~M Gambetta, and Marc
  Ganzhorn.
\newblock Quantum optimization using variational algorithms on near-term
  quantum devices.
\newblock {\em Quantum Science and Technology}, 3(3):030503, 2018.

\bibitem{Cross19}
Andrew~W. Cross, Lev~S. Bishop, Sarah Sheldon, Paul~D. Nation, and Jay~M.
  Gambetta.
\newblock Validating quantum computers using randomized model circuits.
\newblock {\em Phys. Rev. A}, 100:032328, Sep 2019.

\bibitem{Terhal02}
Barbara~M. Terhal and David~P. DiVincenzo.
\newblock Adaptive quantum computation, constant depth quantum circuits and
  arthur-merlin games.
\newblock {\em ArXiv:0205133}, 2002.

\bibitem{Farhi16}
Edward Farhi and Aram~W Harrow.
\newblock Quantum supremacy through the quantum approximate optimization
  algorithm.
\newblock {\em arXiv:1602.07674}, 2016.

\bibitem{lilly2020_modeling}
Megan~N. Lilly and Travis~S. Humble.
\newblock Modeling noisy quantum circuits using experimental characterization.
\newblock arXiv:2001.08653, 2020.

\bibitem{gupta2020_adaptive}
Riddhi~Swaroop Gupta, Claire~L. Edmunds, Alistair~R. Milne, Cornelius Hempel,
  and Michael~J. Biercuk.
\newblock Adaptive characterization of spatially inhomogeneous fields and
  errors in qubit registers.
\newblock {\em npj Quantum Information}, 6(1), 2020.

\bibitem{jiang2007_distributed}
Liang Jiang, Jacob~M. Taylor, Anders~S. Sørensen, and Mikhail~D. Lukin.
\newblock Distributed quantum computation based on small quantum registers.
\newblock {\em Physical Review A}, 76(6), 2007.

\bibitem{rosenberg17}
D.~Rosenberg, D.~Kim, R.~Das, D.~Yost, S.~Gustavsson, D.~Hover, P.~Krantz,
  A.~Melville, L.~Racz, G.~O. Samach, S.~J. Weber, F.~Yan, J.~L. Yoder, A.~J.
  Kerman, and W.~D. Oliver.
\newblock {3D integrated superconducting qubits}.
\newblock {\em npj Quantum Information}, 3(1):1--4, 2017.

\bibitem{Bosman19_MM}
Sal Bosman, Daan Kuitenbrouwer, Wouter Bos, Kiefer Vermeulen, Kelvin Lindeborg,
  Riemer Sorgedrager, Vivien Thiney, and Jakob Kammhuber.
\newblock V26.00012 : Scaling the input/output architecture of quantum
  processors to kqbit, and beyond, size in the nisq era.
\newblock APS March Meeting 2019.

\bibitem{puri2020_bias}
S.~Puri, L.~St-Jean, J.~A. Gross, A.~Grimm, N.~E. Frattini, P.~S. Iyer,
  A.~Krishna, S.~Touzard, L.~Jiang, A.~Blais, S.~T. Flammia, and S.~M. Girvin.
\newblock Bias-preserving gates with stabilized cat qubits.
\newblock {\em Sci Adv}, 6(34), 2020.

\bibitem{michael2016_new}
Marios~H. Michael, Matti Silveri, R.~ T Brierley, Victor~V. Albert, Juha
  Salmilehto, Liang Jiang, and S.~ M Girvin.
\newblock New class of quantum error-correcting codes for a bosonic mode.
\newblock {\em Physical Review X}, 6(3), 2016.

\bibitem{terhal2020_towards}
B.~M. Terhal, J.~Conrad, and C.~Vuillot.
\newblock Towards scalable bosonic quantum error correction.
\newblock {\em Quantum Science and Technology}, 5(4), 2020.

\bibitem{cai2021_bosonic}
Weizhou Cai, Yuwei Ma, Weiting Wang, Chang-Ling Zou, and Luyan Sun.
\newblock Bosonic quantum error correction codes in superconducting quantum
  circuits.
\newblock {\em Fundamental Research}, 1(1):50--67, 2021.

\bibitem{monroe2014_large-scale}
C.~Monroe, R.~Raussendorf, A.~Ruthven, K.~R. Brown, P.~Maunz, L.~M. Duan, and
  J.~Kim.
\newblock Large-scale modular quantum-computer architecture with atomic memory
  and photonic interconnects.
\newblock {\em Physical Review A}, 89(2), 2014.

\bibitem{northup2014_quantum}
T.~E. Northup and R.~Blatt.
\newblock Quantum information transfer using photons.
\newblock {\em Nature Photonics}, 8(5):356--363, 2014.

\bibitem{kimble2001_conversion}
H.~J. Kimble, Yuri Levin, Andrey~B. Matsko, Kip~S. Thorne, and Sergey~P.
  Vyatchanin.
\newblock Conversion of conventional gravitational-wave interferometers into
  quantum nondemolition interferometers by modifying their input and/or output
  optics.
\newblock {\em Physical Review D}, 65(2), 2001.

\bibitem{place2020_new}
Alex P.~M. {Place}, Lila V.~H. {Rodgers}, Pranav {Mundada}, Basil~M. {Smitham},
  Mattias {Fitzpatrick}, Zhaoqi {Leng}, Anjali {Premkumar}, Jacob {Bryon}, Sara
  {Sussman}, Guangming {Cheng}, Trisha {Madhavan}, Harshvardhan~K. {Babla},
  Berthold {Jaeck}, Andras {Gyenis}, Nan {Yao}, Robert~J. {Cava}, Nathalie~P.
  {de Leon}, and Andrew~A. {Houck}.
\newblock {New material platform for superconducting transmon qubits with
  coherence times exceeding 0.3 milliseconds}.
\newblock arXiv:2003.00024, 2020.

\bibitem{romanenko2020_three}
A.~Romanenko, R.~Pilipenko, S.~Zorzetti, D.~Frolov, M.~Awida, S.~Belomestnykh,
  S.~Posen, and A.~Grassellino.
\newblock Three-dimensional superconducting resonators at t<20 mk with photon
  lifetimes up to $\tau=2$\,s.
\newblock {\em Physical Review Applied}, 13(3), 2020.

\bibitem{gertler2020_protecting}
Jeffrey~M. {Gertler}, Brian {Baker}, Juliang {Li}, Shruti {Shirol}, Jens
  {Koch}, and Chen {Wang}.
\newblock {Protecting a Bosonic Qubit with Autonomous Quantum Error
  Correction}.
\newblock arXiv:2004.09322, 2020.

\bibitem{xu2020_demonstration}
Y.~Xu, Y.~Ma, W.~Cai, X.~Mu, W.~Dai, W.~Wang, L.~Hu, X.~Li, J.~Han, H.~Wang,
  Y.~P. Song, Z.~B. Yang, S.~B. Zheng, and L.~Sun.
\newblock Demonstration of controlled-phase gates between two error-correctable
  photonic qubits.
\newblock {\em Phys Rev Lett}, 124(12):120501, 2020.

\bibitem{zhang2019_engineering}
Yaxing Zhang, Brian~J. Lester, Yvonne~Y. Gao, Liang Jiang, R.~J. Schoelkopf,
  and S.~M. Girvin.
\newblock Engineering bilinear mode coupling in circuit qed: Theory and
  experiment.
\newblock {\em Physical Review A}, 99(1), 2019.

\bibitem{petrescu2020_lifetime}
Alexandru Petrescu, Moein Malekakhlagh, and Hakan~E. Türeci.
\newblock Lifetime renormalization of driven weakly anharmonic superconducting
  qubits. ii. the readout problem.
\newblock {\em Physical Review B}, 101(13), 2020.

\bibitem{frattini2017_three}
N.~E. Frattini, U.~Vool, S.~Shankar, A.~Narla, K.~M. Sliwa, and M.~H. Devoret.
\newblock 3-wave mixing josephson dipole element.
\newblock {\em Applied Physics Letters}, 110(22):222603, 2017.

\bibitem{vrajitoarea2019_quantum}
Andrei Vrajitoarea, Ziwen Huang, Peter Groszkowski, Jens Koch, and Andrew~A.
  Houck.
\newblock Quantum control of an oscillator using a stimulated josephson
  nonlinearity.
\newblock {\em Nature Physics}, 16(2):211--217, 2019.

\bibitem{puri2019_stablized}
Shruti Puri, Alexander Grimm, Philippe Campagne-Ibarcq, Alec Eickbusch,
  Kyungjoo Noh, Gabrielle Roberts, Liang Jiang, Mazyar Mirrahimi, Michel~H.
  Devoret, and S.~ M Girvin.
\newblock Stabilized cat in a driven nonlinear cavity: A fault-tolerant error
  syndrome detector.
\newblock {\em Physical Review X}, 9(4), 2019.

\bibitem{pfaff2017_controlled}
W~Pfaff, C.~J. Axline, L.~D. Burkhart, U~Vool, P~Reinhold, L~Frunzio, L~Jiang,
  M.~H. Devoret, and R.~J. Schoelkopf.
\newblock Controlled release of multiphoton quantum states from a microwave
  cavity memory.
\newblock {\em Nature Physics}, 2017.

\bibitem{cirac1997}
J.~I. Cirac, P.~Zoller, H.~J. Kimble, and H.~Mabuchi.
\newblock Quantum state transfer and entanglement distribution among distant
  nodes in a quantum network.
\newblock {\em Physical Review Letters}, 78(16):3221--3224, 1997.

\bibitem{kurpiers2018_deterministic}
P.~Kurpiers, P.~Magnard, T.~Walter, B.~Royer, M.~Pechal, J.~Heinsoo,
  Y.~Salathe, A.~Akin, S.~Storz, J.~C. Besse, S.~Gasparinetti, A.~Blais, and
  A.~Wallraff.
\newblock Deterministic quantum state transfer and remote entanglement using
  microwave photons.
\newblock {\em Nature}, 558(7709):264--267, 2018.

\bibitem{burkhart2020_error-detected}
Luke~D. Burkhart, James~D. Teoh, Yaxing Zhang, Christopher~J. Axline, Luigi
  Frunzio, M.~H. Devoret, Liang Jiang, S.~M. Girvin, and R.~J. Schoelkopf.
\newblock Error-detected state transfer and entanglement in a superconducting
  quantum network.
\newblock {\em PRX Quantum}, 2(3), 2021.

\bibitem{bruzewicz2019_trapped-ion}
Colin~D. Bruzewicz, John Chiaverini, Robert McConnell, and Jeremy~M. Sage.
\newblock Trapped-ion quantum computing: Progress and challenges.
\newblock {\em Applied Physics Reviews}, 6(2), 2019.

\bibitem{brown2016_co-designing}
Kenneth~R. Brown, Jungsang Kim, and Christopher Monroe.
\newblock Co-designing a scalable quantum computer with trapped atomic ions.
\newblock {\em npj Quantum Information}, 2(1), 2016.

\bibitem{wang2019_integrated}
Jianwei Wang, Fabio Sciarrino, Anthony Laing, and Mark~G. Thompson.
\newblock Integrated photonic quantum technologies.
\newblock {\em Nature Photonics}, 14(5):273--284, 2019.

\bibitem{slussarenko2019_photonic}
Sergei Slussarenko and Geoff~J. Pryde.
\newblock Photonic quantum information processing: A concise review.
\newblock {\em Applied Physics Reviews}, 6(4), 2019.

\bibitem{saffman2019_quantum}
Mark Saffman.
\newblock Quantum computing with neutral atoms.
\newblock {\em National Science Review}, 6(1):24--25, 2019.

\bibitem{vandersypen2019_quantum}
Lieven M.~K. Vandersypen and Mark~A. Eriksson.
\newblock Quantum computing with semiconductor spins.
\newblock {\em Physics Today}, 72(8):38--45, 2019.

\bibitem{kloeffel2013_prospects}
Christoph Kloeffel and Daniel Loss.
\newblock Prospects for spin-based quantum computing in quantum dots.
\newblock {\em Annual Review of Condensed Matter Physics}, 4(1):51--81, 2013.

\bibitem{google2020_hartree}
A.~I.~Quantum Google and Collaborators.
\newblock Hartree-fock on a superconducting qubit quantum computer.
\newblock {\em Science}, 369(6507):1084--1089, 2020.

\bibitem{Kandala17}
A.~Kandala, A.~Mezzacapo, K.~Temme, M.~Takita, M.~Brink, J.~M. Chow, and J.~M.
  Gambetta.
\newblock Hardware-efficient variational quantum eigensolver for small
  molecules and quantum magnets.
\newblock {\em Nature}, 549:242--246, 2017.

\end{thebibliography}

\end{document}